\documentclass[10pt,letterpaper]{article}
\usepackage[top=0.85in,left=2.75in,footskip=0.75in]{geometry}

\usepackage{amsmath,amssymb}

\usepackage{changepage}

\usepackage{textcomp,marvosym}

\usepackage{cite}

\usepackage{nameref,hyperref}

\usepackage[right]{lineno}

\usepackage[nopatch=eqnum]{microtype}
\DisableLigatures[f]{encoding = *, family = * }

\usepackage[table]{xcolor}

\usepackage{array}

\newcolumntype{+}{!{\vrule width 2pt}}

\newlength\savedwidth

\raggedright
\usepackage[aboveskip=1pt,labelfont=bf,labelsep=period,justification=raggedright,singlelinecheck=off]{caption}
\renewcommand{\figurename}{Fig}

\makeatletter
\renewcommand{\@biblabel}[1]{\quad#1.}
\makeatother

\usepackage{lastpage,fancyhdr,graphicx}
\usepackage{epstopdf}
\fancyheadoffset[L]{2.25in}
\fancyfootoffset[L]{2.25in}
\usepackage{multirow}

\begin{document}
\vspace*{0.2in}

\begin{flushleft}
{\Large
\textbf\newline{Vulnerable connectivity caused by local communities in spatial networks} % Please use "sentence case" for title and headings (capitalize only the first word in a title (or heading), the first word in a subtitle (or subheading), and any proper nouns).
}
\newline
\\
Yingzhou MOU\textsuperscript{1*},
Yukio HAYASHI\textsuperscript{1},
% with the Lorem Ipsum Consortium\textsuperscript{\textpilcrow}
\\
\bigskip
\textbf{1} Japan Advanced Institute of Science and Technology, Nomi-city, Ishikawa 923-1292, Japan
\\
\bigskip

% Insert additional author notes using the symbols described below. Insert symbol callouts after author names as necessary.
% 
% Remove or comment out the author notes below if they aren't used.
%
% Primary Equal Contribution Note
% \Yinyang These authors contributed equally to this work.

% Additional Equal Contribution Note
% Also use this double-dagger symbol for special authorship notes, such as senior authorship.
% \ddag These authors also contributed equally to this work.

% \textcurrency Current Address: Dept/Program/Center, Institution Name, City, State, Country % change symbol to "\textcurrency a" if more than one current address note
% \textcurrency b Insert second current address 
% \textcurrency c Insert third current address

% \dag Deceased

% Group/Consortium Author Note
% \textpilcrow Membership list can be found in the Acknowledgments section.

% Use the asterisk to denote corresponding authorship and provide email address in note below.
* mouyingzhou@outlook.com

\end{flushleft}
\section*{Abstract}
Local communities by concentration of nodes connected with short links are widely observed in spatial networks. However, how such structure affects robustness of connectivity against malicious attacks remains unclear. This study investigates the impact of local communities on the robustness by modeling planar infrastructure networks whose node's locations are based on statistical population data. Our research reveals that the robustness is weakened by strong local communities in spatial networks. These results highlight the potential of long-distance links in mitigating the negative effects of local community on the robustness.

% Please keep the Author Summary between 150 and 200 words
% Use first person. PLOS ONE authors please skip this step. 
% Author Summary not valid for PLOS ONE submissions.   
% \section*{Author summary}
% Lorem ipsum dolor sit amet, consectetur adipiscing elit. Curabitur eget porta erat. Morbi consectetur est vel gravida pretium. Suspendisse ut dui eu ante cursus gravida non sed sem. Nullam sapien tellus, commodo id velit id, eleifend volutpat quam. Phasellus mauris velit, dapibus finibus elementum vel, pulvinar non tellus. Nunc pellentesque pretium diam, quis maximus dolor faucibus id. Nunc convallis sodales ante, ut ullamcorper est egestas vitae. Nam sit amet enim ultrices, ultrices elit pulvinar, volutpat risus.

% Use "Eq" instead of "Equation" for equation citations.
\section{Introduction}
\label{sec:introduction}
Community structure characterized by dense intra-community links and sparse inter-community bridge link is widely observed in various networks \cite{newman2006modularity}\cite{girvan2002community}, including spatial networks \cite{gastner2006spatial}\cite{guimera2005worldwide}\cite{kaluza2010complex}\cite{wan2023finding}. There are huge methods to detect communities as summarized in survey papers \cite{cherifi2019community}\cite{li2024comprehensive}. One of the most popular approaches is modularity maximization, while it has the limitations for hierarchical structure or overlapping among communities \cite{sales2007extracting}. Thus, other approaches have been proposed such as statistical inference framework that combines the stochastic block model with belief propagation \cite{Decelle2011} \cite{fortunato2010community} \cite{good2010performance}, spin-glass models that minimize an energy \cite{reichardt2006statistical}, and spectral methods \cite{nechaev2018rare} based on the non-backtracking matrix for sparse networks \cite{krzakala2013spectral}. They are applied to real-world networks including chromatin maps \cite{polovnikov2020nonbacktracking}, neural connectomes \cite{onuchin2023communities}, and cryptocurrency markets, where spectral modularity also detects core–periphery structures \cite{polovnikov2020core}. Such extensive researches exist for community detection, however, it has not yet been elucidated how community structure affects the robustness of connectivity against attacks. From only topological point of view, artificially adding community structure is significantly weaken the robustness of connectivity in scale-free (SF) networks and Erdős-Rényi (ER) random graphs \cite{shai2015critical}\cite{nguyen2021modularity}. Although it has been pointed out that \textit{in a topological structure affected by geographical constraints on linking, the connectivity is weakened by constructing local stubs with small cycles} \cite{hayashi2007improvement}, the reason for this weakness remains unclear without considering the influence of community structure.

Unlike only topological SF networks and ER graphs, spatial networks are constrained by physical and geographical factors, which give locally concentration of nodes and relatively short links between nodes. As examples of these factors, in densely populated areas, road networks require more intersections to meet traffic demands \cite{boeing2019urban}, while wireless communication networks \cite{boccaletti2006complex} need more base stations to maintain service quality. However, the detail locations of nodes or links are not open usually for security reason to protect the using as potential targeted attacks. Therefore, we simplify road and communication networks in proper abstraction. Since spatial networks embedded on the surface of Earth have many short links and tend to be planar \cite{boeing2018planarity}, we use relative neighborhood graph (RNG) \cite{toussaint1980relative} and Gabriel graph (GG) \cite{gabriel1969new} as models for road and communication networks, respectively. RNG and GG are known as planar proximity graphs with short links in computer science, and have similarities to road networks \cite{barthelemy2022spatial} or the self-organized biological transport networks such as slime mould with interactions between the growing active zone and chemogradients \cite{adamatzky2012bioevaluation}. On the other hand, node's locations should be set as real as possible. Thus, we apply statistical population data to determine node's locations. 

While robustness usually refers to the ability of a network to maintain connectivity under failures, resilience is a broader concept increasingly recognized as vital for infrastructure systems. It involves not only the capacity to resist and absorb disruptions, but also to recover and adapt afterward \cite{ganin2017resilience} \cite{dong2022editorial}. In spatial networks, resilience has been modeled using cascading failure processes \cite{buldyrev2010catastrophic}, percolation thresholds \cite{fan2018structural}, and more recently, spatial association metrics that capture inter-regional recovery dynamics \cite{wang2023spatial}. Though rich in implications, resilience-oriented approaches often rely on assumptions regarding functional loads, dependency links, or time-evolving recovery speeds. In this study, we limit our scope to static structural robustness as robustness of connectivity shaped by spatial node's locations. However, the study of resilience in our networks will be a promising direction for future work.

This paper aims to investigate how local communities (will be explained later) affect the robustness of connectivity in spatial networks. In particular, we show that spatial concentrations of nodes connected by short links lead to the emergence of local communities, which make spatial networks more vulnerable against node removals.

The remainder of this paper is organized as follows. In Section \ref{sec:methods}, we introduce proximity graphs of RNG and GG with short links, which construct spatial networks whose node's locations are based on population data. We consider three node removal strategies (1) recalculated betweenness attacks (RB) \cite{holme2002attack} by removing high betweenness centrality nodes, (2) initial degree attacks (ID) \cite{barabasi1999emergence} by removing high degree nodes, and (3) random failure (RF) \cite{albert2000error} by removing nodes randomly. We focus on RB, since betweenness centrality-based attacks are the most effective for fragmenting the largest connected component (LCC) in networks with communities, while ID and RF are typical node removals for comparing with the above damages. In Section \ref{sec:results}, we show how local communities emerge and affect the robustness of connectivity. In particular, we conclude that spatial networks with strong community structure weaken the robustness against intentional attacks and random failures.

\section{Spatial Networks with Concentration of Nodes}\label{sec:methods}
We introduce the models of road and communication networks by considering four types of node's location, which give different strengths of local communities for investigating the robustness of connectivity against node removals.

\subsection{Models of Spatial Networks}
\label{sec:spatial_networks}
Most spatial networks such as infrastructure networks can be approximated by planar networks \cite{barthelemy2018morphogenesis}. For example, drivable and walkable street networks in 50 global cities are investigated by using two planarity metrics of spatial planarity ratio $\phi$, which measures the proportion of true intersections to apparent ones, and the edge length ratio $\lambda$, which compares the mean edge lengths in planar and non-planar representations \cite{boeing2018planarity}. The study finds that cities in Europe (e.g., Florence, Paris) and China (e.g., Xi’an, Guangzhou) have higher planarity, while some U.S. cities such as Los Angeles and Dallas are less planar. However, most cases of $\phi$ and $\lambda$ are larger than 0.80. The results suggests that planar models give good approximating for urban street networks. A similar assumption can be extended to parts of backbone communication networks, because the fiber-optic cables are usually embedded under the main trunk roads. 

Although real infrastructure systems may consist of multiple interdependent layers, such as power-grids and communication networks  \cite{dedomenico2014navigability}, \cite{danziger2013interdependent}, we restrict our modeling to single-layer spatial networks. This simplification enables us to isolate the effect of community structures formed by spatial node's locations on the robustness of connectivity. Incorporating multilayer interactions would require assumptions about inter-layer dependencies, which fall beyond the scope of this structural analysis and are left for future exploration.

It has been shown that RNG can effectively approximate the spatial structure of urban road networks in both Japanese and American cities \cite{watanabe2005japan} \cite{watanabe2010usa}. This planarity results from a combination of physical constraints, cost-efficiency considerations, and the growth processes by connecting neighbor nodes in spatial networks. Since spatial networks can be approximated by planar networks, we introduce RNG and GG as models of road and communication networks, respectively, Note that RNG is a subgraph of GG and both of them are planar in proximity graphs with short links. Here, GG has been used for modeling optical or ad-hoc wireless networks, because it resembles to the structure of physical networks \cite{gabriel1969new} \cite{bose1999routing} \cite{karp2000gpsr} partially. While many proximity-based or planar graphs such as Delaunay triangulations \cite{delaunay1934sphere}, k-nearest neighbor (k-NN) graphs \cite{toussaint1980relative}, Voronoi-based constructions \cite{voronoi1908nouvelles} and others exist, we focus on the Gabriel Graph (GG) and its subgraph of Relative Neighborhood Graph (RNG). We emphasize that the obtained results will be applied to a wide class of planar proximity networks, which include power-grids, water supply, gas pipline, and railway networks on the surface of Earth. Moreover, it has been shown that RNG effectively approximates real-world urban street networks \cite{watanabe2005japan, watanabe2010usa}.\\

\begin{figure}
    \centering
    \includegraphics[width=\linewidth]{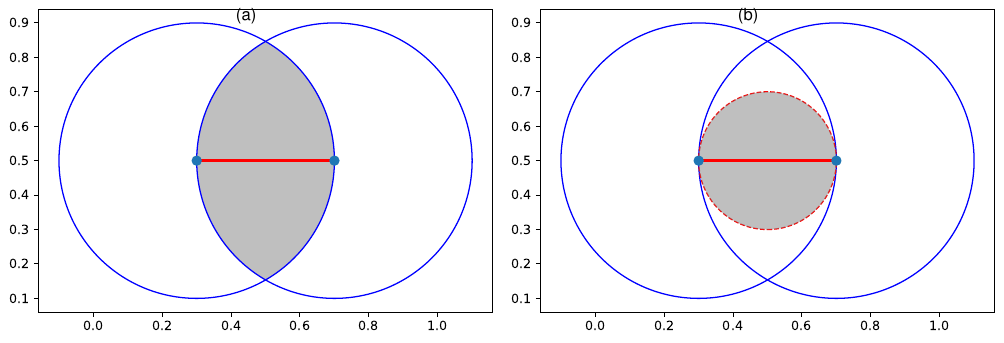}
    \caption{Illustration of connection constraints for (a) RNG and (b) GG. A link colored by red is established between two nodes colored by blue, when no other node exists within the shaded area.}
    \label{fig:rng_gg}
\end{figure}

\textbf{Fig. \ref{fig:rng_gg} RNG and GG}. Illustration of connection constraints for (a) RNG and (b) GG. A link colored by red is established between two nodes colored by blue, when no other node exists within the shaded area.\\

\begin{figure}
    \centering
    \includegraphics[width=0.5\linewidth]{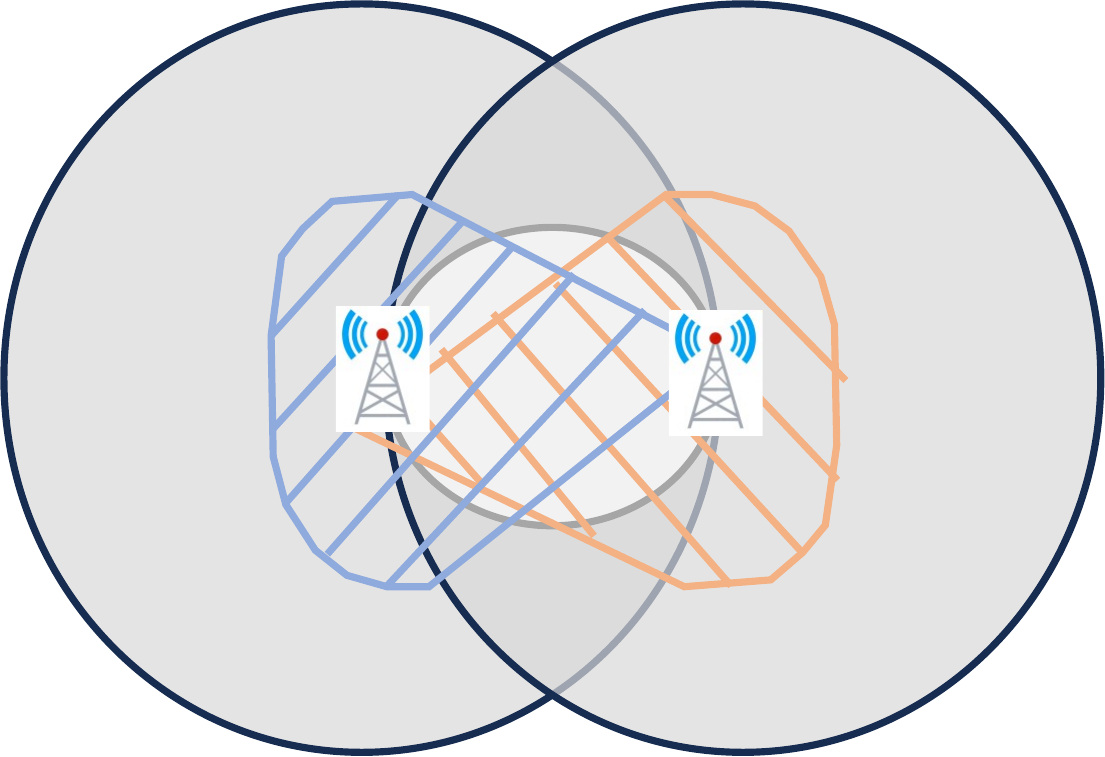}
    \caption{Coverage diagram of radio waves between two base stations in wireless communication. The ranges of strong beams are shown by blue and orange shades. The center circle represents the signal interference area. If other base stations exist within it, the two stations cannot be connected.}
    \label{fig:coverage}
\end{figure}

RNG \cite{toussaint1980relative} and GG \cite{gabriel1969new} are defined as follows. For a given set of nodes on a geographical space, two nodes are connected if no other node exists within a neighboring area of them. As shown in Figs. \ref{fig:rng_gg} (a) (b), each shaded area is the intersection of two circles centered at two nodes with a radius equal to the distance between them. In Fig. \ref{fig:coverage}, blue and orange shades represent strong area of directional beams, while the center circle represents the interference area. If other base station is located in this area with interference by directional beams \cite{andersonmodeling2011}, connecting those two nodes is prohibited. All pairwise distances are computed using Euclidean metrics.\\

\textbf{Fig. \ref{fig:coverage} Signal Coverage}. Coverage diagram of radio waves between two base stations in wireless communication. The ranges of strong beams are shown by blue and orange shades. The center circle represents the signal interference area. If other base stations exist within it, the two stations cannot be connected.\\

\subsection{Node's Locations and Removal Strategies}
\label{sec:nodes_locations}
Some studies focus on the strong correlation between population densities in cities and node's locations \cite{guimera2005worldwide} \cite{kaluza2010complex}. In road networks, each cross point can be considered as a node, areas with higher population densities tend to have more nodes \cite{boeing2019urban}. In communication networks, each base station can be considered as a node, such areas with higher population densities require more nodes, since they maintain adequate signal coverage for communications \cite{singh2015determining} \cite{onim2014optimization} \cite{ons2021}.

Taking into account the correlation between population density and node's locations, we set three types of node's locations: Population (Pop.), Inverse Population (Inv.) and Uniform (Uni.). In a $480 \times 480$ grid of $500\mathrm{m} \times 500\mathrm{m}$ block meshes, each mesh has residential population as mentioned later. For Pop. and Inv., all zero-population blocks are excluded. Here, a node is located by the center of mesh. For Pop., the nodes are located according to densely populated points ranked from highest to lowest, representing an extreme case where investments are exclusively concentrated in the most populated areas. For Inv., the nodes are located according to sparsely populated points ranked from the lowest to the highest, representing a strategic utilization of extremely sparse areas because of lower land prices for constructing networks with links by asphalt pavements or cables and node equipments such as traffic lights or communication devices. This ranking-based node selection is motivated by the observed population distribution of real cities. As shown in Supporting Information \ref{fig:population_per_cell}, when block meshes are ranked in decreasing order by population, the tails of distributions follow exponential decay, where only a small fraction of meshes contain most of the total population. Therefore, selecting the top or bottom $N$ meshes allows us to represent two statistically extreme cases of strong concentration versus wide dispersion without introducing artificial thresholds. For Uni., the nodes are located according to a uniformly at random distribution, which moderate the concentration of nodes in Pop. and Inv. as these extreme cases. In most cities, population density is uneven, there is mixing of sparse and dense node's locations in both Pop. and Inv.. The mixtures also occur in Uni. because of the spatial point process (SPP) theory \cite{keeler2016notes}. Based on the SPP theory, the number of nodes in a given area follows a Poisson distribution, therefore there are mixtures of sparse and dense node's locations. These mixtures give local community structure as shown later. By connecting nodes on the locations of Pop., Inv., and Uni., spatial networks are constructed according to the previously described constraints of short links in RNG and GG shown in Fig. \ref{fig:rng_gg}. Moreover, as completely uniform without mixtures of dense and sparse node's locations, we consider the fourth type of node's locations on a 2D square lattice (2DL) \cite{kittel2018introduction} in matching the condition under a same $P(k)$ for comparing the robustness in 2DL and Pop., Inv., or Uni. networks. According to the slightly different degree distributions $P(k)$ as shown later obtained from Pop., Inv., and Uni. networks, we randomly assign the corresponding number of links to each node on the 2DL. We assume that the other end of assigned link is free (unconnected) to perform the following two trials for linking. The first trial is established by connecting free-ends between neighbor nodes on the lattice. When free-ends are remaining but its neighbor nodes have no available free-ends, the second trial is performed for connecting more distant nodes with free-ends available. Note that the 2DL is not planar with the second trial links.

On the other hand, in practical scenarios, infrastructure networks may be exposed to various types of disruptions, such as natural disasters (e.g., earthquakes, floods), malicious attacks (e.g., cyber-attacks on communication systems, deliberate closure of key intersections), and operational failures (e.g., congestion at major road junctions or overloads in central routers). To investigate the impact of such disruptions on network connectivity, we consider three typical node removals that tend to be chosen by terrorists or to be exposed in disasters. Recalculated betweenness (RB) attacks \cite{holme2002attack} simulate targeted disruptions of nodes that carry high levels of networks flows, like busy intersections in road systems or central equipments in communication infrastructures. Here, for a node $v$, the betweenness centrality $b(v)$ \cite{freeman1977set} is defined as:\\
\[b(v) = \sum_{s \neq v \neq t} \frac{\sigma_{st}(v)}{\sigma_{st}}, \]
where $\sigma_{st}$ is the total number of shortest paths between nodes $s$ and $t$, and $\sigma_{st}(v)$ is the number of those paths that pass through $v$. Initial degree (ID) attacks \cite{barabasi1999emergence} represent failures at nodes with a large number of direct connections, which are more likely to become congested because of their high visibility, like urban traffic hubs or widely connected access routers. Random failure (RF) attacks \cite{albert2000error} model unintentional and spatially uncorrelated disruptions, such as equipment malfunctions, local construction damage, or accidental shutdowns that may occur anywhere in the network without targeting specific node properties. In these strategies, nodes (and links connected to them) are removed one by one until $q/N$ nodes are removed, where $q$ and $N$ denote the fraction of removed nodes and network size, respectively. Removed nodes are selected in decreasing order of degrees for ID attacks or betweenness centrality for RB attacks. Note that in RB attacks, betweenness centrality is recalculated after each node removal, while ID attacks select nodes according to the initial degree. Besides, RF is considered to enhance the difference between intentional attacks and random failures. We emphasize that RB has a strong affect on global fragmentation in networks with communities.

Other attacks, such as cascading failures and localized attacks (LA), are not considered in this study. Cascading failures simulate overload redistribution and recursive node collapses, commonly used to model flow-based infrastructure disruptions \cite{buldyrev2010catastrophic} \cite{xiang2024ecological} \cite{Li2016resilience}. Localized attacks by target nodes within a confined spatial region and may induce system-wide collapse when the affected area exceeds a critical radius. It is shown that LA on spatially embedded networks with dependencies can trigger cascading-like propagation, leading to abrupt global fragmentation even when the initial damage is spatially limited \cite{berezin2015localized}. While these approaches provide valuable insights into failure dynamics, they require further assumptions about flow, thresholds, or spatial radius. Given our focus on structural vulnerability shaped by node placement and short-range connectivity, we restrict our analysis to static removal strategies and leave these dynamic or spatially constrained models for future work.

\section{Effects of Community Structure on the Robustness}
\label{sec:results}
This section numerically shows important results for the robustness in the spatial networks. In subsection \ref{sec:emerging_communities}, we show the emergence of local communities in RNG and GG as modeling road and communication networks whose node's locations are based on Pop., Inv., and Uni.. In subsections \ref{sec:robustness_attack} and \ref{sec:rf_id_attacks}, we show that spatial concentrations of nodes connected by short links make local communities, which affect the robustness of connectivity against both intentional attacks and random failures.

\subsection{Emergence of Local Communities}
\label{sec:emerging_communities}
\begin{figure}
    \centering
   \includegraphics[width=\linewidth]{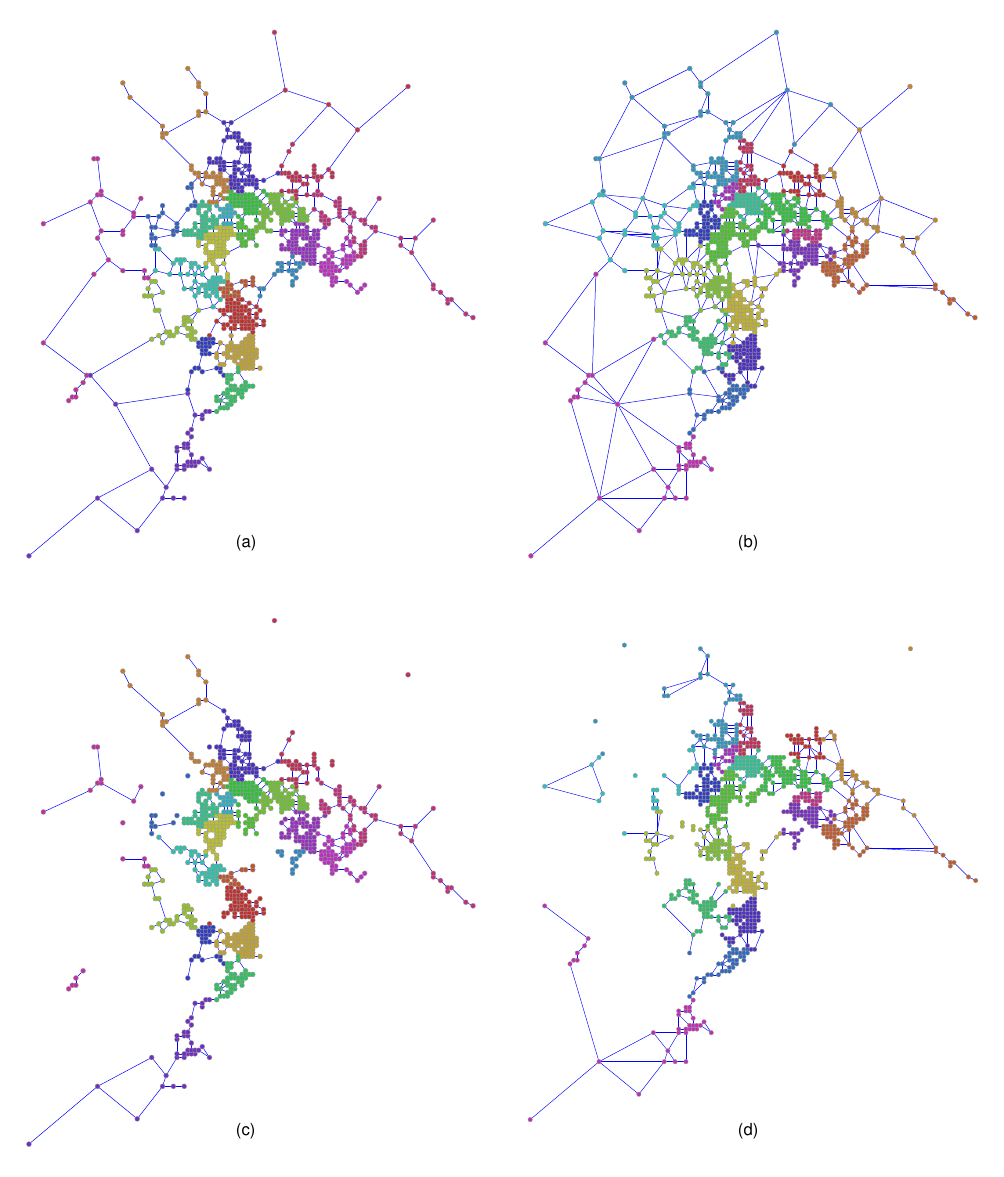}
   \caption{Visualization of community structure in (left) RNG and (right) GG (a)(b) as the original and (c)(d) after 10\% nodes removal for Tokyo with $N = 1024$ nodes located decreasing order of population (Pop.). Note that RNG is a subgraph of GG. Different communities detected by the Louvain method are colored by red, yellow, green, and blue. Each figure depicts a spatial network on a two-dimensional Euclidean plane, where node's locations correspond to the centroid locations of 500m × 500m population mesh blocks. The horizontal and vertical axes represent planar distances derived from these centroid positions.}
    \label{fig:ib_v_pop_tokyo}
\end{figure}

As real population data, the seven Japanese areas used in our study are Fukuoka, Hiroshima, Kyoto-Osaka (referred to as Keihan), Nagoya, Tokyo, Sendai, and Sapporo \cite{japan2010census}. Each of them is divided into $500m$ × $500m$ block meshes with totaling $(8$ × $10$ × $2$ × $3)^2 = 230,400$ blocks. We adopt $500\mathrm{m} \times 500\mathrm{m}$ meshes because this is the finest spatial resolution available in dataset delivered from the Japanese census. Nodes are located at the centers of meshes. We select $N$ nodes whose locations are based on Pop., Inv. and Uni., respectively, as mentioned in subsection \ref{sec:nodes_locations} previously. To ensure that $\sqrt{N}$ is an integer when constructing a 2DL for later comparison, we set $N=100,1024$ and $10000$. Links between nodes are created by connection rules of RNG and GG as shown in Fig. \ref{fig:rng_gg}. Local communities emerge in such spatial networks. In this paper, we refer to local communities as topological modules estimated by community detection methods. For example, Fig. \ref{fig:ib_v_pop_tokyo}a shows the detected community of Tokyo, whose links are constructed using RNG. The blue-colored community covers the area around Chiyoda and Chūō wards with the imperial palace and central government institutions as forming the central administrative and political core in Japan. The green-colored community extends southward into Minato and Shinagawa wards known for their high-end commercial zones and international transit hubs including Shimbashi and Shinagawa stations in Tokyo city. This pattern shows that local communities estimated in proximity-based spatial networks reflect meaningful urban divisions, which may be not shaped exactly by administrative orders but by population distribution and geographic connectivity. To measure the strength of community structure, we introduce modularity $Q$ \cite{newman2004finding} defined as:\\
\[
Q = \frac{1}{2M} \sum_{ij} \left[ A_{ij} - \frac{k_i k_j}{2M} \right] \delta(c_i, c_j),
\]\\
where $A_{ij}$ denotes the $i$, $j$ element of adjacency matrix: \( A_{ij} = 1\) if there is a link between nodes $i$ and $j$, 0 otherwise. \( k_i \) and \( k_j \) are the degrees of them, respectively. $M$ is the total number of links, $c_i$ represents the community identifier to which node $i$ belongs, \( \delta(c_i, c_j) \) is an indicator function equal to 1 if nodes \( i \) and \( j \) belong to a same community, and 0 otherwise. We estimate \(c_i\) by using the popular Louvain method \cite{blondel2008fast} \cite{norton2018detecting} \cite{polovnikov2020core}, which maximizes the modularity $Q$ using only node degrees as local information. While modularity $Q$ provides a well-defined approach to measure the strength of community, it is not suitable for detecting overlapping or hierarchical communities \cite{fortunato2010community}, and its maximization is known to be NP-hard \cite{brandes2008modularity}. However, our goal is not to achieve precise community detection, but to compare the relative strength of community in different spatial networks. As a complementary measure, we also investigate the sparsity of networks, which will be detailed in the following subsection.

\begin{figure}
    \centering
   \includegraphics[width=\linewidth]{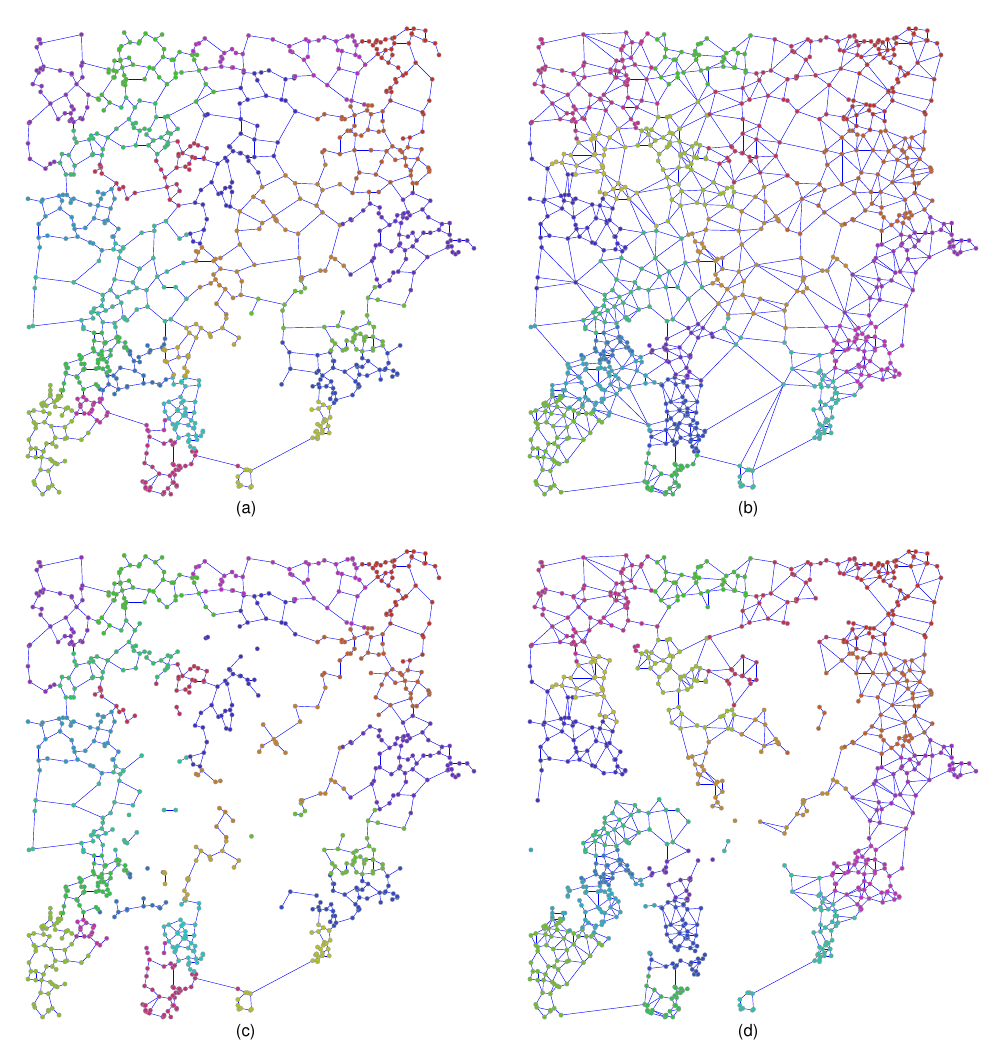}
    \caption{Visualization of community structure in (left) RNG and (right) GG (a)(b)as the original and (c)(d) after 10\% nodes removal for Tokyo with $N = 1024$ nodes located inverse decreasing order of population (Inv.). Note that RNG is a subgraph of GG. Different communities detected by the Louvain method are colored by red, yellow, green, and blue. Each figure depicts a spatial network on a two-dimensional Euclidean plane, where node's locations correspond to the centroid locations of 500m × 500m population mesh blocks. The horizontal and vertical axes represent planar distances derived from these centroid positions.}
    \label{fig:ib_v_inv_tokyo}
\end{figure}

We explain the typical results of estimated communities for Tokyo $N=1024$ node networks by Louvain method. For other cities, the results are shown in Supporting Information \ref{fig:ib_v_pop_fukuoka_1024} to \ref{fig:ib_v_inv_sapporo_1024}. Figs. \ref{fig:ib_v_pop_tokyo}ab and \ref{fig:ib_v_inv_tokyo}ab show the communities of networks constructed by using node's locations based on Pop. and Inv. with short links constrained by RNG and GG. Moreover, Figs. \ref{fig:ib_v_pop_tokyo}cd and \ref{fig:ib_v_inv_tokyo}cd show the communities of networks by using node's locations based on Pop. and Inv. after 10\% nodes (and links connected to them) removals against RB attacks. The removal of nodes causes the network to break into disconnected components, particularly at the boundaries between different communities. This fragmentation is the most noticeable in densely populated areas. For all seven Japanese areas, Table \ref{tab:q_value_combined_1024} shows that, in the original networks, both Pop.-based and Inv.-based networks have slightly higher $Q$ values compared to Uni.-based networks. Similarly, in the 2D lattice, most cases of Pop.-based and Inv.-based networks have slightly higher $Q$ values compared to Uni.-based networks, while exceptions exist in Pop-based networks that links are connected with RNG, where $Q$ values are slightly lower than those of Uni.-based networks (also see community number in \ref{tab:community_number_combined_1024}). A similar ordering of $Q$ values, where Pop.-based and Inv.-based networks higher than Uni.-based networks, is also observed for other networks sizes $N$ as shown in Supporting Information \ref{tab:q_value_100} and \ref{tab:q_value_10000}. Differences in $Q$ of Pop.-based, Inv.-based, and Uni.-based networks suggest that Pop.-based and Inv.-based networks usually have slightly stronger local community than Uni.-based networks, which resulting from the spatial concentrations of nodes. Although Uni.-based networks have slightly lower $Q$ values than that of Pop.-based and Inv.-based networks, they still maintain community structure obviously. Remember that from the SPP theory in subsection \ref{sec:nodes_locations}, there is a mixture of sparse and dense node's locations, and contributes to the emergence of community structure even in Uni.-based networks. Note that modularity $Q$ of Pop.-based networks relocated on a 2D lattice is smaller than that of Inv.-based and Uni.-based networks in RNG. While spatial density of node's locations in Pop.-based, Inv.-based, and Uni.-based networks are weaken in the corresponding 2D lattice networks under the same degree distributions, respectively. We will explain how degree distribution affects the robustness of connectivity later.\\

\textbf{Fig.~\ref{fig:ib_v_pop_tokyo} Community Structure of Tokyo Pop.-based Network}. Visualization of community structure in (left) RNG and (right) GG (a)(b) as the original and (c)(d) after 10\% nodes removal for Tokyo with $N = 1024$ nodes located decreasing order of population (Pop.). Note that RNG is a subgraph of GG. Different communities detected by the Louvain method are colored by red, yellow, green, and blue. Each figure depicts a spatial network on a two-dimensional Euclidean plane, where node's locations correspond to the centroid locations of 500m × 500m population mesh blocks. The horizontal and vertical axes represent planar distances derived from these centroid positions.\\

\textbf{Fig.~\ref{fig:ib_v_inv_tokyo} Community Structure of Tokyo Inv.-based Network}. Visualization of community structure in (left) RNG and (right) GG (a)(b)as the original and (c)(d) after 10\% nodes removal for Tokyo with $N = 1024$ nodes located inverse decreasing order of population (Inv.). Note that RNG is a subgraph of GG. Different communities detected by the Louvain method are colored by red, yellow, green, and blue. Each figure depicts a spatial network on a two-dimensional Euclidean plane, where node's locations correspond to the centroid locations of 500m × 500m population mesh blocks. The horizontal and vertical axes represent planar distances derived from these centroid positions.\\

\begin{table}[!ht]
\centering
\caption{Modularity $Q$ in networks with 1024 nodes for seven major Japanese areas under the original and 2D lattice networks}
\resizebox{\textwidth}{!}{  % 缩放宽度到页面宽度，高度自动缩放
\begin{tabular}{|l|llll|llll|}
\hline
\multirow{2}{*}{Cities} & \multicolumn{4}{c|}{Original} & \multicolumn{4}{c|}{2D lattice} \\
& \multicolumn{2}{c}{RNG} & \multicolumn{2}{c|}{GG} & \multicolumn{2}{c}{RNG} & \multicolumn{2}{c|}{GG} \\
& Inv. & Pop. & Inv. & Pop. & Inv. & Pop. & Inv. & Pop. \\
\hline
Fukuoka   & 0.8905$^\triangle$ & 0.8953$^\triangle$ & 0.8728$^\triangle$ & 0.8857$^\triangle$ & 0.8726$^\triangle$ & \textbf{0.8298} & 0.7677$^\triangle$ & 0.8269$^\triangle$ \\
Hiroshima & 0.8866$^\triangle$ & 0.8993$^\triangle$ & 0.8717$^\triangle$ & 0.8898$^\triangle$ & 0.8704$^\triangle$ & \textbf{0.8332} & 0.7662$^\triangle$ & 0.8203$^\triangle$ \\
Keihan    & 0.8903$^\triangle$ & 0.8924$^\triangle$ & 0.8651$^\triangle$ & 0.8777$^\triangle$ & 0.8693$^\triangle$ & \textbf{0.8302} & 0.7756$^\triangle$ & 0.8149$^\triangle$ \\
Nagoya    & 0.8880$^\triangle$ & 0.8929$^\triangle$ & 0.8612$^\triangle$ & 0.8764$^\triangle$ & 0.8807$^\triangle$ & \textbf{0.8408} & 0.7744$^\triangle$ & 0.8158$^\triangle$ \\
Tokyo     & 0.8846$^\triangle$ & 0.8814$^\triangle$ & 0.8615$^\triangle$ & 0.8610$^\triangle$ & \textbf{0.8644} & \textbf{0.8301} & 0.7548$^\triangle$ & 0.8177$^\triangle$ \\
Sendai    & 0.8853$^\triangle$ & 0.8908$^\triangle$ & 0.8612$^\triangle$ & 0.8801$^\triangle$ & 0.8653$^\triangle$ & \textbf{0.8387} & 0.7648$^\triangle$ & 0.8184$^\triangle$ \\
Sapporo   & 0.9017$^\triangle$ & 0.8775$^\triangle$ & 0.8875$^\triangle$ & 0.8722$^\triangle$ & 0.8745$^\triangle$ & \textbf{0.8187} & 0.7863$^\triangle$ & 0.8144$^\triangle$ \\
\hline
Uniform & \multicolumn{2}{c}{0.8735} & \multicolumn{2}{c|}{0.8557} & \multicolumn{2}{c}{0.8649} & \multicolumn{2}{c|}{0.7450} \\
\hline
\end{tabular}
}
\begin{flushleft}
Modularity $Q$ in networks before (original) and after node relocation on a 2D lattice (relocated). Higher values indicate stronger community structure. Values with upper-triangles ($\triangle$) indicate that Pop.-based or Inv.-based networks have higher modularity than Uni.-based networks in both RNG and GG. Bold values indicate networks have lower modularity than that of Uni.-based networks.
\end{flushleft}
\label{tab:q_value_combined_1024}
\end{table}

\subsection{Robustness of Connectivity Against Recalculated Betweenness Attacks}
\label{sec:robustness_attack}
Robustness of connectivity plays a crucial role in complex networks for quantifying a network's ability to maintain the functionality, when subjected to intentional attacks or random failures such as RB, ID and RF mentioned in subsection \ref{sec:nodes_locations}. As the measures, we use the robustness index $R$ \cite{schneider2011mitigation} and the critical fraction $q_c$ \cite{callaway2000network}. Here, the robustness index $R$ is the area under the line of the LCC's relative size $S^{1st}(q)/N$ against the fraction $q$ of removed nodes. A higher $R$ value indicates greater overall robustness of connectivity. The critical fraction $q_c$ is the value of $q$ at which the relative size $S^{2nd}(q)/N$ of the second LCC reaches its peak. It captures the critical point of structural change in a network. A higher $q_c$ value suggests that the network can withstand more removals of nodes before substantial fragmentation. However, $R$ is a more precise measure because different $R$ values are possible for rapidly and gradually decreasing lines of $S^{1st}(q)/N$ even with a same $q_c$. The combination of these two measures provides us with ways to observe not only how tenaciously a network maintains its connectivity against attacks but also when it reaches its breaking point.

\begin{figure}
    \centering
    \includegraphics[width=\linewidth]{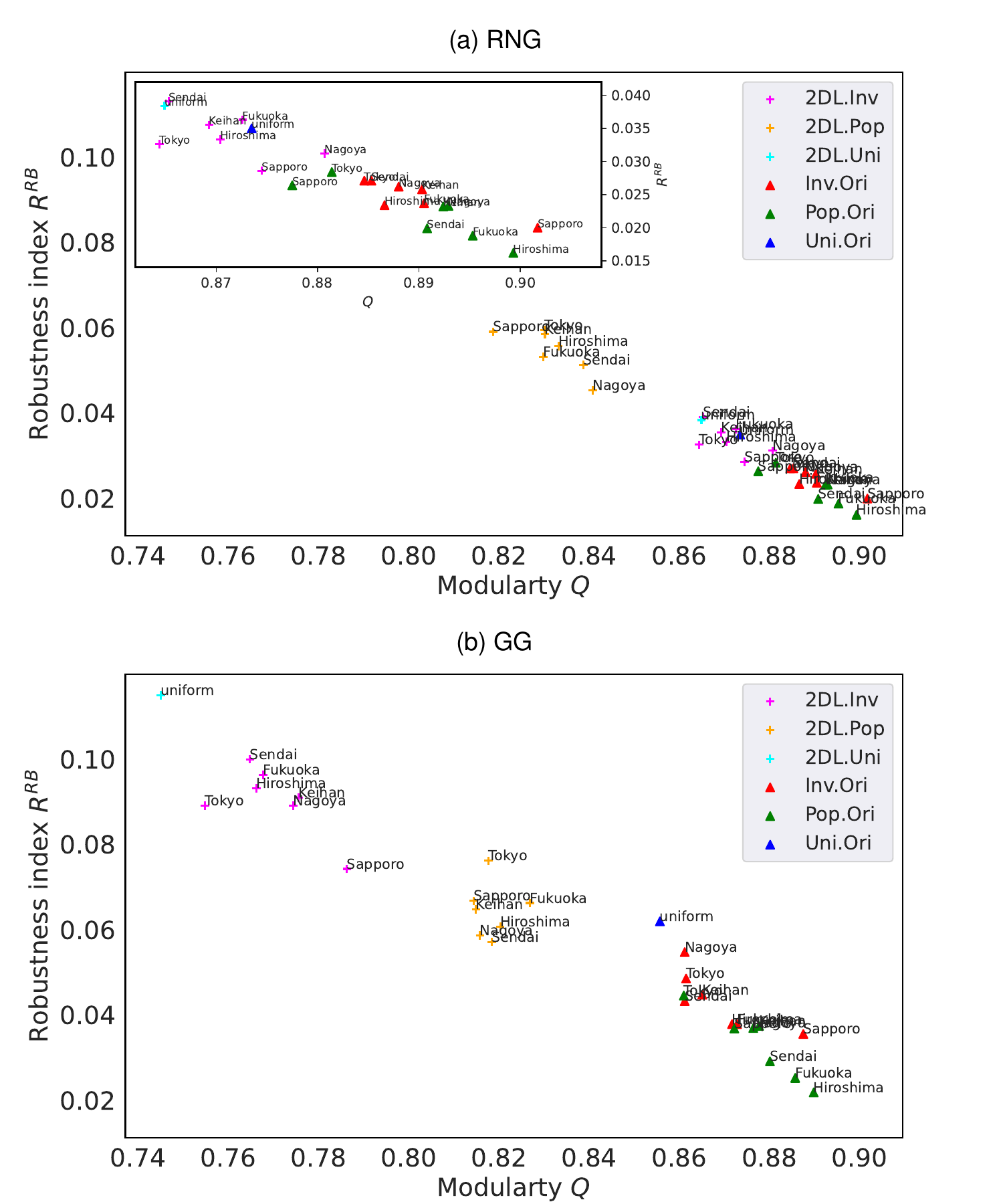}
    \caption{Relation between modularity $Q$ and robustness index $R^{RB}$ in networks with $N = 1024$ nodes of 7 Japanese areas against RB attacks. Each point represents the values of $R^{RB}$ and $Q$ for the original (triangles) and 2D lattice (crosses) networks with node's locations based on Pop. (green/yellow), Inv. (red/magenta), and Uni. (blue/cyan). Links are created by RNG (a) or GG (b). Inset in (a) enlarges the lower-right area with densely points. A monotone decreasing is observed: networks with stronger communities (higher $Q$) tend to have lower robustness index $R^{RB}$. Similar relations are observed in RNG, while the ranges of $Q$ and $R^{RB}$ differ slightly.
    }
    \label{fig:r_to_q_1024}
\end{figure}

\begin{figure}
    \centering
    \includegraphics[width=\linewidth]{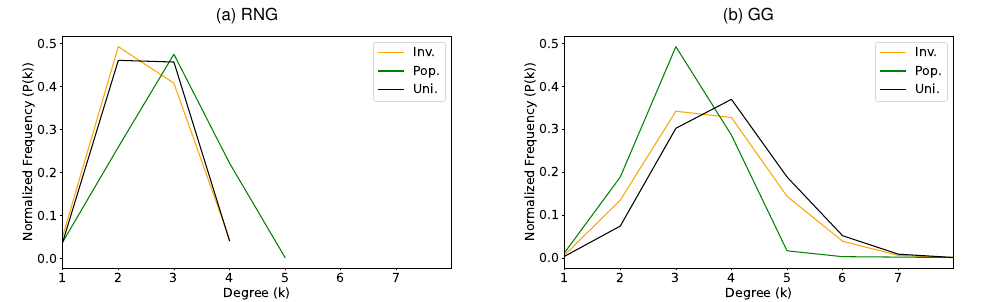}
    \caption{Degree distributions $P(k)$ in the original Tokyo networks with $N=1024$ and with node's locations based on Pop., Inv., and Uni.. All cases have bell-shaped forms with peaks around $k=2$ to $4$.
    }
    \label{fig:degree_1024_tokyo}
\end{figure}

To understand the effect of community structure on robustness of connectivity, we compare the original networks (triangles) with the 2D lattice networks (crosses), which relocate the original networks' nodes on a 2D lattice grid by removing spatial concentrations of nodes but maintaining the degree distributions for node's locations based on Pop. (green and yellow), Inv. (red and magenta), and Uni. (blue and cyan) in 7 Japanese areas. Fig.~\ref{fig:r_to_q_1024} shows clear monotone decreasing between modularity $Q$ and robustness index $R^{RB}$ against RB attacks in these networks. In both GG and RNG, the original Pop.-based and Inv.-based networks show higher $Q$ and lower $R^{RB}$ than those in Uni.-based networks by comparing triangle marks with three colors (green, red, and blue). Such results are caused by that the difference of Pop., Inv., and Uni. is the spatial concentration of nodes under the similar degree distributions $P(k)$ as shown in Fig. \ref{fig:degree_1024_tokyo}. Moreover, by comparing cross and triangle marks for each of the node's locations based on Pop., Inv., and Uni., respectively, in the original and 2D lattice networks under the same degree distributions, we show that 2D lattice networks (cross marks with yellow, magenta, and cyan) have lower $Q$ and higher $R^{RB}$ than the original networks (triangle marks with green, red, and blue) because community in 2D lattice networks have been weakened by relocating nodes on a 2D lattice with longer rewiring links, thus improve the robustness index $R^{RB}$.  We note that small variations for 7 areas in each of the original and 2D lattice networks with the node's locations based on Pop. and Inv. because of slightly different $P(k)$ as shown in Fig. \ref{fig:degree_1024_tokyo}. However, an exception appears in RNG, where 2D lattice Pop.-based $<$ Uni.-based $<$ Inv.-based networks for $Q$, and the relation for $R^{RB}$ has the inverse order. This exception may caused by the slight different degree distribution $P(k)$, as shown in Fig. \ref{fig:degree_1024_tokyo}. In summary, as higher $Q$ values with stronger local community, the robustness becomes more vulnerable in the spatial network because of the limited number of inter-links (especially against RB attacks whose targets are the end-nodes of such bridges between communities). Supporting Informations \ref{fig:r_to_q_100} and \ref{fig:r_to_q_10000} confirm these relations for networks with $N=100$ and $N=10000$.\\

\textbf{Fig. \ref{fig:r_to_q_1024} Modularity $Q$ and Robustness Index $R^{RB}$}. Relation between modularity $Q$ and robustness index $R^{RB}$ in networks with $N = 1024$ nodes of 7 Japanese areas against RB attacks. Each point represents the values of $R^{RB}$ and $Q$ for the original (triangles) and 2D lattice (crosses) networks with node's locations based on Pop. (green/yellow), Inv. (red/magenta), and Uni. (blue/cyan). Links are created by RNG (a) or GG (b). Inset in (a) enlarges the lower-right area with densely points. A monotone decreasing is observed: networks with stronger communities (higher $Q$) tend to have lower robustness index $R^{RB}$. Similar relations are observed in RNG, while the ranges of $Q$ and $R^{RB}$ differ slightly.\\

\textbf{Fig. \ref{fig:degree_1024_tokyo} Degree Distributions}. Degree distributions $P(k)$ in the original Tokyo networks with $N=1024$ and with node's locations based on Pop., Inv., and Uni.. All cases have bell-shaped forms with peaks around $k=2$ to $4$.\\

\begin{figure}
    \centering
    \includegraphics[width=\linewidth]{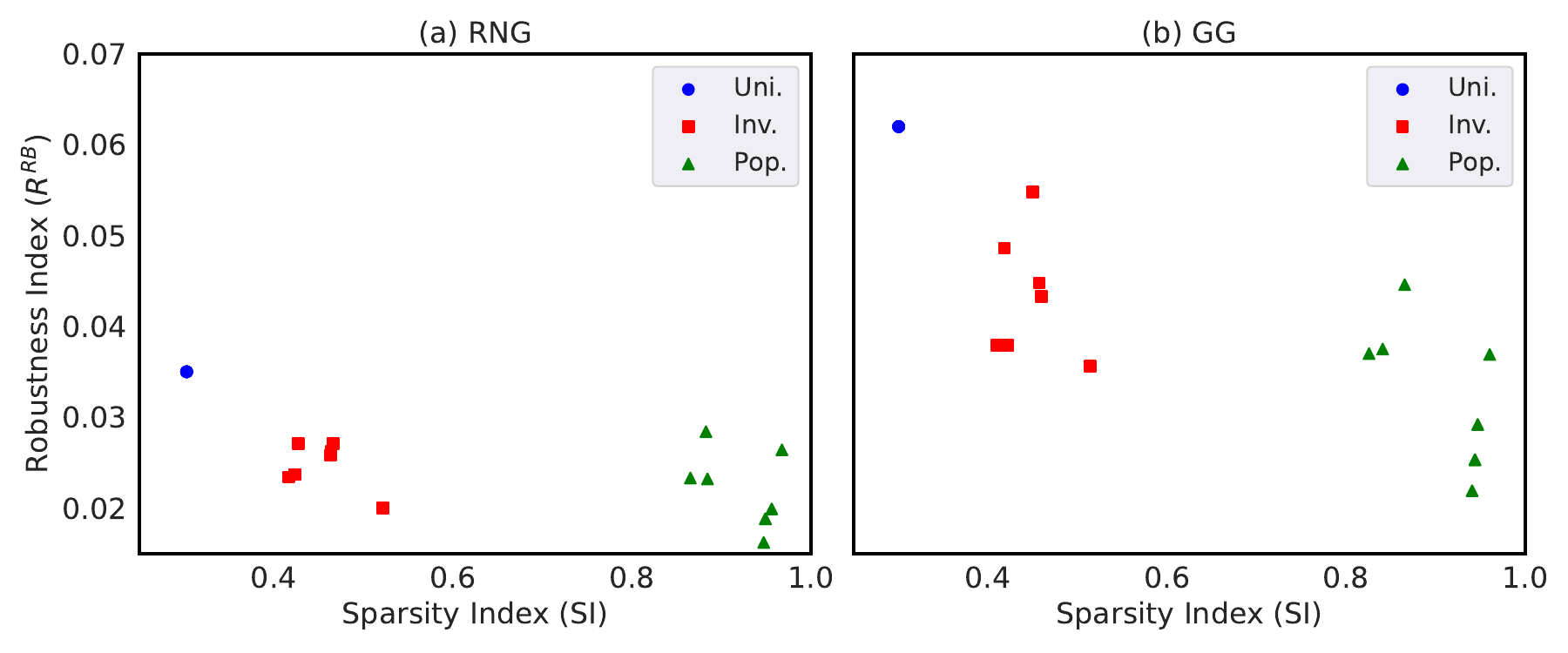}
    \caption{
    Relation between robustness index $R^{RB}$ and sparsity index $SI(G_w)$ for networks with $N=1024$ nodes in seven Japanese areas. Each point represents the results for (a) RNG and (b) GG with node's locations based on Pop. (green), Inv. (red), and Uni. (blue). Colored points show clear monotone decreasing that networks with higher sparsity index $SI(G_w)$ tend to have lower robustness index $R^{RB}$. We note that the robustness index $R^{RB}$ on the vertical axis varies within a narrow range, while the sparsity index $SI(G_w)$ on the horizontal axis spans a much wider range.
    }
    \label{fig:r_to_si_1024}
\end{figure}

For the reason that the robustness of connectivity becomes more vulnerable as stronger communities, we investigate a relation between robustness of connectivity and the following sparsity index $SI(G_w)$ \cite{goswami2018sparsity} \cite{goswami2021sparsity}. The $SI(G_w)$ is defined as follows:\\
\[
SI(G_w) = 1 - \frac{1}{N^2 T_1} \sum_{j=1}^{k} w_j f_j \left(f_j + 2f_{j+1} + 2f_{j+2} + \dots + 2f_k\right)
\]
Here, the total link weight $T_1 = \sum_{e=1}^{M'} w_e f_e$, link weight $w_e$ is defined by normalized Euclidean distances between nodes connected by a link $e$, and $f_e$ is its frequency. The parameter $M'$ denotes the number of different link weights observed in the network. The normalization factor $N^2 T_1$ ensures that $SI(G_w)$ is bounded within the range $[0, 1]$. Fig. \ref{fig:r_to_si_1024} shows that $R^{RB}$ in both RNG and GG decreases monotonically with respect to $SI(G_w)$. In other words, networks with higher spatial sparsity index $SI(G_w)$ tend to have lower robustness index $R^{RB}$. The Pearson's correlation coefficients $r<0$ for all cases against RB attacks confirmed these monotone decreasing as shown in supporting information \ref{tab:pearson_si_robustness}. The blue, red, and green points correspond to Uni.-based, Inv.-based, and Pop.-based networks in 7 Japanese areas. Note that the robustness index $R^{RB}$ on the vertical axis varies within a narrow range, while the sparsity index $SI(G_w)$ on the horizontal axis spans a much wider range. Because in Pop.-based networks, most nodes are densely concentrated in urban regions, but a few nodes are located in peripheral regions. These distant nodes connect to the urban regions by long links, which significantly increase the total link weight $T_1$. Thus, the sparsity index $SI(G_w)$ becomes larger. In Inv.-based networks, nodes are positioned more evenly than in Pop.-based networks (see Fig. \ref{fig:ib_v_pop_tokyo}ab and \ref{fig:ib_v_inv_tokyo}ab for comparison), which results in fewer long links therefore a lower $T_1$ and $SI(G_w)$. In Uni.-based networks, node's locations are the most uniformly positioned compared to Pop.-based and Inv.-based networks, yielding the lowest $T_1$ and the lowest $SI(G_w)$.\\

\textbf{Fig. \ref{fig:r_to_si_1024} Robustness Index $R^{RB}$ and Sparsity Index $SI(G_w)$}. Relation between robustness index $R^{RB}$ and sparsity index $SI(G_w)$ for networks with $N=1024$ nodes in seven Japanese areas. Each point represents the results for (a) RNG and (b) GG with node's locations based on Pop. (green), Inv. (red), and Uni. (blue). Colored points show clear monotone decreasing that networks with higher sparsity index $SI(G_w)$ tend to have lower robustness index $R^{RB}$. We note that the robustness index $R^{RB}$ on the vertical axis varies within a narrow range, while the sparsity index $SI(G_w)$ on the horizontal axis spans a much wider range.\\

\begin{figure}       
\includegraphics[width=\linewidth]{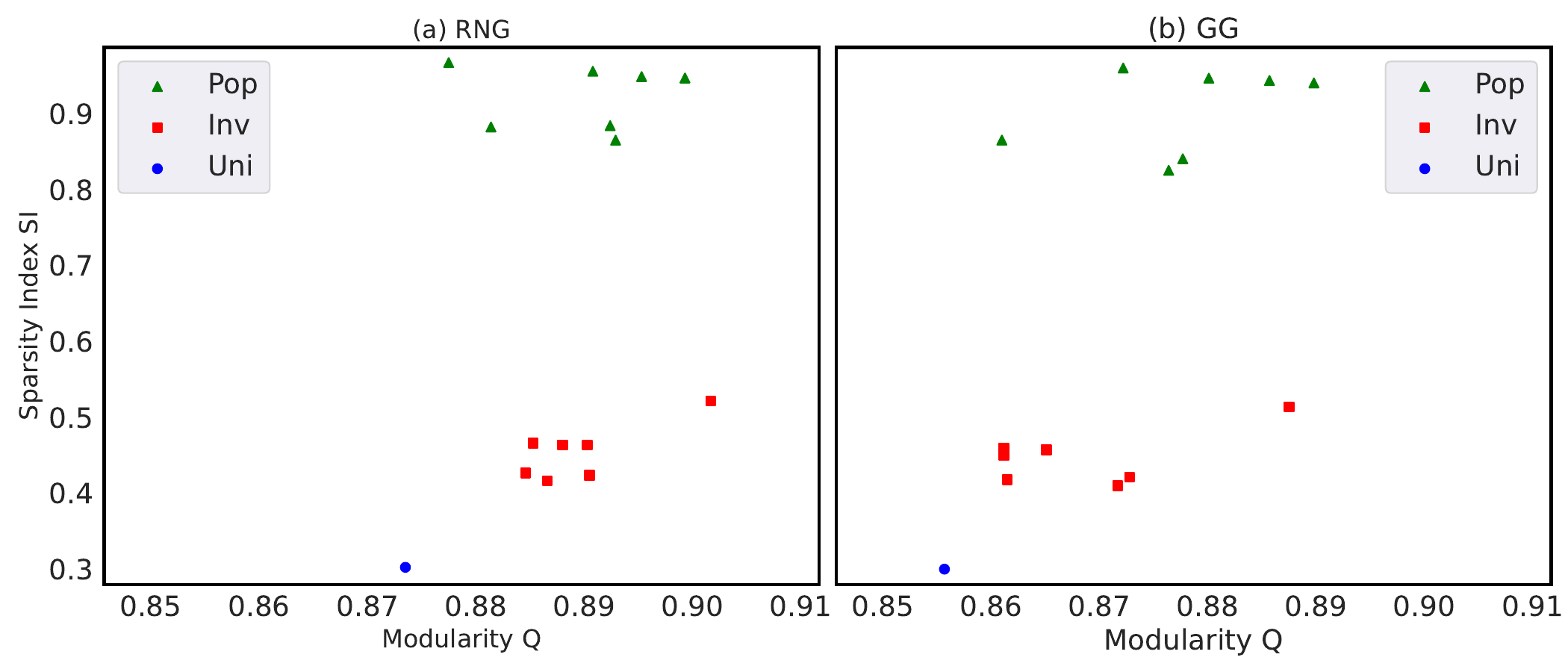}
\caption{
Relation between sparsity index $SI(G_w)$ and modularity $Q$ for the networks with $N = 1024$ nodes in seven Japanese areas. Each point represents the result for (a) RNG and (b) GG with node's locations based on Pop. (green triangles), Inv. (red squares), or Uni. (blue circle). Colored points show clear monotone increasing: networks with  higher spatial sparsity index $SI(G_w)$ tend to have higher $Q$. Pop.-based networks (green triangles) have the highest $SI$ values because of long-range link between spatially concentrated urban regions and peripheral regions, while Uni.-based (blue circle) has the lowest $SI$ because of uniformly positioned node's locations. We note that the scales of vertical and horizontal axes are quite different. In other words, the vertical differences are ignorable.
}
\label{fig:si_q_combine_1024}
\end{figure}

In Fig. \ref{fig:r_to_q_1024}, we observe that higher modularity $Q$ is associated with lower robustness index $R$. Similarly, in Fig. \ref{fig:r_to_si_1024}, we note that networks with higher spatial sparsity index $SI(G_w)$ tend to have lower robustness index $R$. These relations suggest a possible relation between sparsity index $SI(G_w)$ and modularity $Q$. Thus, we investigate the relation between $SI(G_w)$ and $Q$ as shown in Fig. \ref{fig:si_q_combine_1024} with green, red and blue points corresponding to Pop.-based, Inv.-based, and Uni.-based networks with $N=1024$ nodes in seven Japanese areas. There is clear monotone increasing between $SI(G_w)$ and $Q$ of these colored points. In other words, higher sparsity index $SI(G_w)$ is associated with higher modularity $Q$. The Pearson's correlation tests confirm the relation in both RNG and GG with the significance (see Supporting Information \ref{tab:q_si_corr} for the detail). We note that the scales of horizontal ($SI(G_w)$) and vertical ($Q$) axes are quite different. There are a large range of $SI(G_w)$ and a small range of $Q$. In particular, it is remarkable that Pop.-based $>$ Inv.-based $>$ Uni.-based networks for $SI(G_w)$, while the values of $Q$ are nearly same as high. This ranking of $SI(G_w)$ reflects the differences in node's locations, as previously discussed in the analysis of Fig. \ref{fig:r_to_si_1024}.

\textbf{Fig. \ref{fig:si_q_combine_1024} Sparsity Index ($SI(G_w)$) and Modularity $Q$}. Relation between sparsity index $SI(G_w)$ and modularity $Q$ for the networks with $N = 1024$ nodes in seven Japanese areas. Each point represents the result for (a) RNG and (b) GG with node's locations based on Pop. (green triangles), Inv. (red squares), or Uni. (blue circle). Colored points show clear monotone increasing: networks with  higher spatial sparsity index $SI(G_w)$ tend to have higher $Q$. Pop.-based networks (green triangles) have the highest $SI$ values because of long-range link between spatially concentrated urban regions and peripheral regions, while Uni.-based (blue circle) has the lowest $SI$ because of uniformly positioned node's locations. We note that the scales of vertical and horizontal axes are quite different. In other words, the vertical differences are ignorable.\\

\begin{figure}
    \centering
    \includegraphics[width=\linewidth]{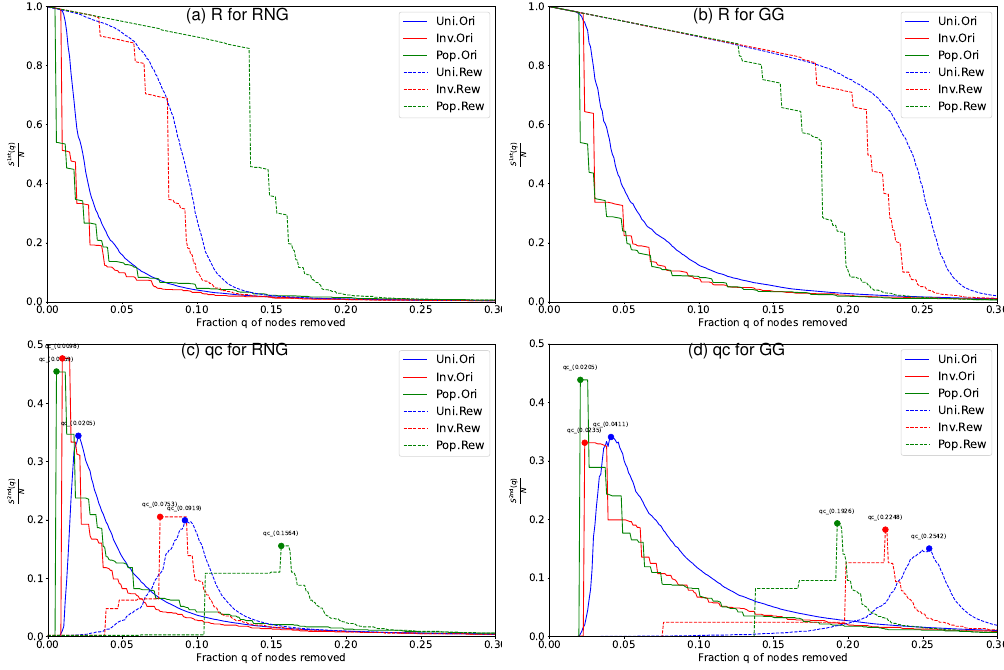}
    \caption{Robustness against recalculated betweenness (RB) attacks in Tokyo networks with $N = 1024$ nodes. For rewired (randomized networks) lines, the rewiring process preserves the original degree distributions. Two measures are applied: (a) (b) the relative size $S^{1st}(q)/N$ of largest connected component, and (c) (d) the critical fraction $q_c$ at the peak of the relative size $S^{2nd}(q)/N$ of second largest component.}
    \label{fig:tokyo_ib_index_1024_ori_rew}
\end{figure}

\begin{figure}
    \centering
    \includegraphics[width=\linewidth]{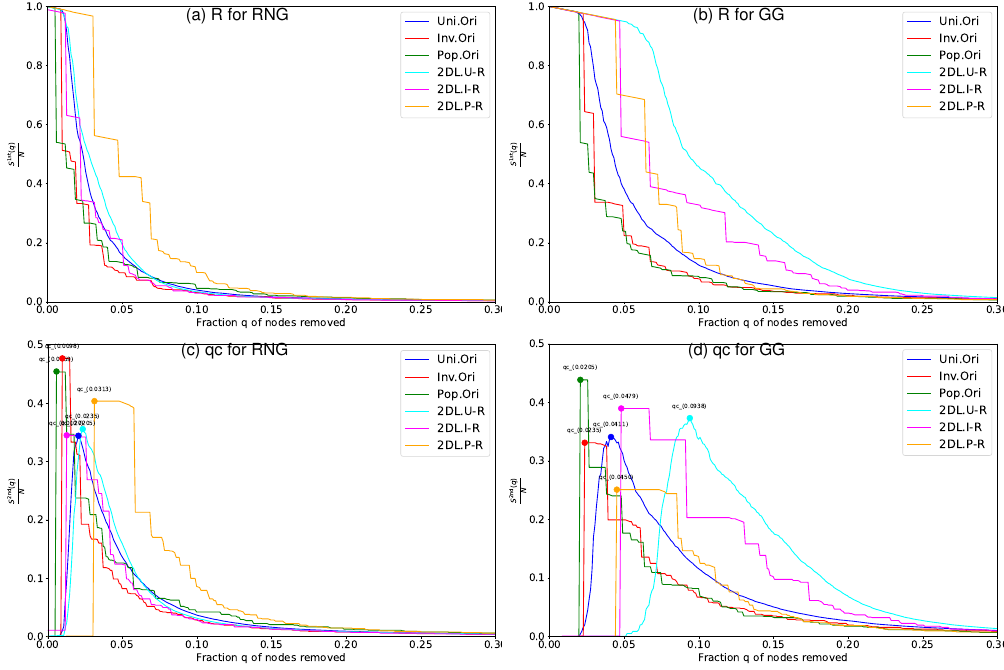}
    \caption{Robustness against recalculated betweenness (RB) attacks in Tokyo networks with $N = 1024$ nodes. For 2DL (relocated networks) lines, the rewiring process preserves the original degree distributions. Two measures are applied: (a) (b) the relative size $S^{1st}(q)/N$ of largest connected component, and (c) (d) the critical fraction $q_c$ at the peak of the relative size $S^{2nd}(q)/N$ of second largest component.}
    \label{fig:tokyo_ib_index_1024_ori_2dl}
\end{figure}

We further investigate how community structure weakens the robustness of connectivity for these networks in explaining the details for examples of Tokyo networks with $N=1024$ nodes. The results show that the spatial concentration of nodes creates strong community structure, which weaken the robustness. In Figs. \ref{fig:tokyo_ib_index_1024_ori_rew} and \ref{fig:tokyo_ib_index_1024_ori_2dl}, it is common that both randomized networks and 2DL networks show rightward shifts in their lines as improving the robustness to the original networks. When $S^{2nd}(q)/N$ reaches its peak at $q_c$, the corresponding $S^{1st}(q)/N$ line is rapidly decreased. We explain the details of comparisons in the following, where the $R$ value is represented by the area under $S^{1st}(q)/N$ line, and that the $q_c$ value is indicated by the peak of $S^{2nd}(q)/N$ line.\\

\textbf{Fig. \ref{fig:tokyo_ib_index_1024_ori_rew} Robustness Against RB for the Original and Randomized Tokyo Networks}. Robustness against recalculated betweenness (RB) attacks in Tokyo networks with $N = 1024$ nodes. For rewired (randomized networks) lines, the rewiring process preserves the original degree distributions. Two measures are applied: (a) (b) the relative size $S^{1st}(q)/N$ of largest connected component, and (c) (d) the critical fraction $q_c$ at the peak of the relative size $S^{2nd}(q)/N$ of second largest component.\\

\textbf{Fig. \ref{fig:tokyo_ib_index_1024_ori_2dl} Robustness Against RB for the Original and Relocated Tokyo Networks}. Robustness against recalculated betweenness (RB) attacks in Tokyo networks with $N = 1024$ nodes. For 2DL (relocated networks) lines, the rewiring process preserves the original degree distributions. Two measures are applied: (a) (b) the relative size $S^{1st}(q)/N$ of largest connected component, and (c) (d) the critical fraction $q_c$ at the peak of the relative size $S^{2nd}(q)/N$ of second largest component.\\

Fig. \ref{fig:tokyo_ib_index_1024_ori_rew} shows the robustness in comparing the original networks with node's locations based on Pop., Inv., and Uni. to their randomized counterparts under the same degree distributions $P(k)$. By comparing RNG (Figs. \ref{fig:tokyo_ib_index_1024_ori_rew}ac) and GG (Figs. \ref{fig:tokyo_ib_index_1024_ori_rew}bd), we confirm that all lines in GG exhibit rightward shifts compared to their corresponding lines for RNG with same colors and line types. This rightward shift indicates higher robustness for GG as shown by both larger areas under the lines of $S^{1st}(q)/N$ (higher $R$ values) and delayed peaks in lines of $S^{2nd}(q)/N$ (higher $q_c$ values).

To investigate the effect of community structure on the robustness of connectivity, we compare the original networks (represented by RGB solid lines) with their randomized counterparts (shown as RGB dashed lines) in Fig. \ref{fig:tokyo_ib_index_1024_ori_rew}a-d. As shown in Fig. \ref{fig:degree_1024_tokyo}, the original networks with three node's locations based on Pop., Inv., and Uni. have similar bell-shaped degree distributions with a peak around $k=2$ to $k=4$. These bell-shaped degree distributions contrast to Scale-Free networks with high-degree hubs by preferentially minimizing the number of hops between nodes in the networks \cite{gastner2006optimal}. However, in our study, the networks are constructed using proximity graphs such as RNG and GG, where nodes are connected to its relatively neighbors without long-range shortcuts. In planar graphs, node degrees are generally bounded and tend not to exceed six \cite[page 10]{barthelemy2018morphogenesis}. Therefore node degrees are constrained in a narrowly range, and the absence of high degree hubs produce a bell-shape degree distribution in contrast to the heavy-tailed degree distributions in Scale-Free networks. Green and red dashed lines (randomized Inv.-based networks) in Fig. \ref{fig:tokyo_ib_index_1024_ori_rew} show significant rightward shifts compared to the corresponding solid lines (the original Pop.-based, Inv.-based networks). These lines indicate that weaker community structure enhances the robustness. Similarly, blue dashed lines (randomized Uni.-based networks) show the improvement of robustness to its solid counterparts. In each of Fig. \ref{fig:tokyo_ib_index_1024_ori_rew}a-d for the original networks (RGB solid lines), both green (Pop.-based networks) and red (Inv.-based networks) solid lines exhibit leftward shifts compared to blue lines (Uni.-based networks). It indicates that Uni.-based networks have higher robustness. These comparisons consistently show that stronger community structure (higher modularity) lead to lower robustness, also as evidenced by increasing modularity $Q$ values (see the differences for the original and randomized networks in Table \ref{tab:q_value_combined_1024}). In other words, the limited number of inter-links cause the potential vulnerabilities against attacks. However, this relation becomes slightly complex in randomized networks with some exceptions: For RNG (Fig. \ref{fig:tokyo_ib_index_1024_ori_rew}a), the green dashed lines (randomized Pop.-based networks) exhibit higher robustness than the blue dashed lines (randomized Uni.-based networks), while the red dashed lines (randomized Inv.-based networks) show lower robustness. For GG (Fig. \ref{fig:tokyo_ib_index_1024_ori_rew}b), both green and red dashed lines have lower robustness than the blue dashed lines. With the exception of randomized networks in RNG, these cases suggest that a uniform positioning of nodes contributes to improved the robustness in Fig. \ref{fig:tokyo_ib_index_1024_ori_rew}ab. \\

\textbf{Fig. \ref{fig:degree_1024_tokyo} Degree Distributions}. Degree distributions $P(k)$ in the original Tokyo networks with $N=1024$ and with node's locations based on Pop., Inv., and Uni.. All cases have bell-shaped forms with peaks around $k=2$ to $4$.\\

Fig. \ref{fig:tokyo_ib_index_1024_ori_2dl} shows the robustness in comparing the original networks with node's locations based on Pop., Inv., and Uni. to their relocated counterparts under the same degree distributions $P(k)$, where nodes in these networks are relocated on a 2D lattice. These networks have three types of node's locations, Pop., Inv., and Uni., to investigate how spatial concentrations of nodes affect the robustness of connectivity. In Fig. \ref{fig:tokyo_ib_index_1024_ori_2dl}a-d, the cyan solid lines (2DL Uni.-based networks or 2DL-Uni.) show higher robustness than the blue solid lines (the original Uni.-based networks or Ori-Uni.) as evidenced by both larger area under the lines of $S^{1st}(q)/N$ (higher $R$ value) and delayed peak in the lines of $S^{2nd}(q)/N$ (higher $q_c$ value). Similar improvements of robustness are also observed, when comparing the yellow solid lines (2DL-Pop.) with the green solid lines (Ori-Pop.), and comparing the magenta solid lines (2DL-Inv.) with the red solid lines (Ori-Inv.). When comparing these networks, we observe a clear hierarchy in the robustness that correlates with the strength of community structure. As shown in Table \ref{tab:q_value_combined_1024}, the original networks show stronger community structure, while 2DL networks show weaker community structure. As exceptions in both randomized and relocated networks, Pop.-based in RNG (in bold) have lower modularity $Q$ than Uni.-based in RNG, which may be attributed to Pop.-based networks have more concentrated degree distributions than Uni.-based networks. In Figs. \ref{fig:tokyo_ib_index_1024_ori_rew} and \ref{fig:tokyo_ib_index_1024_ori_2dl}, randomized networks (RGB dashed lines) without communities have the highest robustness, followed by 2DL networks (Cyan-Yellow-Magenta, CYM solid lines), while the original networks (RGB solid lines) with strong communities have the lowest robustness. For instance, the blue dashed lines (Randomized Uni.) show to be higher robust than the cyan solid lines (2DL-Uni.), which show to be even more robust than the blue solid lines (Ori-Uni.). Similar results can be observed in the comparisons between solid-green and dashed-yellow lines (Pop.-based networks), and between solid-red and dashed-magenta lines (Inv.-based networks). These comparisons reveal that weaker community structure through more homogeneous node's locations improves the robustness as shown by the rightward shifts of CYM solid lines compared to that in the original networks. In summary, spatial networks with weaker community structure have higher robustness against RB attacks.

We also evaluate the significance of differences in the robustness index $R^{RB}$ and the critical fractions ${q_c}^{RB}$ for the original RNG and GG with three node's locations based on Pop., Inv., and Uni. by using IBM SPSS \cite{ibm_spss} statistical tools' one-way ANOVA tests \cite{fisher1925statistical} \cite{laerd_anova}. The results are shown in Supporting Information \ref{tab:rb_anova_summary_1024} which confirm the statistical significance of Uni.-based networks consistently for the superior robustness of connectivity than other networks, because of lower modularity $Q$ and lower sparsity index $SI(G_w)$ as shown in Fig. \ref{fig:si_q_combine_1024}.

\subsection{Robustness of Connectivity Against Initial Degree Attacks and Random Failures}
\label{sec:rf_id_attacks}
We further investigate the robustness of connectivity against typical Initial Degree (ID) attacks and Random Failures (RF). For Uni.-based networks, we confirm that $q_c$ values are almost identical to the analytical values \cite{norrenbrock2016fragmentation} as shown in Tables \ref{tab:combined_id_1024} and \ref{tab:combined_rf_1024}. Table \ref{tab:combined_id_1024} shows the robustness index $R$ and critical fraction $q_c$ against ID attacks, while Table \ref{tab:combined_rf_1024} shows the corresponding $R$ and $q_c$ values against RF. Note that a clear hierarchy exists as following the relation $R^{RB} < R^{ID} < R^{RF}$ and ${q_c}^{RB} < {q_c}^{ID} < {q_c}^{RF}$ (see Supporting Information \ref{tab:combined_rb_1024} for detailed comparison) in the robustness against different attack strategies.

\begin{table}[!ht]
\centering
\caption{$R^{ID}$ and $q_c^{ID}$ against $ID$ attacks in the original networks}
\resizebox{\textwidth}{!}{
\begin{tabular}{|l|ll|ll|ll|ll|}
\hline
\multirow{2}{*}{Cities} & \multicolumn{4}{c|}{$R^{ID}$} & \multicolumn{4}{c|}{$q_c^{ID}$} \\
& \multicolumn{2}{c|}{RNG} & \multicolumn{2}{c|}{GG} & \multicolumn{2}{c|}{RNG} & \multicolumn{2}{c|}{GG} \\
& Inv. & Pop. & Inv. & Pop. & Inv. & Pop. & Inv. & Pop. \\
\hline
Fukuoka & 0.1264 & 0.2541$^\triangle$ & 0.2171 & 0.2693$^\triangle$ & 0.0901 & 0.2172$^\triangle$ & 0.1832 & 0.2202 \\
Hiroshima & 0.1162 & 0.1319$^\triangle$ & 0.2332 & 0.1606 & 0.0821 & 0.0981 & 0.2032 & 0.1051 \\
Keihan & 0.1531$^\triangle$ & 0.1765$^\triangle$ & 0.1998 & 0.1758 & 0.1692$^\triangle$ & 0.0751 & 0.1712 & 0.1782 \\
Nagoya & 0.1404$^\triangle$ & 0.1943$^\triangle$ & 0.2238 & 0.2084 & 0.1552$^\triangle$ & 0.1772$^\triangle$ & 0.2392 & 0.0631 \\
Tokyo & 0.1363$^\triangle$ & 0.2191$^\triangle$ & 0.2250 & 0.2083 & 0.1361$^\triangle$ & 0.2412$^\triangle$ & 0.2362 & 0.1512 \\
Sendai & 0.1240 & 0.1801$^\triangle$ & 0.2132 & 0.1257 & 0.0591 & 0.1061 & 0.1882 & 0.0170 \\
Sapporo & 0.1055 & 0.1646$^\triangle$ & 0.1890 & 0.2042 & 0.0851 & 0.1441$^\triangle$ & 0.1301 & 0.1522 \\
\hline
Uniform & \multicolumn{2}{c|}{0.132} & \multicolumn{2}{c|}{0.247} & \multicolumn{2}{c|}{0.1351} & \multicolumn{2}{c|}{0.2643} \\
Analytical & \multicolumn{4}{c|}{--} & \multicolumn{2}{c|}{0.12} & \multicolumn{2}{c|}{0.263} \\
\hline
\end{tabular}
}
\begin{flushleft}
 Robustness index ($R^{ID}$) and Critical fraction ($q_c^{ID}$) against Initial Degree (ID) attacks in the original networks with $N = 1024$ nodes for seven major Japanese areas. Higher values indicate greater robustness. Values with upper-triangles ($\triangle$) indicate where the cases of Pop.-based and Inv.-based have stronger robustness than the case of the Uni.-based network.
\end{flushleft}
\label{tab:combined_id_1024}
\end{table}

\begin{table}[!ht]
\centering
\caption{$R^{RF}$ and $q_c^{RF}$ against random failure in the original networks}
\resizebox{\textwidth}{!}{
\begin{tabular}{|l|ll|ll|ll|ll|}
\hline
\multirow{2}{*}{Cities} & \multicolumn{4}{c|}{$R^{RF}$} & \multicolumn{4}{c|}{$q_c^{RF}$} \\
& \multicolumn{2}{c|}{RNG} & \multicolumn{2}{c|}{GG} & \multicolumn{2}{c|}{RNG} & \multicolumn{2}{c|}{GG} \\
& Inv. & Pop. & Inv. & Pop. & Inv. & Pop. & Inv. & Pop. \\
\hline
Fukuoka & 0.1573 & 0.1338 & 0.2820 & 0.1902 & 0.1191 & 0.0911 & 0.3724$^\triangle$ & 0.1682 \\
Hiroshima & 0.1476 & 0.1227 & 0.2627 & 0.1510 & 0.1832 & 0.0320 & 0.3093 & 0.0601 \\
Keihan & 0.2040$^\triangle$ & 0.2001$^\triangle$ & 0.2486 & 0.2421 & 0.1522 & 0.2382$^\triangle$ & 0.3844$^\triangle$ & 0.2773 \\
Nagoya & 0.1711 & 0.2060$^\triangle$ & 0.2752 & 0.2068 & 0.1692 & 0.0881 & 0.2653 & 0.3093 \\
Tokyo & 0.1721 & 0.2146$^\triangle$ & 0.2998 & 0.2614 & 0.1031 & 0.2873$^\triangle$ & 0.2332 & 0.2513 \\
Sendai & 0.1771 & 0.1407 & 0.2650 & 0.1806 & 0.1131 & 0.0330 & 0.2843 & 0.0541 \\
Sapporo & 0.1253 & 0.2053$^\triangle$ & 0.2126 & 0.2345 & 0.1391 & 0.0330 & 0.2302 & 0.4014 \\
\hline
Uniform & \multicolumn{2}{c|}{0.1972} & \multicolumn{2}{c|}{0.3083} & \multicolumn{2}{c|}{0.2112} & \multicolumn{2}{c|}{0.3694} \\
Analytical & \multicolumn{4}{c|}{--} & \multicolumn{2}{c|}{0.205} & \multicolumn{2}{c|}{0.365} \\
\hline
\end{tabular}
}
\begin{flushleft}
Robustness index ($R^{RF}$) and Critical fraction ($q_c^{RF}$) against random failure (RF) in the original networks with $N = 1024$ nodes for seven major Japanese areas. Higher values indicate greater robustness. Values with upper-triangles ($\triangle$) indicate where the cases of Pop.-based and Inv.-based have stronger robustness than the case of the Uni.-based network.
\end{flushleft}
\label{tab:combined_rf_1024}
\end{table}

In both ID attacks and RF, networks with node's locations based on Pop. and Inv. show lower $R$ and $q_c$ values than those to Uni.-based networks as similar to the results against RB attacks except some cases. As values with upper-triangles ($\triangle$) in Tables \ref{tab:combined_id_1024} and \ref{tab:combined_rf_1024}, some Pop.-based and Inv.-based networks have higher robustness than those on Uni.-based networks. The IBM SPSS one-way ANOVA tests \cite{ibm_spss} \cite{fisher1925statistical} \cite{laerd_anova} also confirm that most cases show the significant differences of the robustness consistently for the results in the networks constructed by the spatial node's locations against RB attacks. However, only two exceptions are observed: critical fraction ${q_c}^{ID}$ in RNG against ID attacks ($p=0.249>0.05$), and the robustness index $R^{RF}$ in RNG against RF ($p=0.104>0.05$), which are summarized in Supporting Information \ref{tab:id_anova_summary} and \ref{tab:rf_anova_summary}. The reason may be caused by two factors: (1) Higher average degree $\langle k \rangle$ of Pop.-based networks in RNG as shown in Table \ref{tab:average_degree_1024}, which may provide additional alternative paths after node removals; (2) The presence of regular grid-like parts in certain regions of Pop.-based networks contributes to enhance the robustness against RF as shown in Support Information \ref{fig:r_and_qc_to_grid_1024}. The effect of grid-like parts is explained as follows. In our study, all nodes are located limitedly at the centers of mesh blocks (500m × 500m), rather than having arbitrary positions in a geographical space freely. When nodes are densely distributed in neighboring mesh blocks, as often occurs in Pop.-based networks, they tend to form grid-like parts. We quantify these grid-like parts by calculating the ratio of nodes that have degree 4 with four neighbors of degree 4, which represents a typical local grid formation. As shown in Table \ref{tab:degree_4_1024}, Pop.-based networks show notably higher ratio of such grid-like parts (ranging from $2.05\%$ to $15.14\%$ in RNG and $2.25\%$ to $15.33\%$ in GG) compared to Inv.-based networks (mostly $0\%$ in RNG and below $1.17\%$ in GG) and Uni.-based networks ($0\%$ in RNG and $0.79\%$ in GG) as values with upper-triangles ($\triangle$). According to percolation theory \cite{stauffer1992introduction}, a 2D square lattice network maintains its connectivity until the fraction $q_c = 0.4073$ (percolation threshold $p_c = 1 - q_c = 0.5927$) of randomly removed nodes. This critical fraction $q_c = 0.4073$ is larger than the analytical values $q_c = 0.205$ in RNG and $q_c = 0.365$ in GG shown in Table \ref{tab:combined_rf_1024}. The presence of these grid-like parts increases the critical fraction than the above values in RNG and GG on free node's locations at random.  Furthermore, as similar to Section \ref{sec:robustness_attack}, we also investigate the correlation between robustness measures $R$ or $q_c$ and $SI(G_w)$ (see Supporting Information \ref{tab:pearson_si_robustness} for detail). While most cases show monotone decreasing between the robustness and the sparsity, some exceptions in RNG against both ID and RF have been observed. However, the reason of them remains unclear for a future study.

\begin{table}[!ht]
\centering
\caption{Average degree}
\begin{tabular}{|l|ll|ll|}
\hline
\multirow{2}{*}{\textbf{Cities}} & \multicolumn{2}{c|}{RNG} & \multicolumn{2}{c|}{GG} \\
& Inv. & Pop. & Inv. & Pop. \\
\hline
Fukuoka & 2.4492 & 2.8223$^\triangle$ & 3.584 & 2.9844 \\
Hiroshima & 2.4414 & 2.7852$^\triangle$ & 3.5996 & 2.9199 \\
Keihan & 2.4395 & 2.8535$^\triangle$ & 3.5469 & 3.1367 \\
Nagoya & 2.3945 & 2.8535$^\triangle$ & 3.4785 & 3.1367 \\
Tokyo & 2.4473 & 2.8906$^\triangle$ & 3.6113 & 3.1191 \\
Sendai & 2.459 & 2.834$^\triangle$ & 3.582 & 3.0195 \\
Sapporo & 2.3789 & 3.0684$^\triangle$ & 3.3105 & 3.1563 \\
\hline
Uniform & \multicolumn{2}{c|}{2.5017} & \multicolumn{2}{c|}{3.864} \\
\hline
\end{tabular}
\begin{flushleft}
Average degree $\langle k \rangle$ in networks with $N = 1024$ nodes for seven major Japanese areas. Values with upper-triangles ($\triangle$) indicate the cases of Pop.-based and Inv.-based networks show higher $\langle k \rangle$ than the case of Uni.    
\end{flushleft}
\label{tab:average_degree_1024}
\end{table}

\begin{table}[!ht]
\centering
\caption{Proportion of grid-like parts in networks}
\begin{tabular}{|l|ll|ll|}
\hline
\multirow{2}{*}{\textbf{Cities}} & \multicolumn{2}{c|}{RNG} & \multicolumn{2}{c|}{GG} \\
& Inv. & Pop. & Inv. & Pop. \\
\hline
Fukuoka & 0 & 0.0771$^\triangle$ & 0.0039 & 0.0771$^\triangle$ \\
Hiroshima & 0 & 0.0205$^\triangle$ & 0.0078 & 0.0205$^\triangle$ \\
Keihan & 0 & 0.0391$^\triangle$ & 0.0078 & 0.0479$^\triangle$ \\
Nagoya & 0 & 0.0391$^\triangle$ & 0.0088$^\triangle$ & 0.0479$^\triangle$ \\
Tokyo & 0 & 0.0205$^\triangle$ & 0.0068 & 0.0225$^\triangle$ \\
Sendai & 0 & 0.0459$^\triangle$ & 0.0117$^\triangle$ & 0.0498$^\triangle$ \\
Sapporo & 0 & 0.1514$^\triangle$ & 0.001 & 0.1533$^\triangle$ \\
\hline
Uniform & \multicolumn{2}{c|}{0} & \multicolumn{2}{c|}{0.0079} \\
Lattice & \multicolumn{4}{c|}{0.7656} \\
\hline
\end{tabular}
\begin{flushleft}
Proportion of grid-like parts in networks with $N = 1024$ nodes for seven major Japanese areas. A grid-like structure is identified when a node has degree 4 with the neighbors of degree 4. Values with upper-triangles ($\triangle$) indicate cases where the ratio is higher than that of Uni. networks. Note that 2D lattice shows the highest ratio (76.56\%) because of its completely ordered grid arrangement.
\end{flushleft}
\label{tab:degree_4_1024}
\end{table}

\section{Conclusion}
\label{sec:conclusion}
Our study investigates the impact of local community structure on the vulnerability of spatial networks with concentration of nodes. The node's locations are chosen according to the order of real population (Pop.), inverse real population (Inv.), and uniformly at random (Uni.), while the links between nodes are short by using proximity graphs of RNG and GG as modeling road and communication systems, respectively. Remember that the spatial concentrations of nodes connected by short links in Pop.-based, Inv.-based and Uni.-based networks are different for investigating the effect of strengths of community structure on the robustness. We observed the emergence of local communities in Pop.-based and Inv.-based networks connected by short links (see Figs. \ref{fig:ib_v_pop_tokyo}ab and \ref{fig:ib_v_inv_tokyo}ab), which weakens the networks by removing small number of inter-community bridge links (see Figs. \ref{fig:ib_v_pop_tokyo}cd and \ref{fig:ib_v_inv_tokyo}cd). In the details, for the measures of robustness index $R$ and critical fraction $q_c$, Pop.-based and Inv.-based networks are weaker than Uni.-based networks against intentional attacks and random failures in both RNG and GG, because Pop.-based and Inv.-based networks have stronger local communities than Uni.-based networks (see Figs. \ref{fig:tokyo_ib_index_1024_ori_rew} and \ref{fig:tokyo_ib_index_1024_ori_2dl}). In particular, both $R$ and $q_c$ decrease with increasing modularity $Q$ (see \ref{fig:yaxis_tokyo_rng_1024}, \ref{fig:yaxis_tokyo_gg_1024} and Fig. \ref{fig:r_to_q_1024}). Therefore, we conclude that the emergence of local communities weakens the robustness of connectivity in spatial networks.

For this negative impact of local communities on the robustness, it is important to understand how to reduce their effects. Local communities emerge from the spatial concentration of nodes connected by short links. While the strength of local communities could be mitigated by either distributing nodes more evenly or creating long links, the former approach may be impractical because of high land acquisition costs. Therefore, strategically establishing long links offers a more feasible solution to balance reliability and construction costs. In the context of real infrastructure systems, such long links correspond to intercity highways with overpasses and tunnels or urban elevated expressways. In the case of communication systems, they correspond to fiber-optic backbones or long-range satellite links. While implementing such links may brings high costs or land-use conflicts, it is more cost-effective than redistributing node's locations. Beyond scientific discussion, it is a judgment to prioritize either distribution of node's equipments or reinforcing fragile inter-community ties. For example, in urban planning, establishing overpasses or elevated roads between sparsely connected areas, particularly by prioritizing new highway between previously unconnected communities can increase the number of inter-community links, avoid isolation among communities and enhance the robustness of connectivity without necessitating dense local expansions. In communication infrastructure, deploying wireless bridges or satellite relays provide more cost-effective redundancy than laying fiber-optical cables. However, some interventions are necessary for such improvements.

Another important aspect to clarify is that our network modeling focuses on spatially embedded proximity graphs (RNG and GG), rather than space syntax or multilayer network approaches. While we do not aim to replicate urban street structures directly, our results are consistent to reduce the robustness \cite{shai2015critical} \cite{nguyen2021modularity}. Importantly, our model shows community structures emerge from population-based node's locations and geometric constrained linking. This offers a complementary perspective to existing models and may inform robustness-aware infrastructure planning in demographically heterogeneous regions.

However, our study has several limitations. First, while we focus on planar proximity networks of RNG and GG models of for road and communication systems, the applicability of our findings to other types of infrastructure on the surface of Earth such as power-grids or railway systems remains to be explored. Second, although we investigate the effects of three typical node removal strategies (RB, ID, and RF), more destructive attacks such as belief propagation attacks \cite{mugisha2016identifying} or collective influence attacks \cite{morone2015influence} may reveal different vulnerability patterns in spatial networks. 

Third, although the concept of resilience in infrastructure systems usually encompasses redundancy, recovery speed, and adaptive capacity \cite{ouyang2014reliability}, \cite{HENRY2012generic}, \cite{bruneau2003framework}, this study focuses solely on static topological robustness specifically the ability to maintain global connectivity against attacks. We do not consider dynamic behaviors such as cascading failures or temporal recovery processes. However, they are essential to a comprehensive resilience evaluation, and future studies may incorporate them to assess recovery dynamics under more complex disruption scenarios. 

Fourth, we do not also consider multilayer or interdependent network structures. Real-world infrastructure systems such as power-grids, communication networks, and transportation systems are often interconnected across multiple functional layers with dependencies that can significantly affect their vulnerability and recovery dynamics \cite{dedomenico2014navigability}, \cite{danziger2013interdependent}, \cite{cardillo2013modeling}. Multilayer frameworks have been shown to exhibit emergent phenomena such as abrupt phase transitions, increased percolation thresholds, and interlayer-driven cascading failures \cite{danziger2016two}, \cite{danziger2015interdependent}, which cannot be captured by single-layer models. Incorporating such features will require additional assumptions about interlayer coupling and failure propagation, which fall outside the scope of our present static and spatially embedded networks.

Finally, while we explored how sparsity influences the robustness in population-based spatial networks, the effects of sparsity in other spatial configurations from RNG or GG remain unclear. Moreover, the generalizability of our findings beyond Japanese urban areas requires further validation through comparative studies in other countries or spatial settings.

\clearpage
% Either type in your references using
% \begin{thebibliography}{}
% Text
% \end{thebibliography}
%
% or
%
% Compile your BiBTeX database using our plos2015.bst
% style file and paste the contents of your .bbl file
% here. See http://journals.plos.org/plosone/s/latex for 
\bibliography{sn-bibliography}

\clearpage
\section*{Supporting Information}

\textbf{\ref{fig:ib_v_pop_fukuoka_1024} fukuoka\_descend\_1024\_community\_2in1.pdf}: Visualization of community structures in Fukuoka before node removal. $N = 1024$ nodes are located by the decreasing order of population (Pop.). Different colors represent different communities estimated by Louvain method. There are clear community formations particularly in densely populated areas.\\
\textbf{\ref{fig:ib_v_inv_fukuoka_1024} fukuoka\_ascend\_1024\_community\_2in1.pdf}: Visualization of community structures in Fukuoka before node removal. $N = 1024$ nodes are located by the inverse order of population (Inv.). Different colors represent different communities estimated by Louvain method. There are different community formations compared to S1 Fig.\\
\textbf{\ref{fig:ib_v_pop_hiroshima_1024} hiroshima\_descend\_1024\_community\_2in1.pdf}: Visualization of community structures in Hiroshima before node removal. $N = 1024$ nodes are located by the decreasing order of population (Pop.). Different colors represent different communities estimated by Louvain method. There are clear community formations particularly in densely populated areas.\\
\textbf{\ref{fig:ib_v_inv_hiroshima_1024} hiroshima\_ascend\_1024\_community\_2in1.pdf}: Visualization of community structures in Hiroshima before node removal. $N = 1024$ nodes are located by the inverse order of population (Inv.). Different colors represent different communities estimated by Louvain method. There are different community formations compared to S3 Fig.\\
\textbf{\ref{fig:ib_v_pop_keihan_1024} keihan\_descend\_1024\_community\_2in1.pdf}: Visualization of community structures in Keihan before node removal. $N = 1024$ nodes are located by the decreasing order of population (Pop.). Different colors represent different communities estimated by Louvain method. There are clear community formations particularly in densely populated areas.\\
\textbf{\ref{fig:ib_v_inv_keihan_1024} keihan\_ascend\_1024\_community\_2in1.pdf}: Visualization of community structures in Keihan before node removal. $N = 1024$ nodes are located by the inverse order of population (Inv.). Different colors represent different communities estimated by Louvain method. There are different community formations compared to S5 Fig.\\
\textbf{\ref{fig:ib_v_pop_nagoya_1024} nagoya\_descend\_1024\_community\_2in1.pdf}: Visualization of community structures in Nagoya before node removal. $N = 1024$ nodes are located by the decreasing order of population (Pop.). Different colors represent different communities estimated by Louvain method. There are clear community formations particularly in densely populated areas.\\
\textbf{\ref{fig:ib_v_inv_nagoya_1024} nagoya\_ascend\_1024\_community\_2in1.pdf}: Visualization of community structures in Nagoya before node removal. $N = 1024$ nodes are located by the inverse order of population (Inv.). Different colors represent different communities estimated by Louvain method. There are different community formations compared to S7 Fig.\\
\textbf{\ref{fig:ib_v_pop_sendai_1024} sendai\_descend\_1024\_community\_2in1.pdf}: Visualization of community structures in Sendai before node removal. $N = 1024$ nodes are located by the decreasing order of population (Pop.). Different colors represent different communities estimated by Louvain method. There are clear community formations particularly in densely populated areas.\\
\textbf{\ref{fig:ib_v_inv_sendai_1024} sendai\_ascend\_1024\_community\_2in1.pdf}: Visualization of community structures in Sendai before node removal. $N = 1024$ nodes are located by the inverse order of population (Inv.). Different colors represent different communities estimated by Louvain method. There are different community formations compared to S9 Fig.\\
\textbf{\ref{fig:ib_v_pop_sapporo_1024} sapporo\_descend\_1024\_community\_2in1.pdf}: Visualization of community structures in Sapporo before node removal. $N = 1024$ nodes are located by the decreasing order of population (Pop.). Different colors represent different communities estimated by Louvain method. There are clear community formations particularly in densely populated areas.\\
\textbf{\ref{fig:ib_v_inv_sapporo_1024} sapporo\_ascend\_1024\_community\_2in1.pdf}: Visualization of community structures in Sapporo before node removal. $N = 1024$ nodes are located by the inverse order of population (Inv.). Different colors represent different communities estimated by Louvain method. There are different community formations compared to S11 Fig.\\
\textbf{\ref{fig:yaxis_tokyo_rng_1024} tokyo\_rng\_qrqc\_3in1.pdf}: Increasing modularity $Q$ vs. decreasing robustness index $R^{RB}$ or critical fraction ${q_c}^P{RB}$ for varying the size $N$ in Tokyo RNG networks.\\
\textbf{\ref{fig:yaxis_tokyo_gg_1024} tokyo\_gg\_qrqc\_3in1.pdf}: Increasing modularity $Q$ vs. decreasing robustness index $R^{RB}$ or critical fraction ${q_c}^P{RB}$ for varying the size $N$ in Tokyo GG networks.\\
\textbf{\ref{fig:r_and_qc_to_grid_1024} tokyo\_1024\_rnqc2grid\_2in1.pdf}: Scatter plots show relation between robustness measures (a for $R$ and b for $q_c$) and the proportion of grid-like parts against random failures (RF). Networks with N = 1024 nodes are considered, where Pop. networks (green) show notably higher proportions of grid-like parts compared to Inv. (red) and Uni. (blue) networks. See the text at the end of subsection 3.3 for the detail.\\
\textbf{\ref{fig:fukuoka_ib_index_100_3combines} Fukuoka\_100\_3combined\_4in1.pdf}: Robustness against recalculated betweenness (RB) attacks for Fukuoka networks with $N = 100$ nodes. For both Rew (Randomized networks) and 2DL lines, the rewiring process preserves the original degree distributions. Two measures are applied: (a) (b) $S^{1st}(q)/N$ the relative size of largest connected component, and (c) (d) $S^{2nd}(q)/N$ the critical fraction $q_c$ at the peak of the relative size of second largest component.\\
\textbf{\ref{fig:hiroshima_ib_index_100_3combines} Hiroshima\_100\_3combined\_4in1.pdf}: Robustness against recalculated betweenness (RB) attacks for Hiroshima networks with $N = 100$ nodes. For both Rew (Randomized networks) and 2DL lines, the rewiring process preserves the original degree distributions. Two measures are applied: (a) (b) $S^{1st}(q)/N$ the relative size of largest connected component, and (c) (d) $S^{2nd}(q)/N$ the critical fraction $q_c$ at the peak of the relative size of second largest component.\\
\textbf{\ref{fig:keihan_ib_index_100_3combines} Keihan\_100\_3combined\_4in1.pdf}: Robustness against recalculated betweenness (RB) attacks for Keihan networks with $N = 100$ nodes. For both Rew (Randomized networks) and 2DL lines, the rewiring process preserves the original degree distributions. Two measures are applied: (a) (b) $S^{1st}(q)/N$ the relative size of largest connected component, and (c) (d) $S^{2nd}(q)/N$ the critical fraction $q_c$ at the peak of the relative size of second largest component.\\
\textbf{\ref{fig:nagoya_ib_index_100_3combines} Nagoya\_100\_3combined\_4in1.pdf}: Robustness against recalculated betweenness (RB) attacks for Nagoya networks with $N = 100$ nodes. For both Rew (Randomized networks) and 2DL lines, the rewiring process preserves the original degree distributions. Two measures are applied: (a) (b) $S^{1st}(q)/N$ the relative size of largest connected component, and (c) (d) $S^{2nd}(q)/N$ the critical fraction $q_c$ at the peak of the relative size of second largest component.\\
\textbf{\ref{fig:tokyo_ib_index_100_3combines} Tokyo\_100\_3combined\_4in1.pdf}: Robustness against recalculated betweenness (RB) attacks in Tokyo networks with $N = 100$ nodes. For both Rew (Randomized networks) and 2DL lines, the rewiring process preserves the original degree distributions. Two measures are applied: (a) (b) $S^{1st}(q)/N$ the relative size of largest connected component, and (c) (d) $S^{2nd}(q)/N$ the critical fraction $q_c$ at the peak of the relative size of second largest component.\\
\textbf{\ref{fig:sendai_ib_index_100_3combines} Sendai\_100\_3combined\_4in1.pdf}: Robustness against recalculated betweenness (RB) attacks for Sendai networks with $N = 100$ nodes. For both Rew (Randomized networks) and 2DL lines, the rewiring process preserves the original degree distributions. Two measures are applied: (a) (b) $S^{1st}(q)/N$ the relative size of largest connected component, and (c) (d) $S^{2nd}(q)/N$ the critical fraction $q_c$ at the peak of the relative size of second largest component.\\
\textbf{\ref{fig:sapporo_ib_index_100_3combines} Sapporo\_100\_3combined\_4in1.pdf}: Robustness against recalculated betweenness (RB) attacks for Sapporo networks with $N = 100$ nodes. For both Rew (Randomized networks) and 2DL lines, the rewiring process preserves the original degree distributions. Two measures are applied: (a) (b) $S^{1st}(q)/N$ the relative size of largest connected component, and (c) (d) $S^{2nd}(q)/N$ the critical fraction $q_c$ at the peak of the relative size of second largest component.\\
\textbf{\ref{fig:fukuoka_ib_index_1024_3combines} Fukuoka\_1024\_3combined\_4in1.pdf}: Robustness against recalculated betweenness (RB) attacks for Fukuoka networks with $N = 1024$ nodes. For both Rew (Randomized networks) and 2DL lines, the rewiring process preserves the original degree distributions. Two measures are applied: (a) (b) $S^{1st}(q)/N$ the relative size of largest connected component, and (c) (d) $S^{2nd}(q)/N$ the critical fraction $q_c$ at the peak of the relative size of second largest component.\\
\textbf{\ref{fig:hiroshima_ib_index_1024_3combines} Hiroshima\_1024\_3combined\_4in1.pdf}: Robustness against recalculated betweenness (RB) attacks for Hiroshima networks with $N = 1024$ nodes. For both Rew (Randomized networks) and 2DL lines, the rewiring process preserves the original degree distributions. Two measures are applied: (a) (b) $S^{1st}(q)/N$ the relative size of largest connected component, and (c) (d) $S^{2nd}(q)/N$ the critical fraction $q_c$ at the peak of the relative size of second largest component.\\
\textbf{\ref{fig:keihan_ib_index_1024_3combines} Keihan\_1024\_3combined\_4in1.pdf}: Robustness against recalculated betweenness (RB) attacks for Keihan networks with $N = 1024$ nodes. For both Rew (Randomized networks) and 2DL lines, the rewiring process preserves the original degree distributions. Two measures are applied: (a) (b) $S^{1st}(q)/N$ the relative size of largest connected component, and (c) (d) $S^{2nd}(q)/N$ the critical fraction $q_c$ at the peak of the relative size of second largest component.\\
\textbf{\ref{fig:nagoya_ib_index_1024_3combines} Nagoya\_1024\_3combined\_4in1.pdf}: Robustness against recalculated betweenness (RB) attacks for Nagoya networks with $N = 1024$ nodes. For both Rew (Randomized networks) and 2DL lines, the rewiring process preserves the original degree distributions. Two measures are applied: (a) (b) $S^{1st}(q)/N$ the relative size of largest connected component, and (c) (d) $S^{2nd}(q)/N$ the critical fraction $q_c$ at the peak of the relative size of second largest component.\\
\textbf{\ref{fig:sendai_ib_index_1024_3combines} Sendai\_1024\_3combined\_4in1.pdf}: Robustness against recalculated betweenness (RB) attacks for Sendai networks with $N = 1024$ nodes. For both Rew (Randomized networks) and 2DL lines, the rewiring process preserves the original degree distributions. Two measures are applied: (a) (b) $S^{1st}(q)/N$ the relative size of largest connected component, and (c) (d) $S^{2nd}(q)/N$ the critical fraction $q_c$ at the peak of the relative size of second largest component.\\
\textbf{\ref{fig:sapporo_ib_index_1024_3combines} Sapporo\_1024\_3combined\_4in1.pdf}: Robustness against recalculated betweenness (RB) attacks for Sapporo networks with $N = 1024$ nodes. For both Rew (Randomized networks) and 2DL lines, the rewiring process preserves the original degree distributions. Two measures are applied: (a) (b) $S^{1st}(q)/N$ the relative size of largest connected component, and (c) (d) $S^{2nd}(q)/N$ the critical fraction $q_c$ at the peak of the relative size of second largest component.\\
\textbf{\ref{fig:fukuoka_ib_index_10000_3combines} Fukuoka\_10000\_3combined\_4in1.pdf}: Robustness against recalculated betweenness (RB) attacks for Fukuoka networks with $N = 10000$ nodes. For both Rew (Randomized networks) and 2DL lines, the rewiring process preserves the original degree distributions. Two measures are applied: (a) (b) $S^{1st}(q)/N$ the relative size of largest connected component, and (c) (d) $S^{2nd}(q)/N$ the critical fraction $q_c$ at the peak of the relative size of second largest component.\\
\textbf{\ref{fig:hiroshima_ib_index_10000_3combines} Hiroshima\_10000\_3combined\_4in1.pdf}: Robustness against recalculated betweenness (RB) attacks for Hiroshima networks with $N = 10000$ nodes. For both Rew (Randomized networks) and 2DL lines, the rewiring process preserves the original degree distributions. Two measures are applied: (a) (b) $S^{1st}(q)/N$ the relative size of largest connected component, and (c) (d) $S^{2nd}(q)/N$ the critical fraction $q_c$ at the peak of the relative size of second largest component.\\
\textbf{\ref{fig:keihan_ib_index_10000_3combines} Keihan\_10000\_3combined\_4in1.pdf}: Robustness against recalculated betweenness (RB) attacks for Keihan networks with $N = 10000$ nodes. For both Rew (Randomized networks) and 2DL lines, the rewiring process preserves the original degree distributions. Two measures are applied: (a) (b) $S^{1st}(q)/N$ the relative size of largest connected component, and (c) (d) $S^{2nd}(q)/N$ the critical fraction $q_c$ at the peak of the relative size of second largest component.\\
\textbf{\ref{fig:nagoya_ib_index_10000_3combines} Nagoya\_10000\_3combined\_4in1.pdf}: Robustness against recalculated betweenness (RB) attacks for Nagoya networks with $N = 10000$ nodes. For both Rew (Randomized networks) and 2DL lines, the rewiring process preserves the original degree distributions. Two measures are applied: (a) (b) $S^{1st}(q)/N$ the relative size of largest connected component, and (c) (d) $S^{2nd}(q)/N$ the critical fraction $q_c$ at the peak of the relative size of second largest component.\\
\textbf{\ref{fig:tokyo_ib_index_10000_3combines} Tokyo\_10000\_3combined\_4in1.pdf}: Robustness against recalculated betweenness (RB) attacks in Tokyo networks with $N = 10000$ nodes. For both Rew (Randomized networks) and 2DL lines, the rewiring process preserves the original degree distributions. Two measures are applied: (a) (b) $S^{1st}(q)/N$ the relative size of largest connected component, and (c) (d) $S^{2nd}(q)/N$ the critical fraction $q_c$ at the peak of the relative size of second largest component.\\
\textbf{\ref{fig:sendai_ib_index_10000_3combines} Sendai\_10000\_3combined\_4in1.pdf}: Robustness against recalculated betweenness (RB) attacks for Sendai networks with $N = 10000$ nodes. For both Rew (Randomized networks) and 2DL lines, the rewiring process preserves the original degree distributions. Two measures are applied: (a) (b) $S^{1st}(q)/N$ the relative size of largest connected component, and (c) (d) $S^{2nd}(q)/N$ the critical fraction $q_c$ at the peak of the relative size of second largest component.\\
\textbf{\ref{fig:sapporo_ib_index_10000_3combines} Sapporo\_10000\_3combined\_4in1.pdf}: Robustness against recalculated betweenness (RB) attacks for Sapporo networks with $N = 10000$ nodes. For both Rew (Randomized networks) and 2DL lines, the rewiring process preserves the original degree distributions. Two measures are applied: (a) (b) $S^{1st}(q)/N$ the relative size of largest connected component, and (c) (d) $S^{2nd}(q)/N$ the critical fraction $q_c$ at the peak of the relative size of second largest component.\\
\textbf{\ref{fig:r_to_q_100} r2q\_100\_2in1.pdf} Relation between robustness index $R^{RB}$ and modularity $Q$ in networks with $N$ = 100 nodes.\\
\textbf{\ref{fig:r_to_q_10000} r2q\_10000\_2in1.pdf} Relation between robustness index $R^{RB}$ and modularity $Q$ in networks with $N$ = 10000 nodes\\
\textbf{\ref{fig:population_per_cell} population\_per\_cell.pdf} Semi-logarithmic plot of population per $500\mathrm{m} \times 500\mathrm{m}$ block meshes in seven major Japanese areas, with meshes sorted in decreasing order of population. Each curve represents an area. The linear decay on the logarithmic scale indicates that a small number of meshes concentrate the majority of the urban population. This heavy-tailed distribution supports the use of rank-based node selections for the Pop. and Inv. networks.\\
\textbf{\ref{tab:combined_rb_100} r\_qc\_rb\_100}: Robustness index ($R^{RB}$) and Critical fraction ($q_c^{RB}$) against Recalculated Betweenness (RB) attacks in networks with $N = 100$ nodes for seven major Japanese areas. Higher values indicate greater robustness of connectivity. For $R^{RB}$, values with upper-triangles ($\triangle$) indicate higher robustness than Uni. case, while values with lower-triangles ($\triangledown$) indicate lower robustness than Uni. case. For $q_c^{RB}$, all values are marked with lower-triangles ($\triangledown$) as they show lower robustness than Uni. case.\\
\textbf{\ref{tab:combined_rb_1024} r\_qc\_rb\_1024}: Robustness index ($R^{RB}$) and Critical fraction ($q_c^{RB}$) against Recalculated Betweenness (RB) attacks in networks with $N = 1024$ nodes for seven major Japanese areas. Higher values indicate greater robustness of connectivity. Values with lower-triangles ($\triangledown$) indicate where the cases of Pop. and Inv. have lower robustness of connectivity than the cases of Uni. for both RNG and GG.\\
\textbf{\ref{tab:combined_rb_10000} r\_qc\_rb\_10000}: Robustness index ($R^{RB}$) and Critical fraction ($q_c^{RB}$) against Recalculated Betweenness (RB) attacks in networks with $N = 10000$ nodes for seven major Japanese areas. Higher values indicate greater robustness of connectivity. Values with lower-triangles ($\triangledown$) indicate where the cases of Pop. and Inv. have lower robustness of connectivity than the cases of Uni. for both RNG and GG.\\
\textbf{\ref{tab:average_degree_100} average\_degree\_100}: Average degree $\langle k \rangle$ in networks with (N) = 100 nodes for seven major Japanese areas. Higher average degrees mean more links per node in the network. Values with upper-triangles ($\triangle$) indicate where the cases of Pop. and Inv. have higher $<k>$ than the cases of Uni. for both RNG and GG.\\
\textbf{\ref{tab:average_degree_10000} average\_degree\_10000}: Average degree $\langle k \rangle$ in networks with (N) = 10000 nodes for seven major Japanese areas. Higher average degrees mean more links per node in the network. Values with upper-triangles ($\triangle$) indicate where the cases of Pop. and Inv. have higher $<k>$ than the cases of Uni. for both RNG and GG.\\
\textbf{\ref{tab:q_value_100} Table q\_values\_100}: Modularity $Q$ in networks with 100 nodes for seven major Japanese areas. Higher values indicate stronger community structures. Values with upper-triangles ($\triangle$) or lower-triangles ($\triangledown$) indicate where the cases of Pop. and Inv. have higher or lower modularity than the case of Uni. for both RNG and GG. Note the generally higher modularity in Pop. and Inv. compared to Uni. networks for both RNG and GG.\\
\textbf{\ref{tab:q_value_10000} q\_values\_10000}: Modularity $Q$ in networks with 10000 nodes for seven major Japanese areas. Higher values indicate stronger community structures. Values with upper-triangles ($\triangle$) indicate where the cases of Pop. and Inv. have higher modularity than the case of Uni. for both RNG and GG. Note the generally higher modularity in Pop. and Inv. compared to Uni. networks for both RNG and GG.\\
\textbf{\ref{tab:rb_anova_summary_1024} rb\_anova\_1024}: One-way ANOVA results evaluating the influence of node distribution (Population-based, Inverse, and Uniform) on the robustness index ($R$) and the critical fraction ($q_c$) against Recalculate Betweenness (RB) attacks ($N=1024$ nodes). \textbf{F-value} indicates the ratio of between-group variance to within-group variance. \textbf{p-value} shows the probability that the observed group differences are due to chance (typically, $p < 0.05$ is considered statistically significant). \boldmath{$\eta^2$} denotes the effect size, indicating the proportion of variance explained by the group factor; conventionally, $\eta^2 > 0.14$ is considered a large effect. \\
\textbf{\ref{tab:id_anova_summary} id\_anova\_1024}: One-way ANOVA results evaluating the influence of node distribution (Population-based, Inverse population-based, and Uniform) on the robustness index ($R$) and the critical fraction ($q_c$) against Initial Degree (ID) attacks ($N = 1024$ nodes). \textbf{F-value} indicates the ratio of between-group variance to within-group variance. \textbf{p-value} shows the probability that the observed group differences are due to chance (typically, $p < 0.05$ is considered statistically significant). \boldmath{$\eta^2$} denotes the effect size, indicating the proportion of variance explained by the group factor. In this table, the $q_c$ result for RNG networks shows no significant difference ($p = 0.249$). This may be due to two factors: (1) the higher average degree $\langle k \rangle$ in RNG Pop. networks (see Table~\ref{tab:average_degree_1024}), which offers more alternative paths after node removal; and (2) the presence of grid-like substructures in Pop. networks, which can enhance local community and delay critical fragmentation.\\
\textbf{\ref{tab:rf_anova_summary} rf\_anova\_1024}: One-way ANOVA results evaluating the influence of node distribution (Population-based, Inverse population-based, and Uniform) on the robustness index ($R$) and the critical fraction ($q_c$) against Random Failures (RF) ($N = 1024$ nodes). \textbf{F-value} indicates the ratio of between-group variance to within-group variance. \textbf{p-value} shows the probability that the observed group differences are due to chance (typically, $p < 0.05$ is considered statistically significant). \boldmath{$\eta^2$} denotes the effect size, indicating the proportion of variance explained by the group factor; conventionally, $\eta^2 > 0.14$ is considered a large effect. Notably, the $R$ result for RNG networks shows no significant difference ($p = 0.104$), and the $q_c$ result is marginally significant ($p = 0.028$). These results may be explained by two factors: (1) higher average degree $\langle k \rangle$, which provides alternative paths after node removal (see Table~\ref{tab:average_degree_1024}); and (2) the presence of grid-like substructures in certain area of the network, which enhance local community and delay critical fragmentation.\\
\textbf{\ref{tab:community_number_combined_1024} community\_number\_1024}: Estimated number of communities in original networks and after node relocation on a 2D lattice (relocated). Values with upper-triangles ($\triangle$) indicate that Pop.-based or Inv.-based has more communities than Uni.-based networks. Values in bold indicate fewer communities than Uni.-based networks in both RNG and GG.
\textbf{\ref{tab:combined_sparsity_1024} sparsity\_1024}: Sparsity Index (SI($G_w$)) of RNG and GG for different Japanese cities  ($N=1024$). Values are calculated based on normalized edge distances, representing the spatial sparsity of each configuration. Uniform and 2D lattice cases serve as comparative baselines for sparsity. 2D lattice havs the highest SI($G_w$), which may result from long edges between faraway nodes. \\
\textbf{\ref{tab:pearson_si_robustness} pearson\_si\_robustness}: The Pearson's correlation coefficients ($r$) and significance levels ($p$-values) between the sparsity index (SI) and robustness measures ($R$ and $q_c$) in RNG and GG against three attack strategies (RB, ID, and RF). \textbf{Bold values} indicate statistically significant results ($p < 0.05$). A negative correlation ($r < 0$) suggests that higher sparsity (larger SI) is associated with lower robustness, implying that spatially sparser networks are more vulnerable. Conversely, a positive correlation ($r > 0$) indicates that sparsity is associated with higher robustness, which may occur in certain configurations where redundant links or grid-like structures compensates for sparsity. \\
\textbf{\ref{tab:q_si_corr} correlation\_q\_si}: The Pearson's correlation coefficient ($r$) and significance level ($p$-value) between modularity $Q$ and sparsity index SI($G_w$) for RNG and GG. The correlation is statistically significant only for GG ($p < 0.05$), indicating monotone increasing trends association between modularity $Q$ and spatial sparsity index $SI(G_w)$ in GG. \\

\clearpage
\section*{Acknowledgments}
\subsection*{Availability of data and materials}
The population data used in this study are available from the ``Regional Mesh Statistics: First Regional Division, 2010 Population Census (World Geodetic System)'' provided by the Statistical Information Institute for Consulting and Analysis (SINFONICA). The data were purchased under license from SINFONICA (located at Nogaku Shorin Building 5F, 3-6 Kanda-Jinbocho, Chiyoda-ku, Tokyo 101-0051, Japan). These data are subject to usage restrictions. Researchers interested in accessing these data should contact SINFONICA (\url{https://www.sinfonica.or.jp/}) for purchase and licensing information.

The code used to generate the results is also available from the corresponding author upon reasonable request.

\subsection*{Competing interests}
The authors declare that they have no competing interests.

\subsection*{Funding}
JSPS KAKENHI Grant Number JP.21H03425

\subsection*{Authors' contributions}
\textbf{Conceptualization}: Yukio Hayashi.\\
\textbf{Funding acquistion}: Yukio Hayashi.\\
\textbf{Investigation}: Yingzhou Mou, Yukio Hayashi.\\
\textbf{Methodology}: Yingzhou Mou, Yukio Hayashi.\\
\textbf{Supervision}: Yukio Hayashi.\\
\textbf{Visualization}: Yingzhou Mou.\\
\textbf{Writing - original draft}: Yingzhou Mou.\\
\textbf{Writing - review \& editing}: Yingzhou Mou, Yukio Hayashi.\\

\subsection*{Acknowledgements}
This research is supported in part by JSPS KAKENHI Grant Number JP.21H03425

\subsection*{Authors' information}
\begin{enumerate}
\item Yingzhou MOU (Corresponding author)\\
      Japan Advanced Institute of Science and Technology, Nomi-city,\\
      Ishikawa 923-1292, Japan\\
      E-mail: \href{mailto:mouyingzhou@outlook.com}{mouyingzhou@outlook.com}

\item Yukio HAYASHI\\
      Japan Advanced Institute of Science and Technology, Nomi-city,\\
      Ishikawa 923-1292, Japan\\
      E-mail: \href{mailto:yhayashi@jaist.ac.jp}{yhayashi@jaist.ac.jp}
\end{enumerate}

\clearpage

\appendix

% 重置图和表格计数器
\setcounter{figure}{0}
\setcounter{table}{0}

% 保存原始的图表名称命令
\let\origfigurename\figurename
\let\origtablename\tablename

% 将图表名称设为空
\renewcommand{\figurename}{}
\renewcommand{\tablename}{}

% 重新定义图的编号格式
\renewcommand{\thefigure}{S\arabic{figure} \origfigurename}
\renewcommand{\thetable}{S\arabic{table} \origtablename}

\clearpage
\begin{figure}
    \centering
    \includegraphics[width=\linewidth]{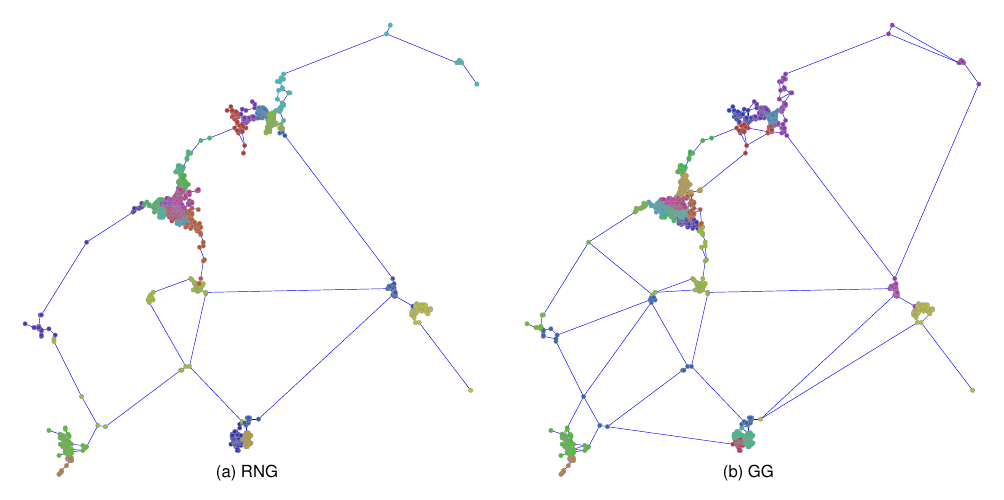}  
    \caption{Visualization of community structures in Fukuoka before node removal. $N=1024$ nodes are located by the decreasing order of population (Pop.). Different colors represent different communities estimated by Louvain method. There are clear community formations particularly in densely populated areas.}
    \label{fig:ib_v_pop_fukuoka_1024}
\end{figure}

\begin{figure}
    \centering
   \includegraphics[width=\linewidth]{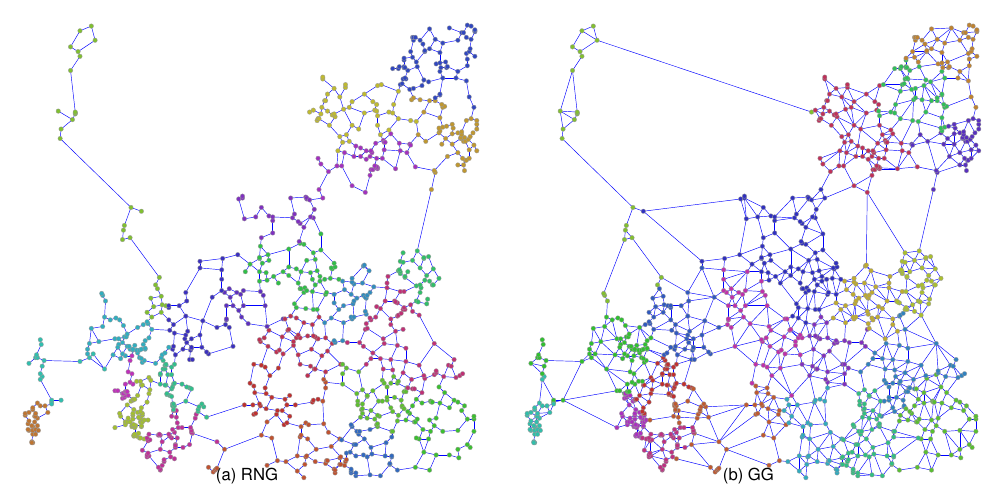}
    \caption{Visualization of community structures in Fukuoka before node removal. $N=1024$ nodes are located by the inverse order of population (Inv.). Different colors represent different communities estimated by Louvain method. There are different community formations compared to Figure \ref{fig:ib_v_pop_fukuoka_1024}.}
     \label{fig:ib_v_inv_fukuoka_1024}
\end{figure}

\begin{figure}
    \centering
    \includegraphics[width=\linewidth]{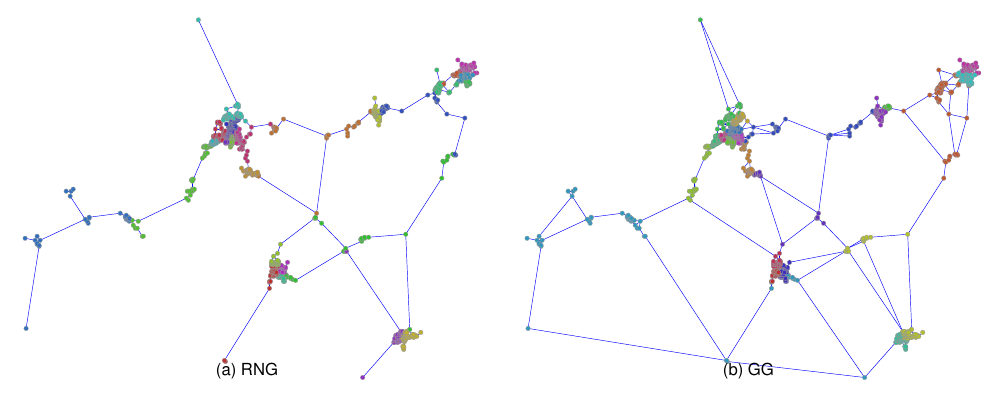}  
    \caption{Visualization of community structures in Hiroshima before node removal. $N=1024$ nodes are located by the decreasing order of population (Pop.). Different colors represent different communities estimated by Louvain method. There are clear community formations particularly in densely populated areas.}
    \label{fig:ib_v_pop_hiroshima_1024}
\end{figure}

\begin{figure}
    \centering
   \includegraphics[width=\linewidth]{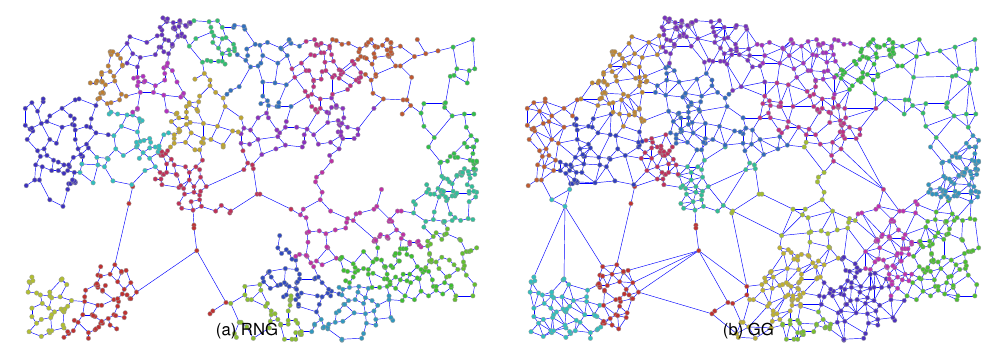}
    \caption{Visualization of community structures in Hiroshima before node removal. $N=1024$ nodes are located by the inverse order of population (Inv.). Different colors represent different communities estimated by Louvain method. There are different community formations compared to Figure \ref{fig:ib_v_pop_hiroshima_1024}.}
     \label{fig:ib_v_inv_hiroshima_1024}
\end{figure}

\begin{figure}
    \centering
\includegraphics[width=\linewidth]{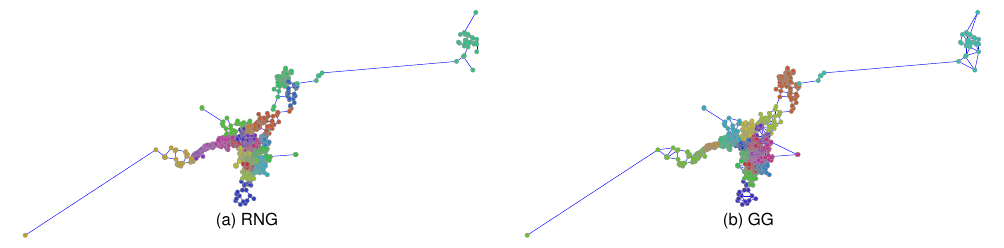}  
    \caption{Visualization of community structures in Keihan before node removal. $N=1024$ nodes are located by the decreasing order of population (Pop.). Different colors represent different communities estimated by Louvain method. There are clear community formations particularly in densely populated areas.}
    \label{fig:ib_v_pop_keihan_1024}
\end{figure}

\begin{figure}
    \centering
   \includegraphics[width=\linewidth]{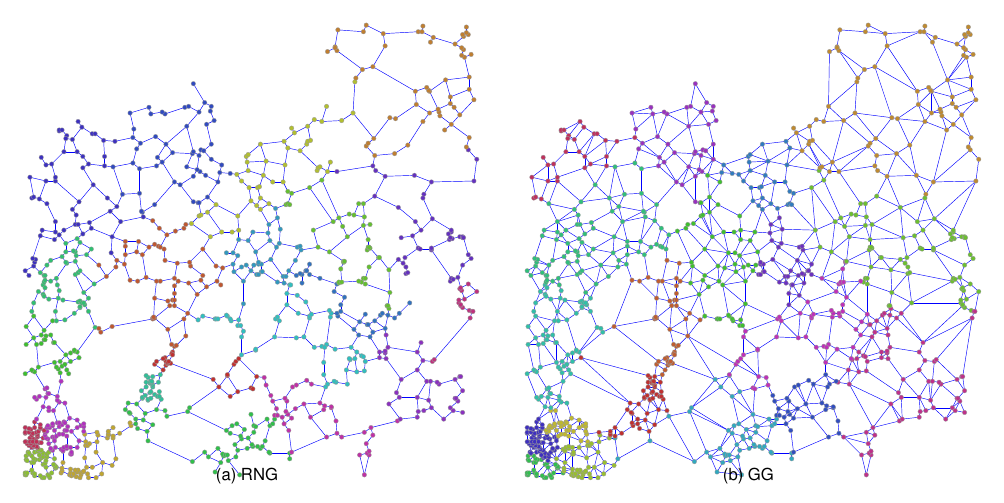}
    \caption{Visualization of community structures in Keihan before node removal. $N=1024$ nodes are located by the inverse order of population (Inv.). Different colors represent different communities estimated by Louvain method. There are different community formations compared to Figure \ref{fig:ib_v_pop_keihan_1024}.}
     \label{fig:ib_v_inv_keihan_1024}
\end{figure}

\begin{figure}
    \centering
\includegraphics[width=\linewidth]{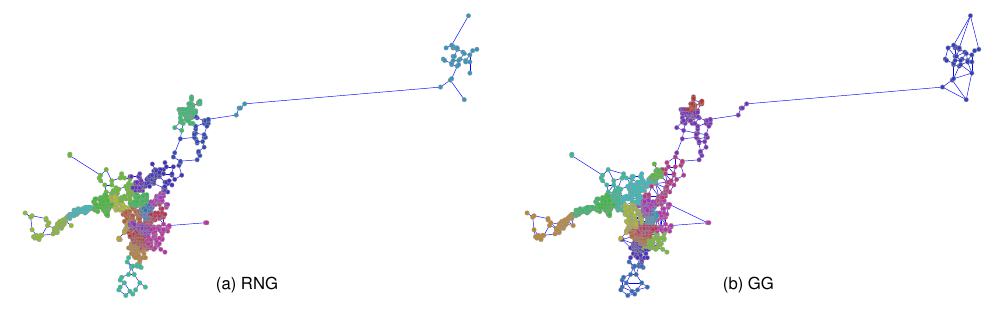}  
    \caption{Visualization of community structures in Nagoya before node removal. $N=1024$ nodes are located by the decreasing order of population (Pop.). Different colors represent different communities estimated by Louvain method. There are clear community formations particularly in densely populated areas.}
    \label{fig:ib_v_pop_nagoya_1024}
\end{figure}

\begin{figure}
    \centering
   \includegraphics[width=\linewidth]{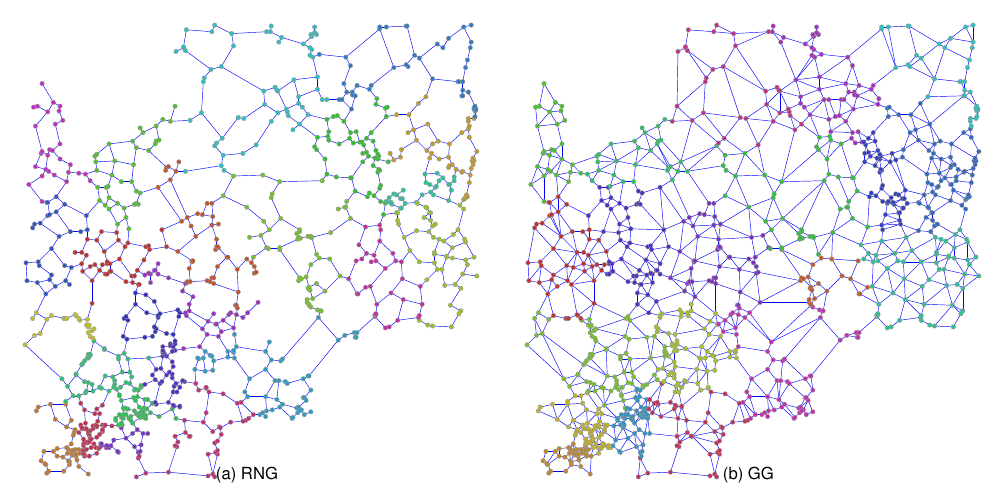}
    \caption{Visualization of community structures in Nagoya before node removal. $N=1024$ nodes are located by the inverse order of population (Inv.). Different colors represent different communities estimated by Louvain method. There are different community formations compared to Figure \ref{fig:ib_v_pop_nagoya_1024}.}
     \label{fig:ib_v_inv_nagoya_1024}
\end{figure}

\begin{figure}
    \centering
\includegraphics[width=\linewidth]{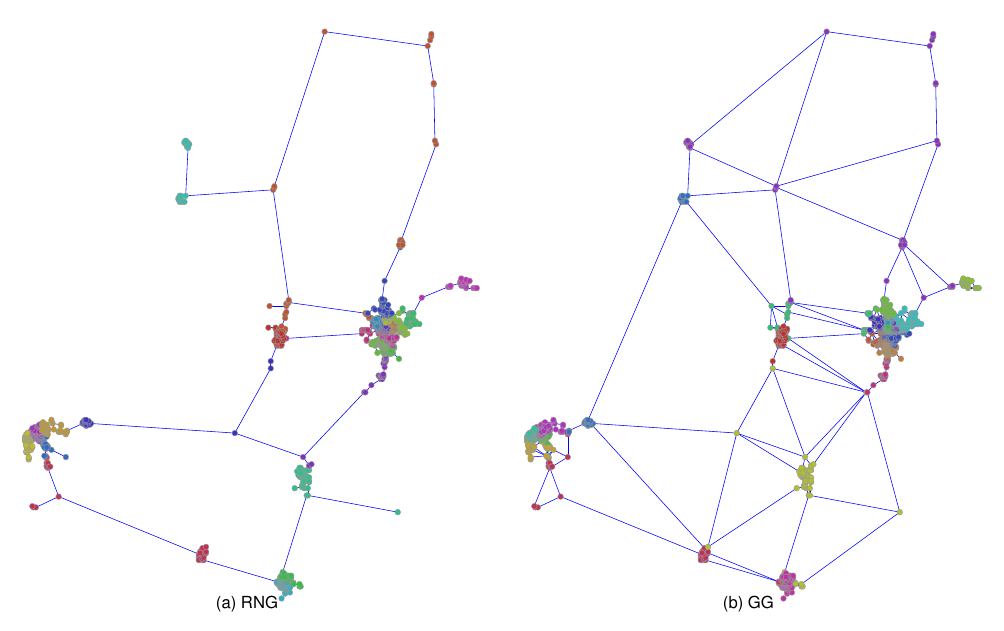}  
    \caption{Visualization of community structures in Sendai before node removal. $N=1024$ nodes are located by the decreasing order of population (Pop.). Different colors represent different communities estimated by Louvain method. There are clear community formations particularly in densely populated areas.}
    \label{fig:ib_v_pop_sendai_1024}
\end{figure}

\begin{figure}
    \centering
   \includegraphics[width=\linewidth]{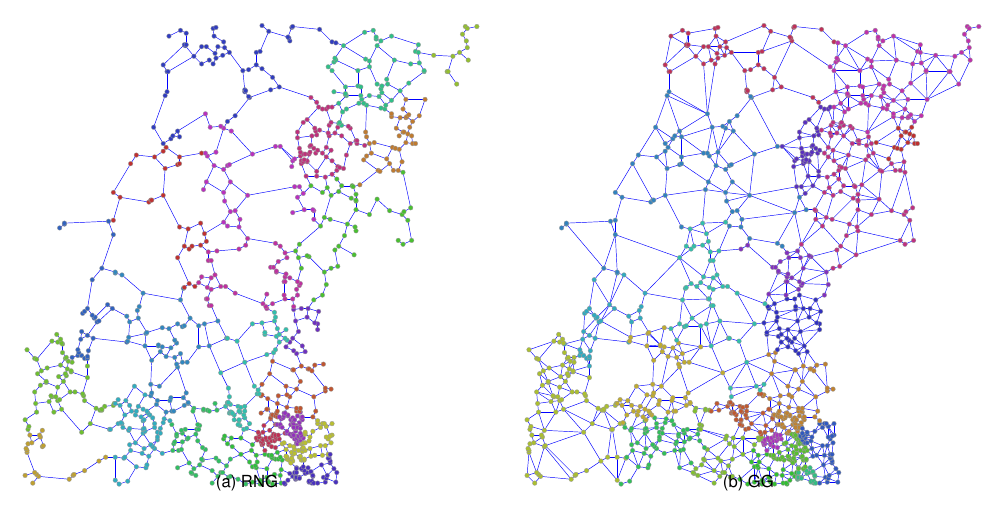}
    \caption{Visualization of community structures in Sendai before node removal. $N=1024$ nodes are located by the inverse order of population (Inv.). Different colors represent different communities estimated by Louvain method. There are different community formations compared to Figure \ref{fig:ib_v_pop_sendai_1024}.}
     \label{fig:ib_v_inv_sendai_1024}
\end{figure}

\begin{figure}
    \centering
\includegraphics[width=\linewidth]{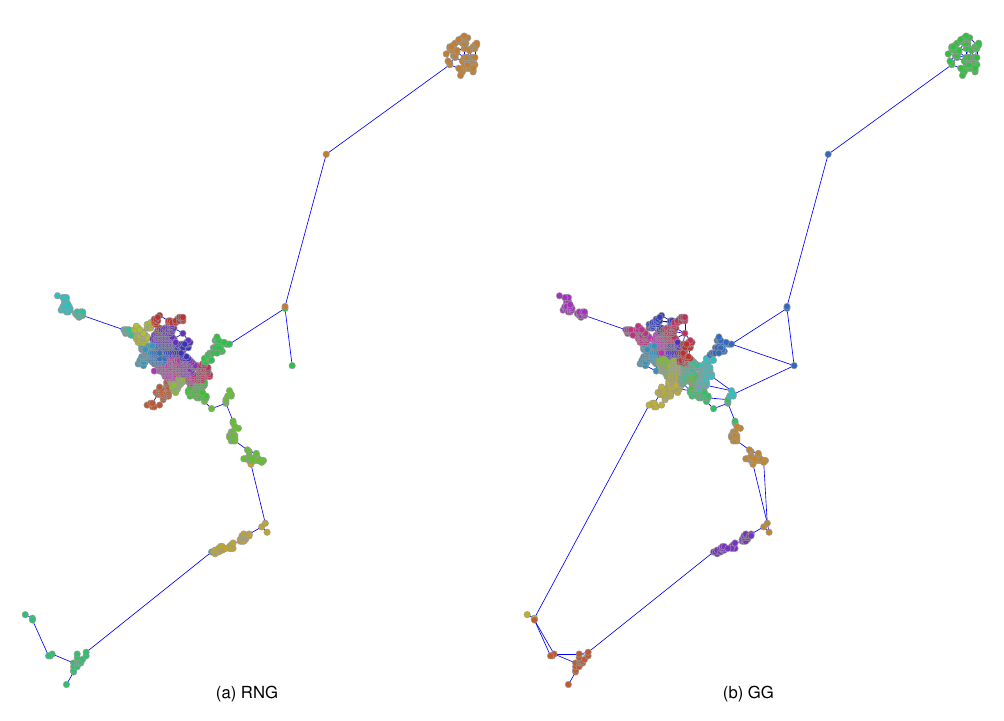}  
    \caption{Visualization of community structures in Sapporo before node removal. $N=1024$ nodes are located by the decreasing order of population (Pop.). Different colors represent different communities estimated by Louvain method. There are clear community formations particularly in densely populated areas.}
    \label{fig:ib_v_pop_sapporo_1024}
\end{figure}

\begin{figure}
    \centering
   \includegraphics[width=\linewidth]{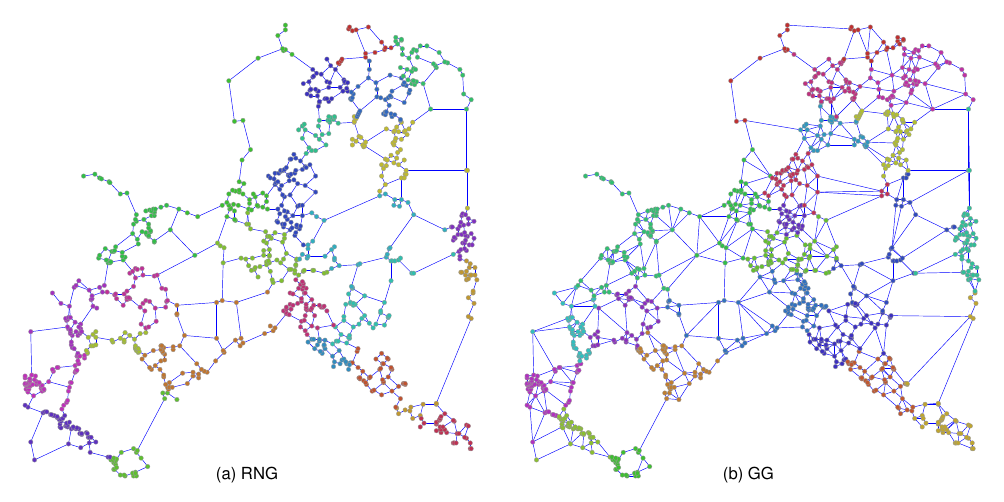}
    \caption{Visualization of community structures in Sapporo before node removal. $N=1024$ nodes are located by the inverse order of population (Inv.). Different colors represent different communities estimated by Louvain method. There are different community formations compared to Figure \ref{fig:ib_v_pop_sapporo_1024}.}
     \label{fig:ib_v_inv_sapporo_1024}
\end{figure}

\begin{figure}
    \centering
    \includegraphics[width=\linewidth]{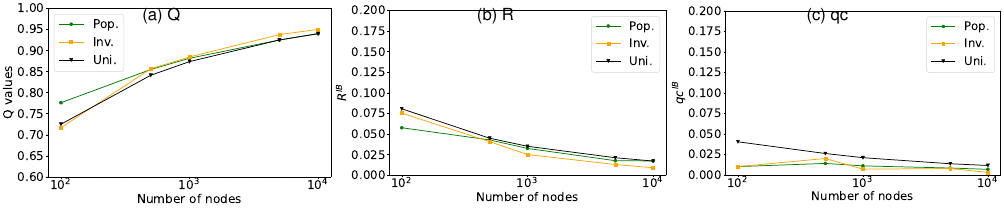}
    \caption{Increasing modularity $Q$ vs. decreasing robustness index $R^{RB}$ or critical fraction $q_c^{RB}$ for varying the size $N$ in Tokyo RNG networks.}
    \label{fig:yaxis_tokyo_rng_1024}
\end{figure}

\begin{figure}
    \centering
    \includegraphics[width=\linewidth]{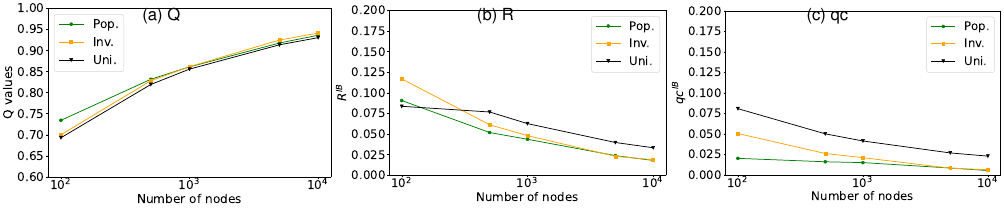}
    \caption{Increasing modularity $Q$ vs. decreasing robustness index $R^{RB}$ or critical fraction $q_c^{RB}$ for varying the size $N$ in Tokyo GG networks.}
    \label{fig:yaxis_tokyo_gg_1024}
\end{figure}

% \subsection{scatters of Q vs R or qc}

\begin{figure}
    \centering
    \includegraphics[width=\linewidth]{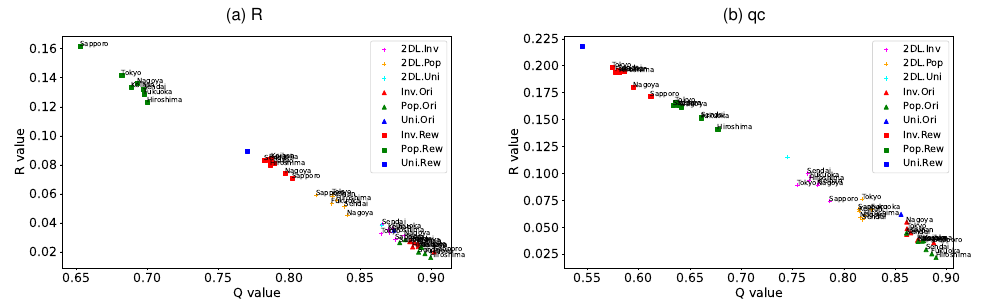}
    \caption{Scatter plots show relation between network robustness measures (a for $R$ and b for $q_c$) and the proportion of grid-like parts against random failures (RF). Networks with $N=1024$ nodes are considered, where Pop. networks (green) show notably higher proportions of grid-like parts compared to Inv. (red) and Uni. (blue) networks. See the text at the end of subsection \ref{sec:rf_id_attacks} for the detail.}
    \label{fig:r_and_qc_to_grid_1024}
\end{figure}

% \subsection{Robustness against RB attacks}
\begin{figure}
\includegraphics[width=\textwidth]{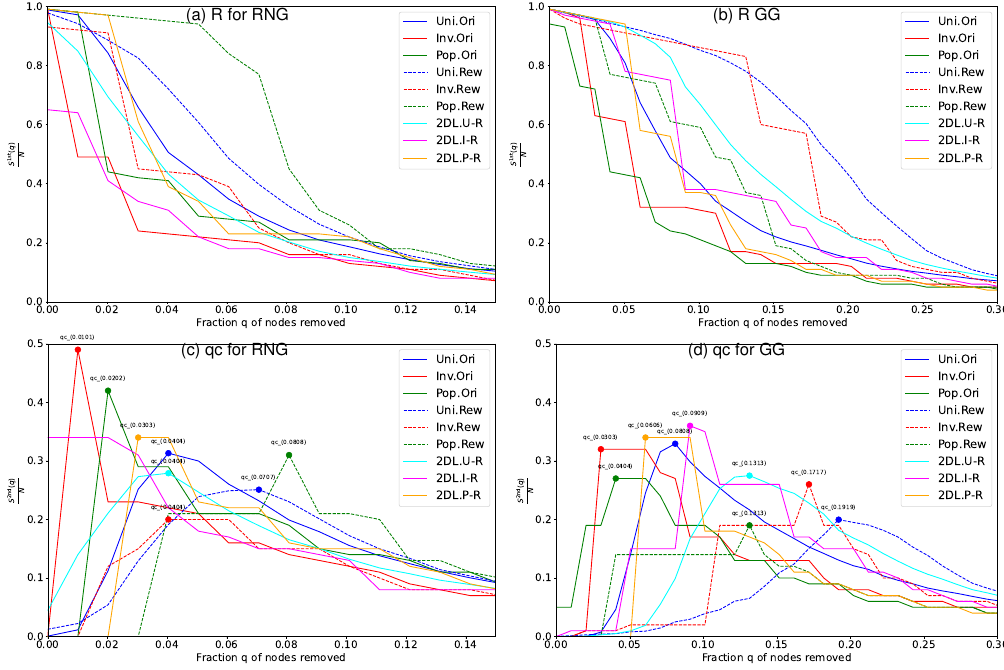}
\caption{Robustness against recalculated betweenness (RB) attacks for Fukuoka networks with $N = 100$ nodes. For both Rew (Randomized networks) and 2DL lines, the rewiring process preserves the original degree distributions. Two measures are applied: (a) (b) $S^{1st}(q)/N$ the relative size of largest connected component, and (c) (d) $S^{2nd}(q)/N$ the critical fraction $q_c$ at the peak of the relative size of second largest component.}
    \label{fig:fukuoka_ib_index_100_3combines}
\end{figure}

\begin{figure}
\includegraphics[width=\textwidth]{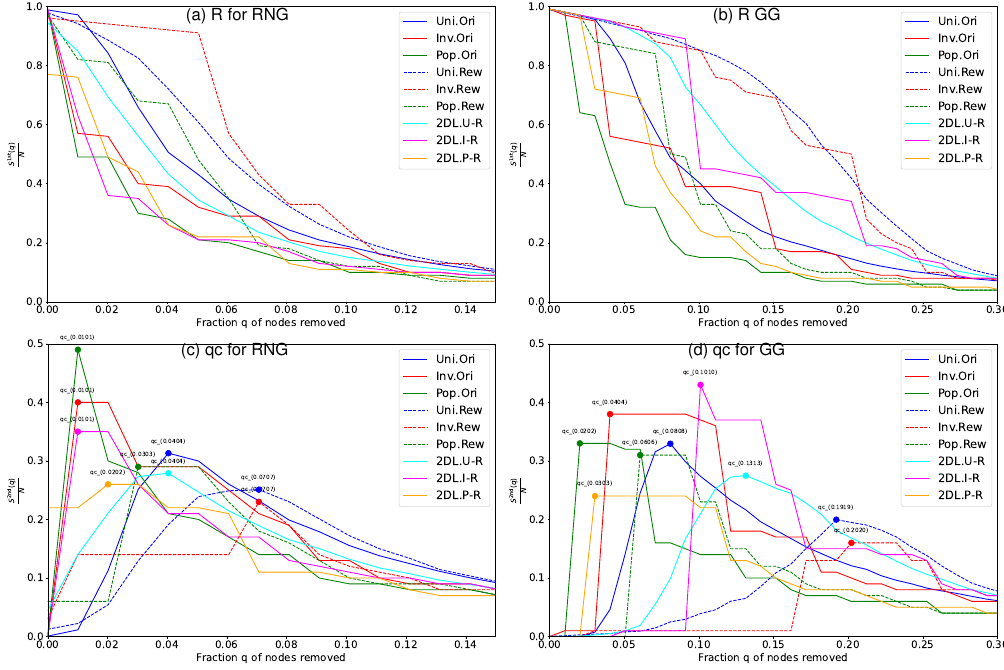}
\caption{Robustness against recalculated betweenness (RB) attacks for Hiroshima networks with $N = 100$ nodes. For both Rew (Randomized networks) and 2DL lines, the rewiring process preserves the original degree distributions. Two measures are applied: (a) (b) $S^{1st}(q)/N$ the relative size of largest connected component, and (c) (d) $S^{2nd}(q)/N$ the critical fraction $q_c$ at the peak of the relative size of second largest component.}
    \label{fig:hiroshima_ib_index_100_3combines}
\end{figure}

\begin{figure}
\includegraphics[width=\textwidth]{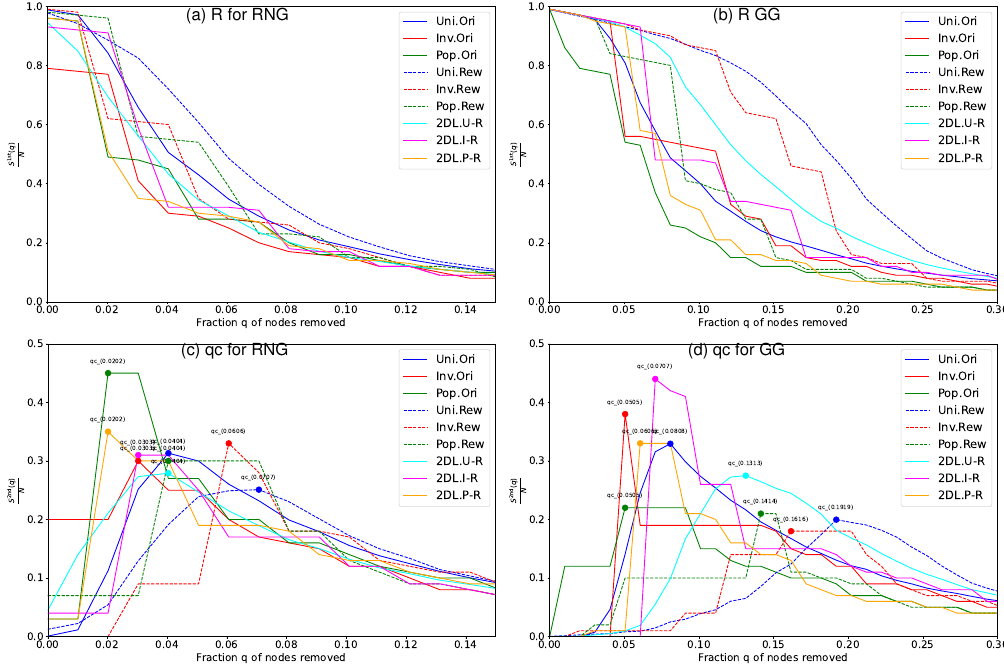}
\caption{Robustness against recalculated betweenness (RB) attacks for Keihan networks with $N = 100$ nodes. For both Rew (Randomized networks) and 2DL lines, the rewiring process preserves the original degree distributions. Two measures are applied: (a) (b) $S^{1st}(q)/N$ the relative size of largest connected component, and (c) (d) $S^{2nd}(q)/N$ the critical fraction $q_c$ at the peak of the relative size of second largest component.}
    \label{fig:keihan_ib_index_100_3combines}
\end{figure}

\begin{figure}
\includegraphics[width=\textwidth]{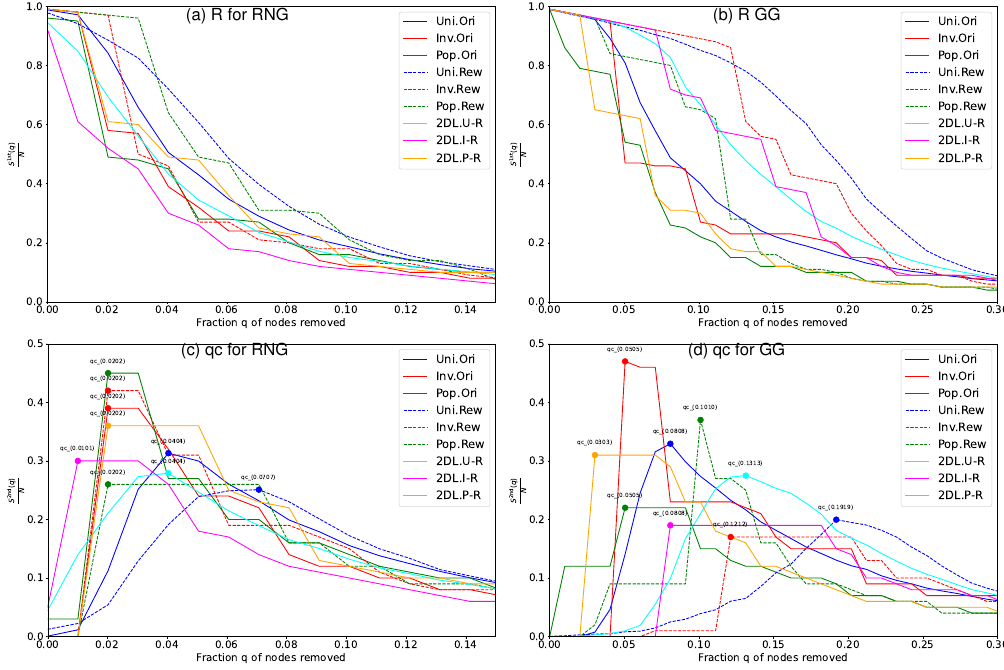}
\caption{Robustness against recalculated betweenness (RB) attacks for Nagoya networks with $N = 100$ nodes. For both Rew (Randomized networks) and 2DL lines, the rewiring process preserves the original degree distributions. Two measures are applied: (a) (b) $S^{1st}(q)/N$ the relative size of largest connected component, and (c) (d) $S^{2nd}(q)/N$ the critical fraction $q_c$ at the peak of the relative size of second largest component.}
    \label{fig:nagoya_ib_index_100_3combines}
\end{figure}

\begin{figure}
\includegraphics[width=\textwidth]{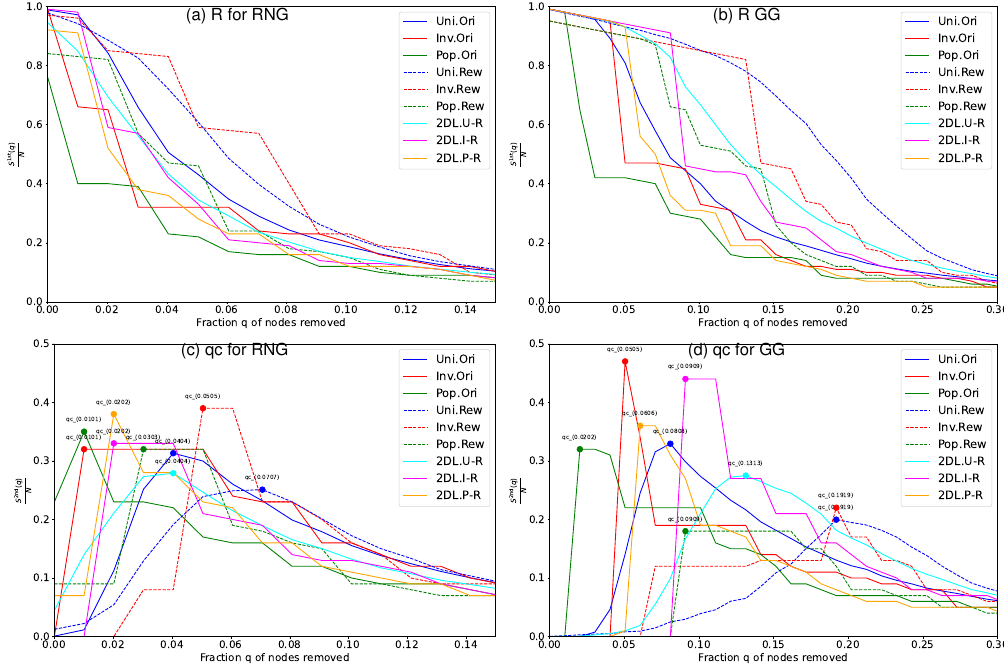}
\caption{Robustness against recalculated betweenness (RB) attacks for Tokyo networks with $N = 100$ nodes. For both Rew (Randomized networks) and 2DL lines, the rewiring process preserves the original degree distributions. Two measures are applied: (a) (b) $S^{1st}(q)/N$ the relative size of largest connected component, and (c) (d) $S^{2nd}(q)/N$ the critical fraction $q_c$ at the peak of the relative size of second largest component.}
    \label{fig:tokyo_ib_index_100_3combines}
\end{figure}

\begin{figure}
\includegraphics[width=\textwidth]{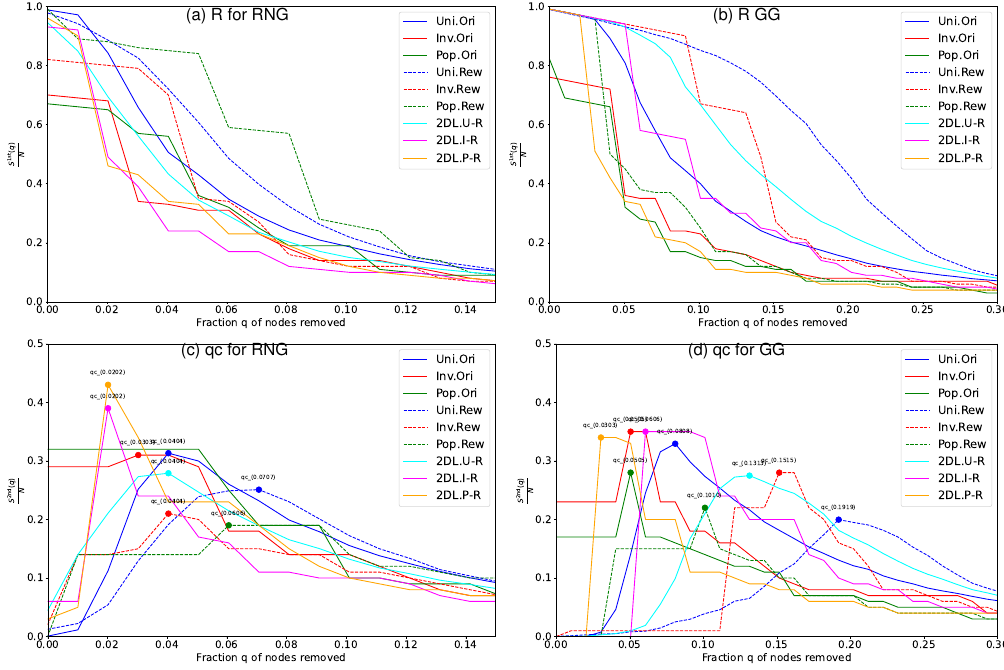}
\caption{Robustness against recalculated betweenness (RB) attacks for Sendai networks with $N = 100$ nodes. For both Rew (Randomized networks) and 2DL lines, the rewiring process preserves the original degree distributions. Two measures are applied: (a) (b) $S^{1st}(q)/N$ the relative size of largest connected component, and (c) (d) $S^{2nd}(q)/N$ the critical fraction $q_c$ at the peak of the relative size of second largest component.}
    \label{fig:sendai_ib_index_100_3combines}
\end{figure}

\begin{figure}
\includegraphics[width=\textwidth]{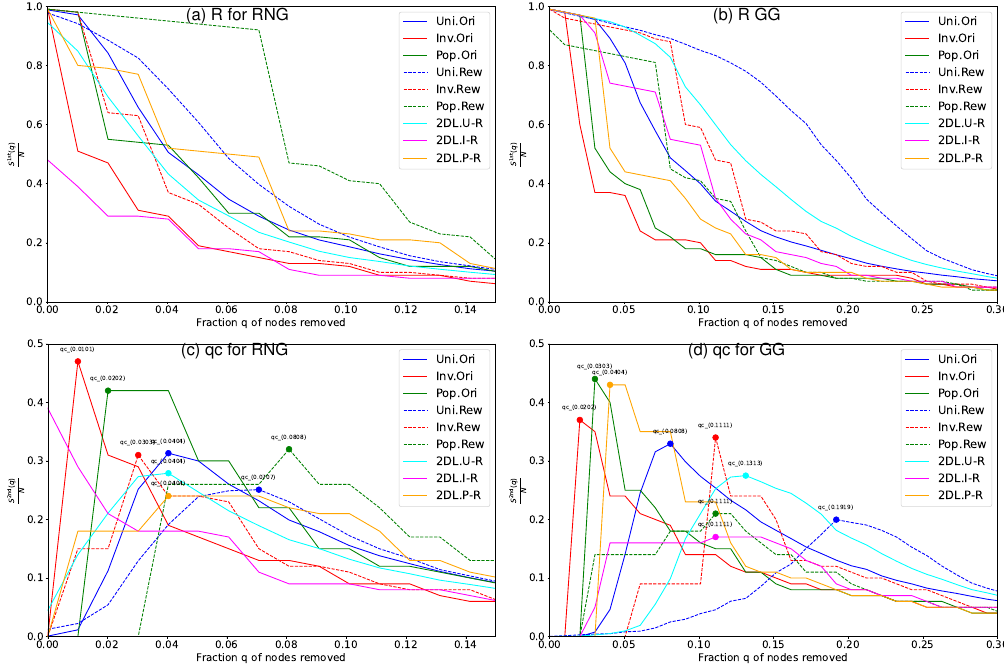}
\caption{Robustness against recalculated betweenness (RB) attacks for Sapporo networks with $N = 100$ nodes. For both Rew (Randomized networks) and 2DL lines, the rewiring process preserves the original degree distributions. Two measures are applied: (a) (b) $S^{1st}(q)/N$ the relative size of largest connected component, and (c) (d) $S^{2nd}(q)/N$ the critical fraction $q_c$ at the peak of the relative size of second largest component.}
    \label{fig:sapporo_ib_index_100_3combines}
\end{figure}

\begin{figure}
\includegraphics[width=\textwidth]{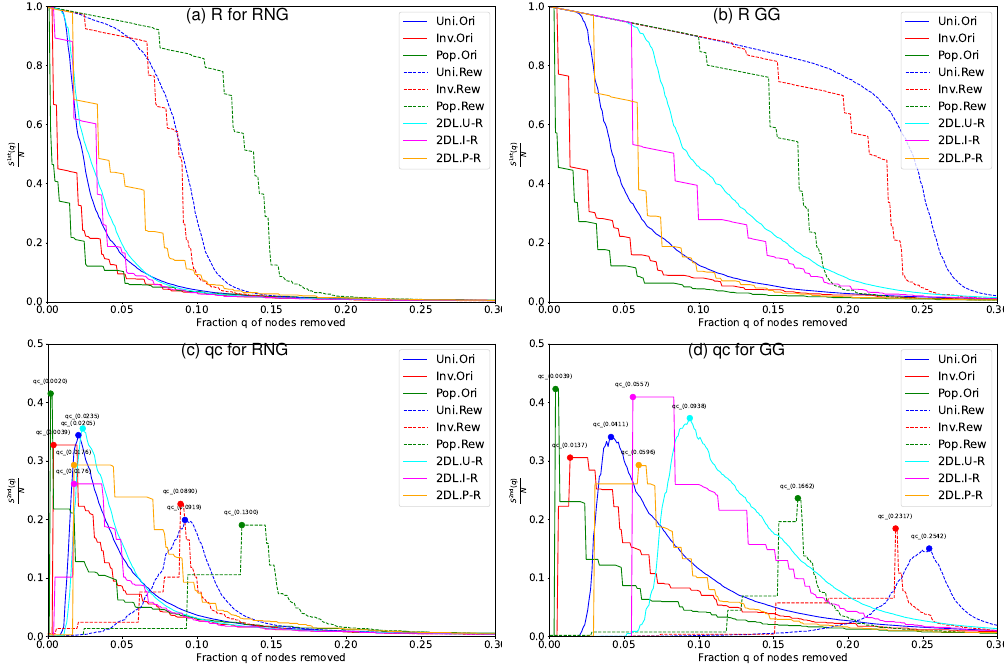}
\caption{Robustness against recalculated betweenness (RB) attacks for Fukuoka networks with $N = 1024$ nodes. For both Rew (Randomized networks) and 2DL lines, the rewiring process preserves the original degree distributions. Two measures are applied: (a) (b) $S^{1st}(q)/N$ the relative size of largest connected component, and (c) (d) $S^{2nd}(q)/N$ the critical fraction $q_c$ at the peak of the relative size of second largest component.}
    \label{fig:fukuoka_ib_index_1024_3combines}
\end{figure}

\begin{figure}
\includegraphics[width=\textwidth]{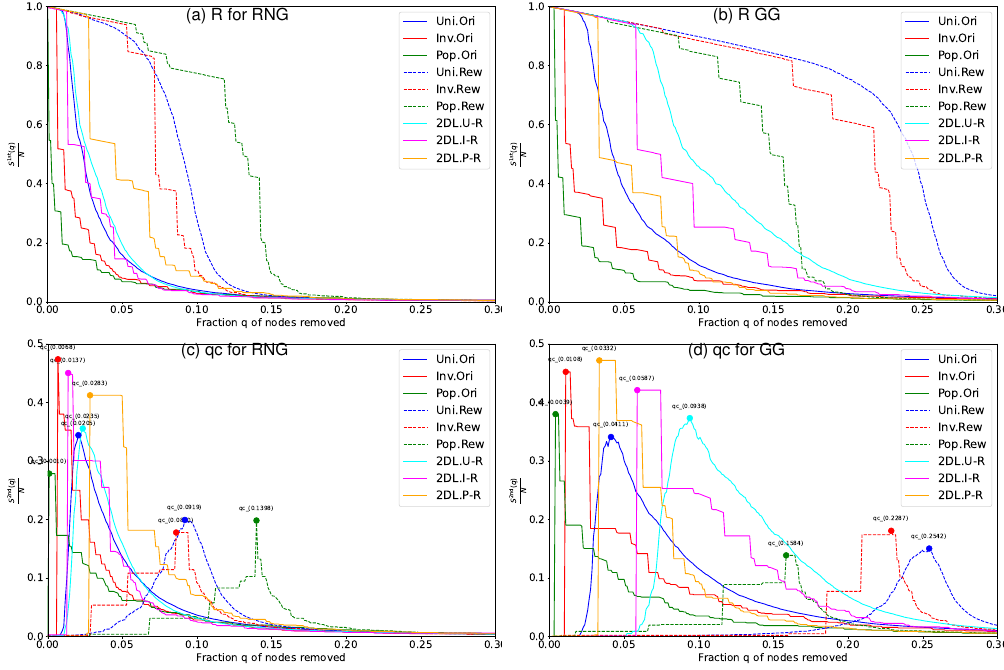}
\caption{Robustness against recalculated betweenness (RB) attacks for Hiroshima networks with $N = 1024$ nodes. For both Rew (Randomized networks) and 2DL lines, the rewiring process preserves the original degree distributions. Two measures are applied: (a) (b) $S^{1st}(q)/N$ the relative size of largest connected component, and (c) (d) $S^{2nd}(q)/N$ the critical fraction $q_c$ at the peak of the relative size of second largest component.}
    \label{fig:hiroshima_ib_index_1024_3combines}
\end{figure}

\begin{figure}
\includegraphics[width=\textwidth]{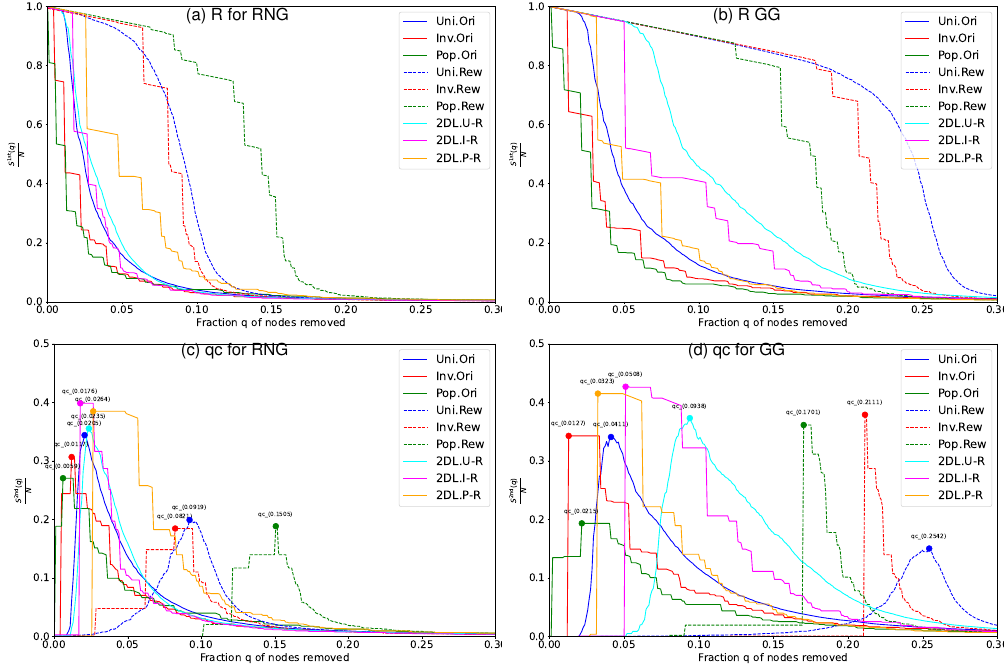}
\caption{Robustness against recalculated betweenness (RB) attacks for Keihan networks with $N = 1024$ nodes. For both Rew (Randomized networks) and 2DL lines, the rewiring process preserves the original degree distributions. Two measures are applied: (a) (b) $S^{1st}(q)/N$ the relative size of largest connected component, and (c) (d) $S^{2nd}(q)/N$ the critical fraction $q_c$ at the peak of the relative size of second largest component.}
    \label{fig:keihan_ib_index_1024_3combines}
\end{figure}

\begin{figure}
\includegraphics[width=\textwidth]{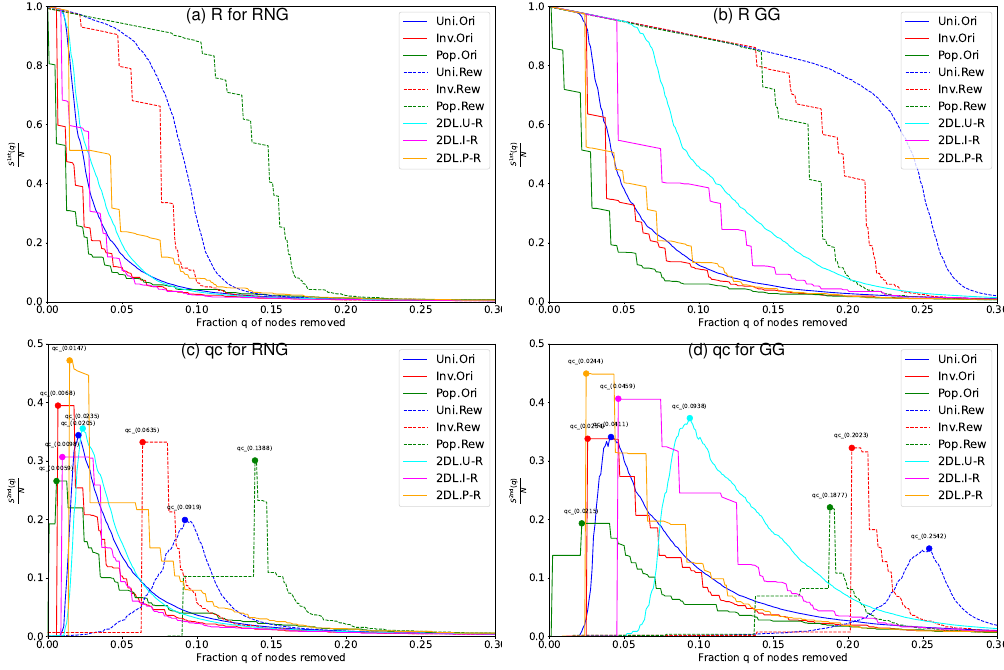}
\caption{Robustness against recalculated betweenness (RB) attacks for Nagoya networks with $N = 1024$ nodes. For both Rew (Randomized networks) and 2DL lines, the rewiring process preserves the original degree distributions. Two measures are applied: (a) (b) $S^{1st}(q)/N$ the relative size of largest connected component, and (c) (d) $S^{2nd}(q)/N$ the critical fraction $q_c$ at the peak of the relative size of second largest component.}
    \label{fig:nagoya_ib_index_1024_3combines}
\end{figure}

\begin{figure}
\includegraphics[width=\textwidth]{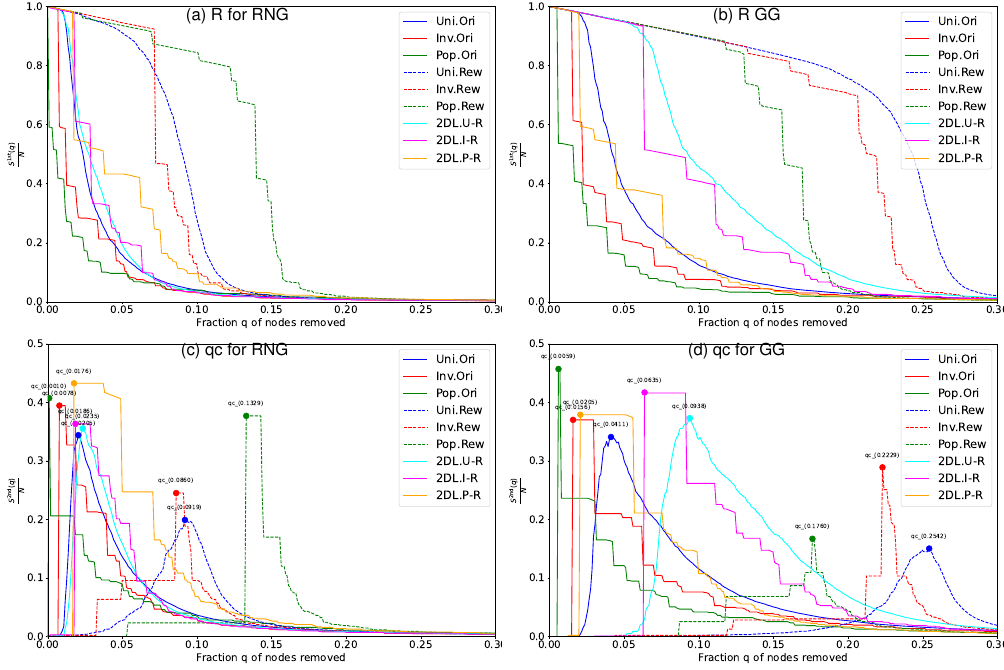}
\caption{Robustness against recalculated betweenness (RB) attacks for Sendai networks with $N = 1024$ nodes. For both Rew (Randomized networks) and 2DL lines, the rewiring process preserves the original degree distributions. Two measures are applied: (a) (b) $S^{1st}(q)/N$ the relative size of largest connected component, and (c) (d) $S^{2nd}(q)/N$ the critical fraction $q_c$ at the peak of the relative size of second largest component.}
    \label{fig:sendai_ib_index_1024_3combines}
\end{figure}

\begin{figure}
\includegraphics[width=\textwidth]{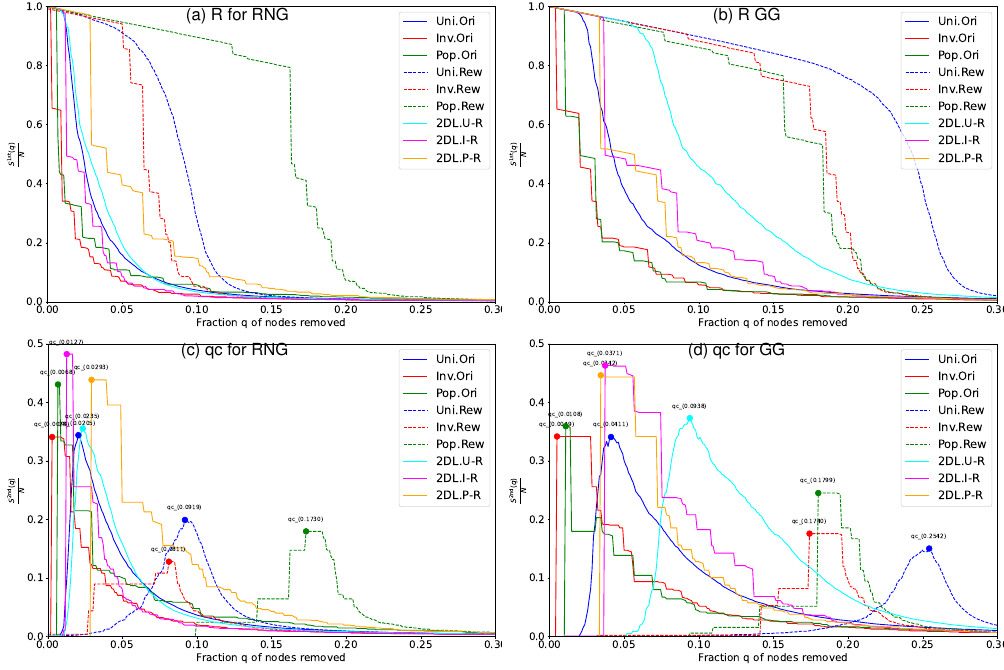}
\caption{Robustness against recalculated betweenness (RB) attacks for Sapporo networks with $N = 1024$ nodes. For both Rew (Randomized networks) and 2DL lines, the rewiring process preserves the original degree distributions. Two measures are applied: (a) (b) $S^{1st}(q)/N$ the relative size of largest connected component, and (c) (d) $S^{2nd}(q)/N$ the critical fraction $q_c$ at the peak of the relative size of second largest component.}
    \label{fig:sapporo_ib_index_1024_3combines}
\end{figure}

\clearpage

\begin{figure}
\includegraphics[width=\textwidth]{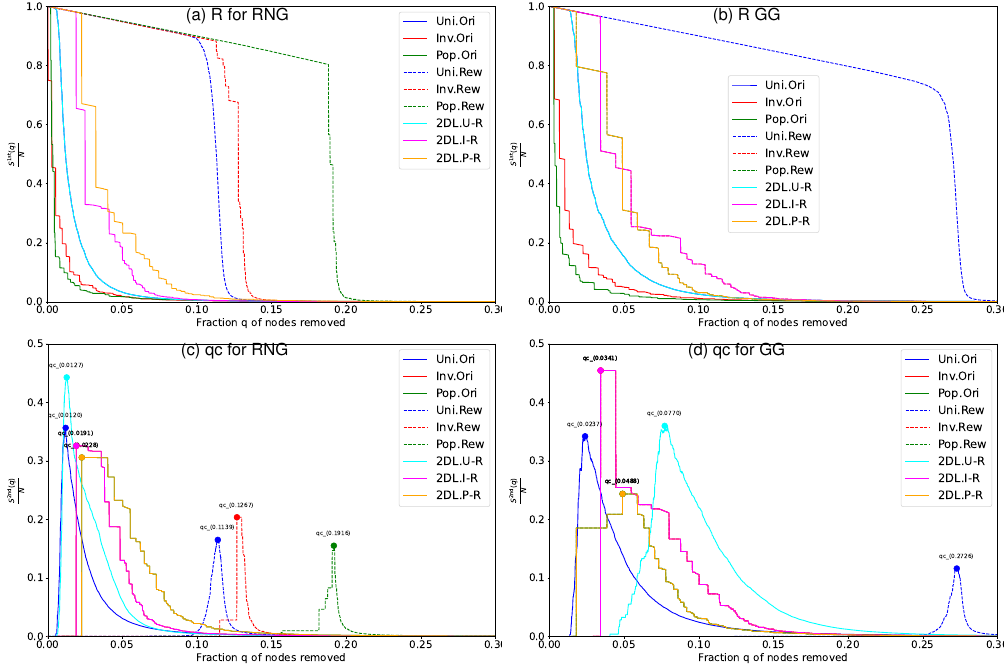}
\caption{Robustness against recalculated betweenness (RB) attacks for Fukuoka networks with $N = 10000$ nodes. For both Rew (Randomized networks) and 2DL lines, the rewiring process preserves the original degree distributions. Two measures are applied: (a) (b) $S^{1st}(q)/N$ the relative size of largest connected component, and (c) (d) $S^{2nd}(q)/N$ the critical fraction $q_c$ at the peak of the relative size of second largest component.}
    \label{fig:fukuoka_ib_index_10000_3combines}
\end{figure}

\begin{figure}
\includegraphics[width=\textwidth]{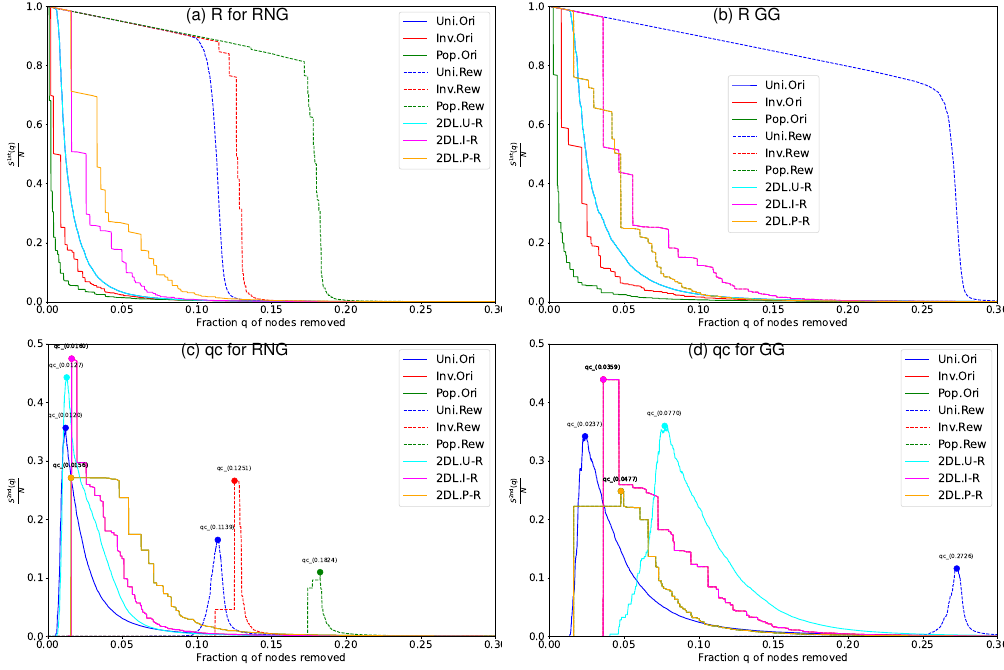}
\caption{Robustness against recalculated betweenness (RB) attacks for Hiroshima networks with $N = 10000$ nodes. For both Rew (Randomized networks) and 2DL lines, the rewiring process preserves the original degree distributions. Two measures are applied: (a) (b) $S^{1st}(q)/N$ the relative size of largest connected component, and (c) (d) $S^{2nd}(q)/N$ the critical fraction $q_c$ at the peak of the relative size of second largest component.}
    \label{fig:hiroshima_ib_index_10000_3combines}
\end{figure}

\begin{figure}
\includegraphics[width=\textwidth]{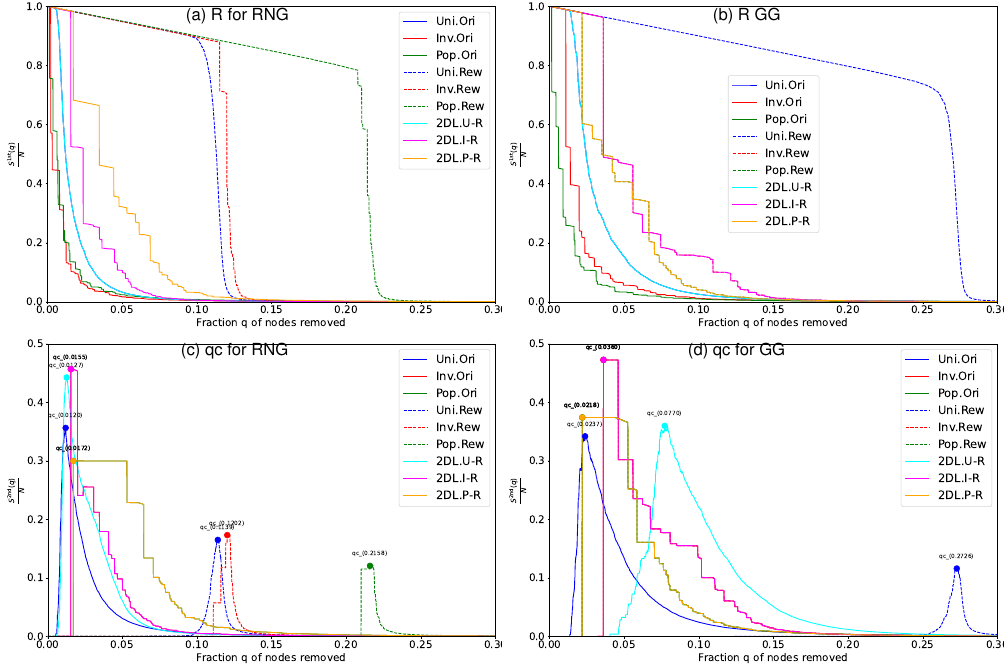}
\caption{Robustness against recalculated betweenness (RB) attacks for Keihan networks with $N = 10000$ nodes. For both Rew (Randomized networks) and 2DL lines, the rewiring process preserves the original degree distributions. Two measures are applied: (a) (b) $S^{1st}(q)/N$ the relative size of largest connected component, and (c) (d) $S^{2nd}(q)/N$ the critical fraction $q_c$ at the peak of the relative size of second largest component.}
    \label{fig:keihan_ib_index_10000_3combines}
\end{figure}

\begin{figure}
\includegraphics[width=\textwidth]{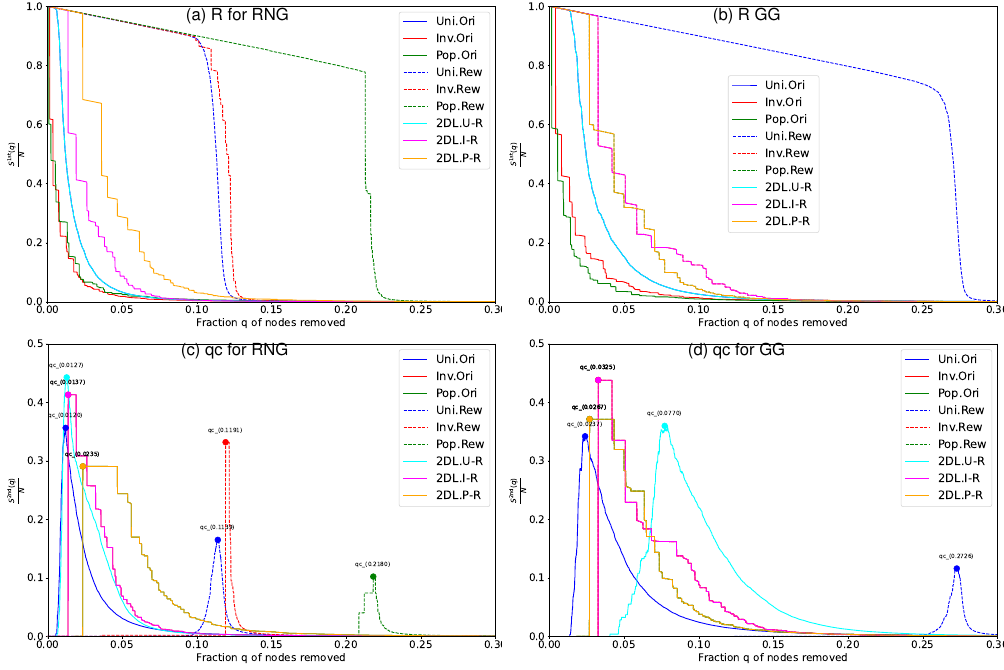}
\caption{Robustness against recalculated betweenness (RB) attacks for Nagoya networks with $N = 10000$ nodes. For both Rew (Randomized networks) and 2DL lines, the rewiring process preserves the original degree distributions. Two measures are applied: (a) (b) $S^{1st}(q)/N$ the relative size of largest connected component, and (c) (d) $S^{2nd}(q)/N$ the critical fraction $q_c$ at the peak of the relative size of second largest component.}
    \label{fig:nagoya_ib_index_10000_3combines}
\end{figure}

\begin{figure}
\includegraphics[width=\textwidth]{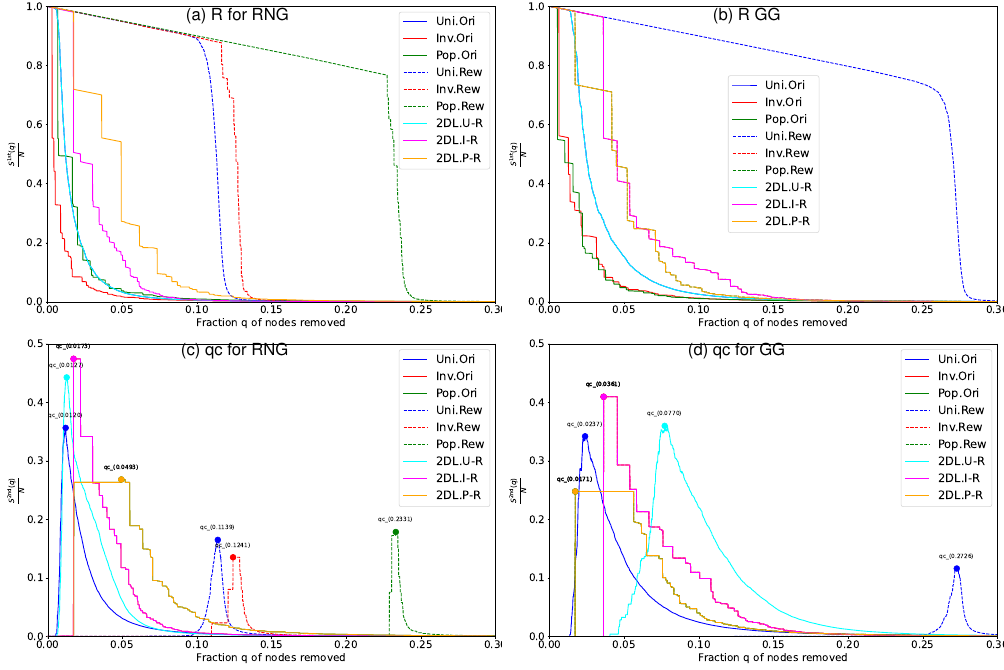}
\caption{Robustness against recalculated betweenness (RB) attacks for Tokyo networks with $N = 10000$ nodes. For both Rew (Randomized networks) and 2DL lines, the rewiring process preserves the original degree distributions. Two measures are applied: (a) (b) $S^{1st}(q)/N$ the relative size of largest connected component, and (c) (d) $S^{2nd}(q)/N$ the critical fraction $q_c$ at the peak of the relative size of second largest component.}
    \label{fig:tokyo_ib_index_10000_3combines}
\end{figure}

\begin{figure}
\includegraphics[width=\textwidth]{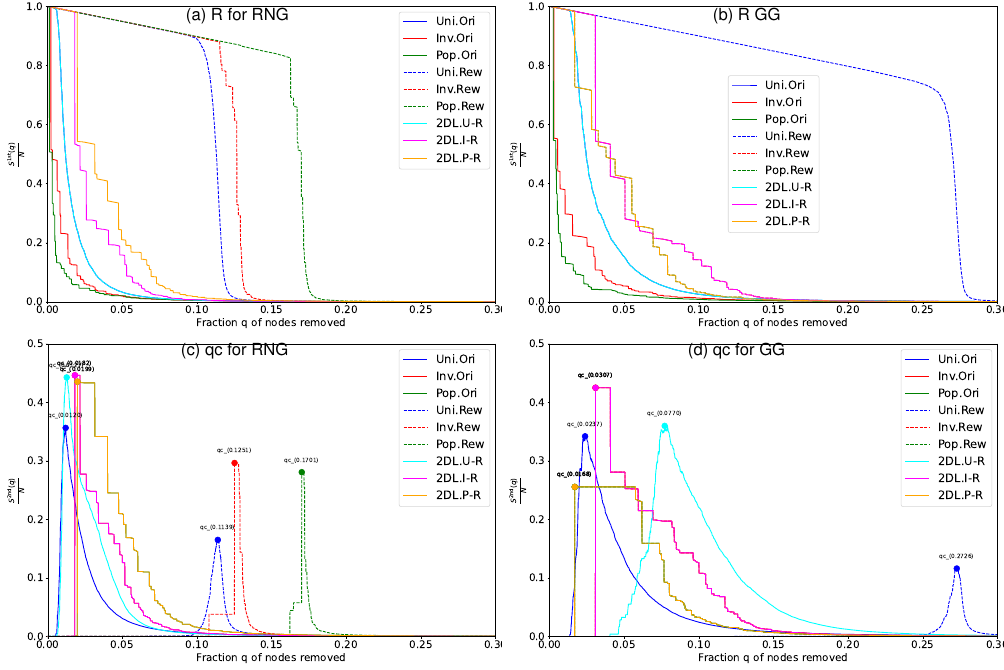}
\caption{Robustness against recalculated betweenness (RB) attacks for Sendai networks with $N = 10000$ nodes. For both Rew (Randomized networks) and 2DL lines, the rewiring process preserves the original degree distributions. Two measures are applied: (a) (b) $S^{1st}(q)/N$ the relative size of largest connected component, and (c) (d) $S^{2nd}(q)/N$ the critical fraction $q_c$ at the peak of the relative size of second largest component.}
    \label{fig:sendai_ib_index_10000_3combines}
\end{figure}

\begin{figure}
\includegraphics[width=\textwidth]{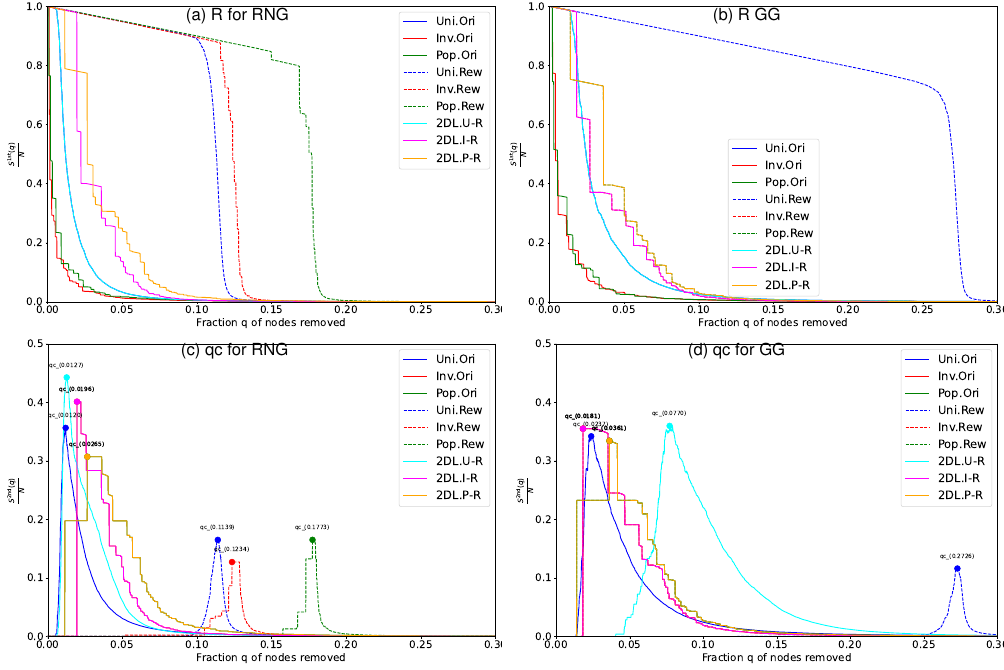}
\caption{Robustness against recalculated betweenness (RB) attacks for Sapporo networks with $N = 10000$ nodes. For both Rew (Randomized networks) and 2DL lines, the rewiring process preserves the original degree distributions. Two measures are applied: (a) (b) $S^{1st}(q)/N$ the relative size of largest connected component, and (c) (d) $S^{2nd}(q)/N$ the critical fraction $q_c$ at the peak of the relative size of second largest component.}
    \label{fig:sapporo_ib_index_10000_3combines}
\end{figure}

\begin{figure}    
\includegraphics[width=\linewidth]{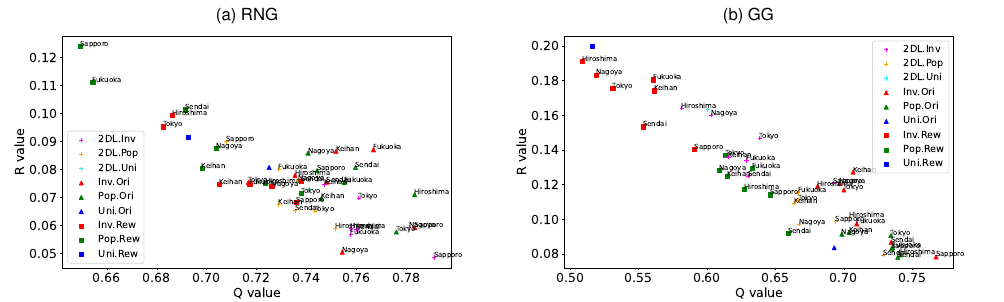}
\caption{Relation between robustness index $R^{RB}$ and modularity $Q$ in networks with $N$ = 100 nodes}
\label{fig:r_to_q_100}
\end{figure}

\begin{figure}   
\includegraphics[width=\linewidth]{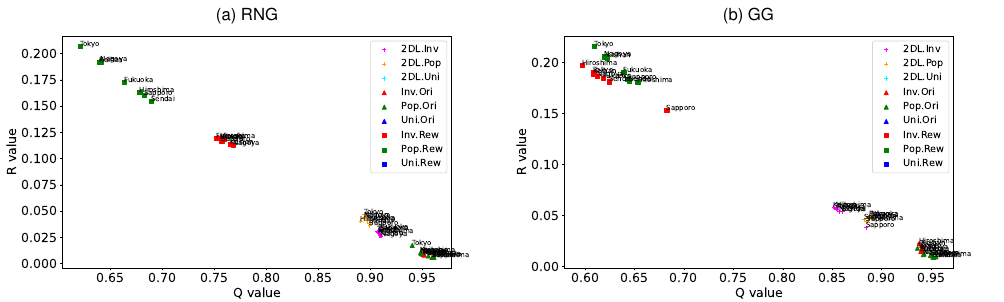}
\caption{Relation between robustness index $R^{RB}$ and modularity $Q$ in networks with $N$ = 10000 nodes}
\label{fig:r_to_q_10000}
\end{figure}

\begin{figure}        
\includegraphics[width=\linewidth]{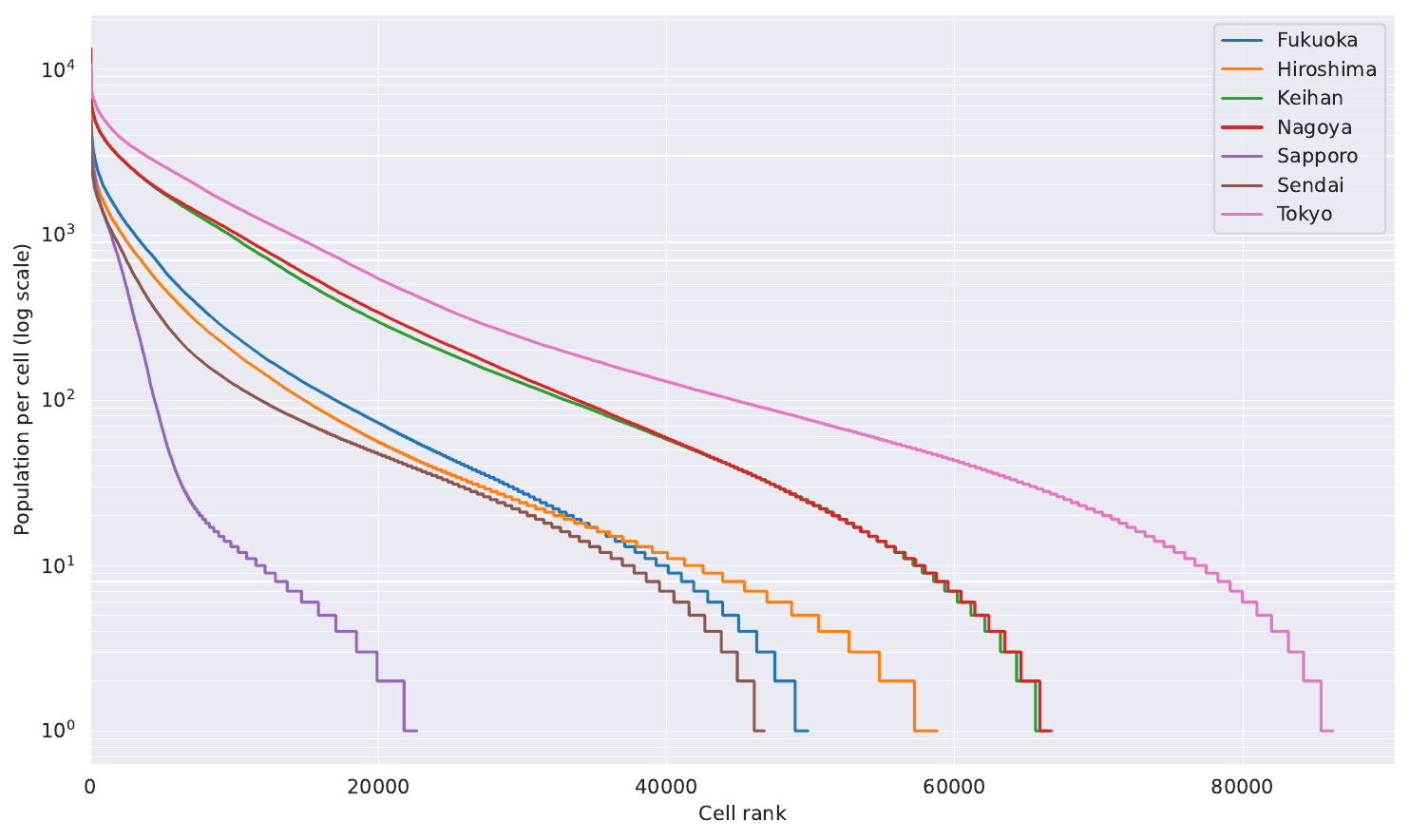}
\caption{Semi-logarithmic plot of population per $500\mathrm{m} \times 500\mathrm{m}$ block meshes in seven major Japanese areas, with meshes sorted in decreasing order of population. Each curve represents an area. The linear decay on the logarithmic scale indicates that a small number of meshes concentrate the majority of the urban population. This heavy-tailed distribution supports the use of rank-based node selections for the Pop. and Inv. networks.}
\label{fig:population_per_cell}
\end{figure}

\clearpage
\begin{table}
\small
\centering
\resizebox{\textwidth}{!}{
\begin{tabular}{|l|ll|ll|ll|ll|}
\hline
\multirow{2}{*}{Cities} & \multicolumn{4}{c|}{$R^{RB}$} & \multicolumn{4}{c|}{$q_c^{RB}$} \\
& \multicolumn{2}{c|}{RNG} & \multicolumn{2}{c|}{GG} & \multicolumn{2}{c|}{RNG} & \multicolumn{2}{c|}{GG} \\
& Inv. & Pop. & Inv. & Pop. & Inv. & Pop. & Inv. & Pop. \\
\hline
Fukuoka & 0.0870$^\triangle$ & 0.0754$^\triangledown$ & 0.0975$^\triangle$ & 0.0837 & 0.0101$^\triangledown$ & 0.0202$^\triangledown$ & 0.0303$^\triangledown$ & 0.0404$^\triangledown$ \\
Hiroshima & 0.0779$^\triangledown$ & 0.0710$^\triangledown$ & 0.1191$^\triangle$ & 0.0796$^\triangledown$ & 0.0101$^\triangledown$ & 0.0101$^\triangledown$ & 0.0404$^\triangledown$ & 0.0202$^\triangledown$ \\
Keihan & 0.0865$^\triangle$ & 0.0697$^\triangledown$ & 0.1273$^\triangle$ & 0.0924$^\triangle$ & 0.0303$^\triangledown$ & 0.0202$^\triangledown$ & 0.0505$^\triangledown$ & 0.0505$^\triangledown$ \\
Nagoya & 0.0505$^\triangledown$ & 0.0858$^\triangle$ & 0.1208$^\triangle$ & 0.0915$^\triangle$ & 0.0202$^\triangledown$ & 0.0202$^\triangledown$ & 0.0505$^\triangledown$ & 0.0505$^\triangledown$ \\
Tokyo & 0.0754$^\triangledown$ & 0.0577$^\triangledown$ & 0.1170$^\triangle$ & 0.0907$^\triangle$ & 0.0101$^\triangledown$ & 0.0101$^\triangledown$ & 0.0505$^\triangledown$ & 0.0202$^\triangledown$ \\
Sendai & 0.0755$^\triangledown$ & 0.0807 & 0.0869$^\triangle$ & 0.0781$^\triangledown$ & 0.0303$^\triangledown$ & 0.0000$^\triangledown$ & 0.0505$^\triangledown$ & 0.0505$^\triangledown$ \\
Sapporo & 0.0592$^\triangledown$ & 0.0796$^\triangledown$ & 0.0784$^\triangledown$ & 0.0832$^\triangledown$ & 0.0101$^\triangledown$ & 0.0202$^\triangledown$ & 0.0202$^\triangledown$ & 0.0303$^\triangledown$ \\
\hline
Uniform & \multicolumn{2}{c|}{0.0807} & \multicolumn{2}{c|}{0.0837} & \multicolumn{2}{c|}{0.0404} & \multicolumn{2}{c|}{0.0808} \\
2D Lattice & \multicolumn{4}{c|}{0.1636} & \multicolumn{4}{c|}{0.0909} \\
\hline
\end{tabular}
}
\caption{Robustness index ($R^{RB}$) and Critical fraction ($q_c^{RB}$) against Recalculated Betweenness (RB) attacks in networks with $N = 100$ nodes for seven major Japanese areas. Higher values indicate greater robustness of connectivity. For $R^{RB}$, values with upper-triangles ($\triangle$) indicate higher robustness than Uni.-based case, while values with lower-triangles ($\triangledown$) indicate lower robustness than Uni.-based case. For $q_c^{RB}$, all values are marked with lower-triangles ($\triangledown$) as they show lower robustness than Uni.-based case.}
\label{tab:combined_rb_100}
\end{table}

\begin{table}
\small
\centering
\resizebox{\textwidth}{!}{
\begin{tabular}{|l|ll|ll|ll|ll|}
\hline
\multirow{2}{*}{Cities} & \multicolumn{4}{c|}{$R^{RB}$} & \multicolumn{4}{c|}{$q_c^{RB}$} \\
& \multicolumn{2}{c|}{RNG} & \multicolumn{2}{c|}{GG} & \multicolumn{2}{c|}{RNG} & \multicolumn{2}{c|}{GG} \\
& Inv. & Pop. & Inv. & Pop. & Inv. & Pop. & Inv. & Pop. \\
\hline
Fukuoka & 0.0237$^\triangledown$ & 0.0188$^\triangledown$ & 0.0379$^\triangledown$ & 0.0253$^\triangledown$ & 0.0039$^\triangledown$ & 0.0020$^\triangledown$ & 0.0137$^\triangledown$ & 0.0039$^\triangledown$ \\
Hiroshima & 0.0234$^\triangledown$ & 0.0162$^\triangledown$ & 0.0379$^\triangledown$ & 0.0219$^\triangledown$ & 0.0068$^\triangledown$ & 0.0010$^\triangledown$ & 0.0108$^\triangledown$ & 0.0039$^\triangledown$ \\
Keihan & 0.0258$^\triangledown$ & 0.0232$^\triangledown$ & 0.0448$^\triangledown$ & 0.0375$^\triangledown$ & 0.0117$^\triangledown$ & 0.0059$^\triangledown$ & 0.0127$^\triangledown$ & 0.0215$^\triangledown$ \\
Nagoya & 0.0262$^\triangledown$ & 0.0233$^\triangledown$ & 0.0548$^\triangledown$ & 0.0370$^\triangledown$ & 0.0068$^\triangledown$ & 0.0059$^\triangledown$ & 0.0254$^\triangledown$ & 0.0215$^\triangledown$ \\
Tokyo & 0.0271$^\triangledown$ & 0.0284$^\triangledown$ & 0.0486$^\triangledown$ & 0.0446$^\triangledown$ & 0.0098$^\triangledown$ & 0.0059$^\triangledown$ & 0.0235$^\triangledown$ & 0.0205$^\triangledown$ \\
Sendai & 0.0271$^\triangledown$ & 0.0199$^\triangledown$ & 0.0433$^\triangledown$ & 0.0292$^\triangledown$ & 0.0078$^\triangledown$ & 0.0010$^\triangledown$ & 0.0156$^\triangledown$ & 0.0059$^\triangledown$ \\
Sapporo & 0.0200$^\triangledown$ & 0.0264$^\triangledown$ & 0.0356$^\triangledown$ & 0.0369$^\triangledown$ & 0.0029$^\triangledown$ & 0.0068$^\triangledown$ & 0.0049$^\triangledown$ & 0.0108$^\triangledown$ \\
\hline
Uniform & \multicolumn{2}{c|}{0.0350} & \multicolumn{2}{c|}{0.0620} & \multicolumn{2}{c|}{0.0205} & \multicolumn{2}{c|}{0.0411} \\
2D Lattice & \multicolumn{4}{c|}{0.0677} & \multicolumn{4}{c|}{0.0303} \\
\hline
\end{tabular}
}
\caption{Robustness index ($R^{RB}$) and Critical fraction ($q_c^{RB}$) against Recalculated Betweenness (RB) attacks in networks with $N = 1024$ nodes for seven major Japanese areas. Higher values indicate greater robustness of connectivity. Values with lower-triangles ($\triangledown$) indicate where the cases of Pop.-based and Inv.-based networks have lower robustness of connectivity than the cases of Uni.-based networks for both RNG and GG.}
\label{tab:combined_rb_1024}
\end{table}

\begin{table}
\small
\centering
\resizebox{\textwidth}{!}{
\begin{tabular}{|l|ll|ll|ll|ll|}
\hline
\multirow{2}{*}{Cities} & \multicolumn{4}{c|}{$R^{RB}$} & \multicolumn{4}{c|}{$q_c^{RB}$} \\
& \multicolumn{2}{c|}{RNG} & \multicolumn{2}{c|}{GG} & \multicolumn{2}{c|}{RNG} & \multicolumn{2}{c|}{GG} \\
& Inv. & Pop. & Inv. & Pop. & Inv. & Pop. & Inv. & Pop. \\
\hline
Fukuoka & 0.0082$^\triangledown$ & 0.0070$^\triangledown$ & 0.0149$^\triangledown$ & 0.0091$^\triangledown$ & 0.0028$^\triangledown$ & 0.0045$^\triangledown$ & 0.0032$^\triangledown$ & 0.0029$^\triangledown$ \\
Hiroshima & 0.0106$^\triangledown$ & 0.0059$^\triangledown$ & 0.0227$^\triangledown$ & 0.0095$^\triangledown$ & 0.0021$^\triangledown$ & 0.0011$^\triangledown$ & 0.0079$^\triangledown$ & 0.0052$^\triangledown$ \\
Keihan & 0.0090$^\triangledown$ & 0.0105$^\triangledown$ & 0.0208$^\triangledown$ & 0.0123$^\triangledown$ & 0.0021$^\triangledown$ & 0.0012$^\triangledown$ & 0.0109$^\triangledown$ & 0.0014$^\triangledown$ \\
Nagoya & 0.0086$^\triangledown$ & 0.0095$^\triangledown$ & 0.0166$^\triangledown$ & 0.0116$^\triangledown$ & 0.0014$^\triangledown$ & 0.0008$^\triangledown$ & 0.0040$^\triangledown$ & 0.0014$^\triangledown$ \\
Tokyo & 0.0092$^\triangledown$ & 0.0174$^\triangledown$ & 0.0189$^\triangledown$ & 0.0180$^\triangledown$ & 0.0030$^\triangledown$ & 0.0068$^\triangledown$ & 0.0063$^\triangledown$ & 0.0053$^\triangledown$ \\
Sendai & 0.0094$^\triangledown$ & 0.0063$^\triangledown$ & 0.0146$^\triangledown$ & 0.0089$^\triangledown$ & 0.0022$^\triangledown$ & 0.0012$^\triangledown$ & 0.0029$^\triangledown$ & 0.0025$^\triangledown$ \\
Sapporo & 0.0059$^\triangledown$ & 0.0074$^\triangledown$ & 0.0108$^\triangledown$ & 0.0112$^\triangledown$ & 0.0011$^\triangledown$ & 0.0019$^\triangledown$ & 0.0039$^\triangledown$ & 0.0019$^\triangledown$ \\
\hline
Uniform & \multicolumn{2}{c|}{0.0173} & \multicolumn{2}{c|}{0.0334} & \multicolumn{2}{c|}{0.0115} & \multicolumn{2}{c|}{0.0229} \\
2D Lattice & \multicolumn{4}{c|}{0.0647} & \multicolumn{4}{c|}{0.0679} \\
\hline
\end{tabular}
}
\caption{Robustness index ($R^{RB}$) and Critical fraction ($q_c^{RB}$) against Recalculated Betweenness (RB) attacks in networks with $N = 10000$ nodes for seven major Japanese areas. Higher values indicate greater robustness of connectivity. Values with lower-triangles ($\triangledown$) indicate where the cases of Pop.-based and Inv.-based have lower robustness of connectivity than the cases of Uni.-based for both RNG and GG.}
\label{tab:combined_rb_10000}
\end{table}

% \subsection{average degree table}

\begin{table}
\small
\centering
\begin{tabular}{|l|ll|ll|}
\hline
\multirow{2}{*}{\textbf{Cities}} & \multicolumn{2}{c|}{RNG} & \multicolumn{2}{c|}{GG} \\
& Inv. & Pop. & Inv. & Pop. \\
\hline
Fukuoka & 2.28 & 2.5 & 3.14 & 2.72 \\
Hiroshima & 2.34 & 2.24 & 3.52 & 2.6 \\
Keihan & 2.3 & 2.34 & 3.34 & 2.82 \\
Nagoya & 2.26 & 2.34 & 3.36 & 2.82 \\
Tokyo & 2.36 & 2.24 & 3.38 & 2.8 \\
Sendai & 2.24 & 2.38$^\triangle$ & 3.14 & 2.5 \\
Sapporo & 2.18 & 2.54 & 2.92 & 2.62 \\
\hline
Uniform & \multicolumn{2}{c|}{2.37} & \multicolumn{2}{c|}{3.56} \\
2D Lattice & \multicolumn{4}{c|}{3.6} \\
\hline
\end{tabular}
\caption{Average degree $\langle k \rangle$ in networks with (N) = 100 nodes for seven major Japanese areas. Higher average degrees mean more links per node in the network. Values with upper-triangles ($\triangle$) indicate where the cases of Pop.-based and Inv.-based networks have higher $<k>$ than the cases of Uni.-based for both RNG and GG.}
\label{tab:average_degree_100}
\end{table}

\begin{table}
\small
\centering
\begin{tabular}{|l|ll|ll|}
\hline
\multirow{2}{*}{\textbf{Cities}} & \multicolumn{2}{c|}{RNG} & \multicolumn{2}{c|}{GG} \\
& Inv. & Pop. & Inv. & Pop. \\
\hline
Fukuoka & 2.67$^\triangle$ & 3.07$^\triangle$ & 3.35 & 3.21 \\
Hiroshima & 2.66$^\triangle$ & 3.01$^\triangle$ & 3.49 & 3.14 \\
Keihan & 2.62$^\triangle$ & 3.21$^\triangle$ & 3.39 & 3.32 \\
Nagoya & 2.61$^\triangle$ & 3.21$^\triangle$ & 3.38 & 3.33 \\
Tokyo & 2.65$^\triangle$ & 3.32$^\triangle$ & 3.42 & 3.41 \\
Sendai & 2.66$^\triangle$ & 2.95$^\triangle$ & 3.31 & 3.19 \\
Sapporo & 2.66$^\triangle$ & 2.99$^\triangle$ & 2.99 & 3.19 \\
\hline
Uniform & \multicolumn{2}{c|}{2.54} & \multicolumn{2}{c|}{3.96} \\
2D Lattice & \multicolumn{4}{c|}{3.96} \\
\hline
\end{tabular}
\caption{Average degree $\langle k \rangle$ in networks with (N) = 10000 nodes for seven major Japanese areas. Higher average degrees mean more links per node in the network. Values with upper-triangles ($\triangle$) indicate where the cases of Pop.-based and Inv.-based networks have higher $<k>$ than the cases of Uni. for both RNG and GG.}
\label{tab:average_degree_10000}
\end{table}

% \subsection{q values table}

\begin{table}
\small
\centering
\begin{tabular}{|l|ll|ll|}
\hline
\multirow{2}{*}{\textbf{Cities}} & \multicolumn{2}{c|}{RNG} & \multicolumn{2}{c|}{GG} \\
& Inv. & Pop. & Inv. & Pop. \\
\hline
Fukuoka & 0.7668$^\triangle$ & 0.7551$^\triangle$ & 0.7094$^\triangle$ & 0.7356$^\triangle$ \\
Hiroshima & 0.7354$^\triangle$ & 0.7833$^\triangle$ & 0.6808$^\triangledown$ & 0.7414$^\triangle$ \\
Keihan & 0.7518$^\triangle$ & 0.746$^\triangle$ & 0.7067$^\triangle$ & 0.7039$^\triangle$ \\
Nagoya & 0.7543$^\triangle$ & 0.7406$^\triangle$ & 0.6968$^\triangle$ & 0.6985$^\triangle$ \\
Tokyo & 0.7167$^\triangledown$ & 0.776$^\triangle$ & 0.7$^\triangle$ & 0.7343$^\triangle$ \\
Sendai & 0.7481$^\triangle$ & 0.7596$^\triangle$ & 0.7345$^\triangle$ & 0.7393$^\triangle$ \\
Sapporo & 0.7834$^\triangle$ & 0.7442$^\triangle$ & 0.7673$^\triangle$ & 0.7348$^\triangle$ \\
\hline
Uniform & \multicolumn{2}{c|}{0.725} & \multicolumn{2}{c|}{0.6929} \\
\hline
\end{tabular}
\caption{Modularity $Q$ in networks with 100 nodes for seven major Japanese areas. Higher values indicate stronger community structures. Values with upper-triangles ($\triangle$) or lower-triangles ($\triangledown$) indicate where the cases of Pop.-based and Inv.-based networks have higher or lower modularity than the case of Uni.-based networks for both RNG and GG. Note the generally higher modularity in Pop.-based and Inv.-based networks compared to Uni.-based networks for both RNG and GG.}
\label{tab:q_value_100}
\end{table}

\begin{table}
\small
\centering
\begin{tabular}{|l|ll|ll|}
\hline
\multirow{2}{*}{\textbf{Cities}} & \multicolumn{2}{c|}{RNG} & \multicolumn{2}{c|}{GG} \\
& Inv. & Pop. & Inv. & Pop. \\
\hline
Fukuoka & 0.9514$^\triangle$ & 0.9603$^\triangle$ & 0.9395$^\triangle$ & 0.9516$^\triangle$ \\
Hiroshima & 0.948$^\triangle$ & 0.9609$^\triangle$ & 0.9379$^\triangle$ & 0.9549$^\triangle$ \\
Keihan & 0.9509$^\triangle$ & 0.9481$^\triangle$ & 0.9406$^\triangle$ & 0.9429$^\triangle$ \\
Nagoya & 0.9506$^\triangle$ & 0.9485$^\triangle$ & 0.9406$^\triangle$ & 0.942$^\triangle$ \\
Tokyo & 0.9499$^\triangle$ & 0.9399$^\triangle$ & 0.9411$^\triangle$ & 0.9361$^\triangle$ \\
Sendai & 0.9508$^\triangle$ & 0.9594$^\triangle$ & 0.9406$^\triangle$ & 0.9525$^\triangle$ \\
Sapporo & 0.9589$^\triangle$ & 0.9553$^\triangle$ & 0.9492$^\triangle$ & 0.9495$^\triangle$ \\
\hline
Uniform & \multicolumn{2}{c|}{0.9397} & \multicolumn{2}{c|}{0.9303} \\
\hline
\end{tabular}
\caption{Modularity $Q$ in networks with 10000 nodes for seven major Japanese areas. Higher values indicate stronger community structures. Values with upper-triangles ($\triangle$) indicate where the cases of Pop.-based and Inv.-based networks have higher modularity than the case of Uni.-based networks for both RNG and GG. Note the generally higher modularity in Pop.-based and Inv.-based networks compared to Uni.-based networks for both RNG and GG.}
\label{tab:q_value_10000}
\end{table}

\begin{table}[!ht]
\centering
\resizebox{\textwidth}{!}{
\begin{tabular}{|l|l|l|l|l|p{5.8cm}|}
\hline
\textbf{Network} & \textbf{Metric} & \textbf{F-value} & \textbf{p-value} & $\eta^2$ & \textbf{Distributional Differences} \\
\hline
RNG & $R^{RB}$ & 37.86 & \boldmath{$3.56 \times 10^{-7}$} & 0.808 & Uni. $>$ Inv. = Pop. \\
\hline
GG & $R^{RB}$ & 40.60 & \boldmath{$2.13 \times 10^{-7}$} & 0.819 & Uni. $>$ Inv. $>$ Pop.\\
\hline
RNG & ${q_c}^{RB}$ & 98.65 & \boldmath{$2.00 \times 10^{-10}$} & 0.916 & Uni. $>$ Inv. $\geq$ Pop. \\
\hline
GG & ${q_c}^{RB}$ & 43.03 & \boldmath{$1.39 \times 10^{-7}$} & 0.827 & Uni. $>$ Inv. = Pop. \\
\hline
\end{tabular}
}
\caption{
One-way ANOVA results evaluating the influence of node distribution (Population-based, Inverse population-based, and Uniform) on the robustness index ($R$) and the critical fraction ($q_c$) against Recalculate Betweenness (RB) attacks ($N=1024$ nodes). \textbf{F-value} indicates the ratio of between-group variance to within-group variance. \textbf{p-value} shows the probability that the observed group differences are due to chance (typically, $p < 0.05$ is considered statistically significant). \boldmath{$\eta^2$} denotes the effect size, indicating the proportion of variance explained by the group factor; conventionally, $\eta^2 > 0.14$ is considered a large effect.
}
\label{tab:rb_anova_summary_1024}
\end{table}

\begin{table}[!ht]
\centering
\resizebox{\textwidth}{!}{
\begin{tabular}{|l|l|l|l|l|p{7.5cm}|}
\hline
\textbf{Network} & \textbf{Metric} & \textbf{F-value} & \textbf{p-value} & $\eta^2$ & \textbf{Distributional Differences} \\
\hline
RNG & $R^{ID}$ & 13.25 & \boldmath{$0.00029$} & 0.595 & Uni. is significantly stronger than Pop. and Inv. \\
\hline
GG & $R^{ID}$ & 6.75 & \boldmath{$0.0065$} & 0.429 & Uni. $>$ Inv. $>$ Pop. \\
\hline
RNG & ${q_c}^{ID}$ & 1.50 & 0.249 & 0.143 & \textit{Not significant; inconsistent rankings across cities} \\
\hline
GG & ${q_c}^{ID}$ & 15.76 & \boldmath{$0.00011$} & 0.637 & Uni. is significantly stronger than Pop. and Inv. \\
\hline
\end{tabular}
}
\caption{
One-way ANOVA results evaluating the influence of node's locations (Population-based, Inverse population-based, and Uniform) on the robustness index ($R$) and the critical fraction ($q_c$) against Initial Degree (ID) attacks ($N = 1024$ nodes). \textbf{F-value} indicates the ratio of between-group variance to within-group variance. \textbf{p-value} shows the probability that the observed group differences are due to chance (typically, $p < 0.05$ is considered statistically significant). \boldmath{$\eta^2$} denotes the effect size, indicating the proportion of variance explained by the group factor. $\eta^2 > 0.14$ is considered a large effect. In this table, the $q_c$ result for networks in RNG shows no significant difference ($p = 0.249$). This may be due to two factors: (1) the higher average degree $\langle k \rangle$ of Pop.-based networks in RNG (see Table~\ref{tab:average_degree_1024}), which offers more alternative paths after node removal; and (2) the presence of grid-like substructures in Pop.-based networks, which can enhance local community and delay critical fragmentation.
}
\label{tab:id_anova_summary}
\end{table}

\begin{table}[!ht]
\centering
\resizebox{\textwidth}{!}{
\begin{tabular}{|l|l|l|l|l|p{7.5cm}|}
\hline
\textbf{Network} & \textbf{Metric} & \textbf{F-value} & \textbf{p-value} & $\eta^2$ & \textbf{Distributional Differences} \\
\hline
RNG & $R^{RF}$ & 2.58 & 0.104 & 0.223 & \textit{Not significant; Pop. and Uni. each ranked highest in 3 cities} \\
\hline
GG & $R^{RF}$ & 22.62 & \boldmath{$0.000012$} & 0.715 & Uni. $>$ Inv. $>$ Pop.\\
\hline
RNG & ${q_c}^{RF}$ & 4.39 & \boldmath{$0.028$} & 0.328 & Uni. is significantly stronger than Pop. and Inv. \\
\hline
GG & ${q_c}^{RF}$ & 5.87 & \boldmath{$0.0109$} & 0.395 & Uni. is significantly stronger than Pop. and Inv. \\
\hline
\end{tabular}
}
\caption{
One-way ANOVA results evaluating the influence of node's locations (Population-based, Inverse population-based, and Uniform) on the robustness index ($R$) and the critical fraction ($q_c$) against Random Failures (RF) ($N = 1024$ nodes). \textbf{F-value} indicates the ratio of between-group variance to within-group variance. \textbf{p-value} shows the probability that the observed group differences are due to chance (typically, $p < 0.05$ is considered statistically significant). \boldmath{$\eta^2$} denotes the effect size, indicating the proportion of variance explained by the group factor; conventionally, $\eta^2 > 0.14$ is considered a large effect. Notably, the $R$ result for networks in RNG shows no significant difference ($p = 0.104$), and the $q_c$ result is marginally significant ($p = 0.028$). These results may be explained by two factors: (1) higher average degree $\langle k \rangle$, which provides alternative paths after node removal (see Table~\ref{tab:average_degree_1024}); and (2) the presence of grid-like substructures in certain area of the network, which enhance local community and delay critical fragmentation.
}

\label{tab:rf_anova_summary}
\end{table}

\begin{table}[!ht]
\centering
\caption{Number of detected communities in networks with 1024 nodes for seven major Japanese areas under Original and 2D Lattice conditions}
\begin{tabular}{|l|llll|llll|}
\hline
\multirow{2}{*}{\textbf{Cities}} & \multicolumn{4}{c|}{\textbf{Original}} & \multicolumn{4}{c|}{\textbf{2D Lattice}} \\
& \multicolumn{2}{c}{RNG} & \multicolumn{2}{c|}{GG} & \multicolumn{2}{c}{RNG} & \multicolumn{2}{c|}{GG} \\
& Inv. & Pop. & Inv. & Pop. & Inv. & Pop. & Inv. & Pop. \\
\hline
Fukuoka   & 25$^\triangle$ & 24$^\triangle$ & 21$^\triangle$ & 23$^\triangle$ & 25 & \textbf{24} & 17$^\triangle$ & 19$^\triangle$ \\
Hiroshima & 22 & 27$^\triangle$ & 20 & 22$^\triangle$ & 28$^\triangle$ & \textbf{23} & 17$^\triangle$ & 20$^\triangle$ \\
Keihan    & 22 & 23$^\triangle$ & \textbf{19} & \textbf{19} & 27$^\triangle$ & \textbf{22} & 16 & 19$^\triangle$ \\
Nagoya    & 24$^\triangle$ & 22 & 20 & 20 & 28$^\triangle$ & \textbf{23} & 16 & 20$^\triangle$ \\
Tokyo     & 23$^\triangle$ & 24$^\triangle$ & \textbf{19} & 21$^\triangle$ & 26$^\triangle$ & \textbf{22} & 16 & 21$^\triangle$ \\
Sendai    & 23$^\triangle$ & 25$^\triangle$ & 21$^\triangle$ & 22$^\triangle$ & 25 & \textbf{23} & 20$^\triangle$ & 19$^\triangle$ \\
Sapporo   & 25$^\triangle$ & 22 & 22$^\triangle$ & 20 & 25 & \textbf{20} & 19$^\triangle$ & 22$^\triangle$ \\
\hline
Uniform & \multicolumn{2}{c}{22} & \multicolumn{2}{c|}{20} & \multicolumn{2}{c}{25} & \multicolumn{2}{c|}{16} \\
\hline
\end{tabular}
\begin{flushleft}
Estimated number of communities in original networks and after node relocation on a 2D lattice (relocated). Values with upper-triangles ($\triangle$) indicate that Pop.-based or Inv.-based has more communities than Uni.-based networks. Values in bold indicate fewer communities than Uni.-based networks in both RNG and GG.
\end{flushleft}
\label{tab:community_number_combined_1024}
\end{table}

\begin{table}[!ht]
\small
\centering
\begin{tabular}{|l|ll|ll|}
\hline
\multirow{2}{*}{Cities} & \multicolumn{2}{c|}{RNG} & \multicolumn{2}{c|}{GG} \\
& Inv. & Pop. & Inv. & Pop. \\
\hline
Fukuoka   & 0.4235 & 0.9489 & 0.4214 & 0.9437 \\
Hiroshima & 0.4163 & 0.9470 & 0.4099 & 0.9405 \\
Keihan    & 0.4634 & 0.8842 & 0.4572 & 0.8406 \\
Nagoya    & 0.4638 & 0.8651 & 0.4500 & 0.8254 \\
Tokyo     & 0.4270 & 0.8825 & 0.4181 & 0.8652 \\
Sendai    & 0.4660 & 0.9559 & 0.4594 & 0.9468 \\
Sapporo   & 0.5217 & 0.9673 & 0.5139 & 0.9602 \\
\hline
Uniform       & \multicolumn{2}{c|}{0.3025} & \multicolumn{2}{c|}{0.3001} \\
2D Lattice    & \multicolumn{2}{c|}{0.9738} & \multicolumn{2}{c|}{0.9200} \\
\hline
\end{tabular}
\caption{
Sparsity Index ($SI(G_w)$) of RNG and GG for different Japanese cities  ($N=1024$). Values are calculated based on normalized edge distances, representing the spatial sparsity of each configuration. Uniform and 2D Lattice cases serve as comparative baselines for sparsity. 2D Lattice has the highest $SI(G_w)$, which may result from long edges between remote nodes.
}
\label{tab:combined_sparsity_1024}
\end{table}

\begin{table}[!ht]
\centering
\small
\begin{tabular}{|l|cc|cc|cc|}
\hline
\textbf{Measures} & \multicolumn{2}{c|}{\textbf{RB}} & \multicolumn{2}{c|}{\textbf{ID}} & \multicolumn{2}{c|}{\textbf{RF}} \\
\textbf{}       & RNG & GG & RNG & GG & RNG & GG \\
\hline
\textbf{Pearson $r$ ($R$)}    & -0.4071 & \textbf{-0.6514} & \textbf{+0.7017} & -0.3532 & +0.0786 & \textbf{-0.7300} \\
\textbf{Pearson $r$ ($q_c$)}  & \textbf{-0.5468} & -0.3181 & +0.3498 & \textbf{-0.5543} & -0.2504 & -0.4458 \\
\hline
\textbf{$p$-value ($R$)}      & 0.1486 & \textbf{0.0116}   & \textbf{0.0052}  & 0.2155  & 0.7894  & \textbf{0.0030}  \\
\textbf{$p$-value ($q_c$)}    & \textbf{0.0430}  & 0.2678  & 0.2202  & \textbf{0.0397}  & 0.3879  & 0.1101  \\
\hline
\end{tabular}
\caption{
Pearson correlation coefficients ($r$) and significance levels ($p$-values) between the sparsity index (SI) and robustness measures ($R$ and $q_c$) in RNG and GG against three attack strategies (RB, ID, and RF). \textbf{Bold values} indicate statistically significant results ($p < 0.05$). A negative correlation ($r < 0$) suggests that higher sparsity (larger SI) is associated with lower robustness, implying that spatially sparser networks are more vulnerable. Conversely, a positive correlation ($r > 0$) indicates that sparsity is associated with higher robustness, which may occur in certain cases where redundant links or grid-like structures compensates for sparsity.
}
\label{tab:pearson_si_robustness}
\end{table}

\begin{table}[!ht]
\centering
\begin{tabular}{|l|p{3.5cm}<{\centering}|p{3.5cm}<{\centering}|}
\hline
\textbf{Metric} & \textbf{RNG} & \textbf{GG} \\
\hline
$r$ & 0.2694 & 0.5930 \\
$p$-value & 0.3315 & \textbf{0.0198} \\
\hline
\end{tabular}
\caption{
Pearson correlation coefficient ($r$) and significance level ($p$-value) between modularity $Q$ and sparsity index SI($G_w$) for RNG and GG. The correlation coefficient $r > 0$ indicating a monotonic increasing relation between modularity $Q$ and spatial sparsity $SI(G_w)$ in RNG and GG, while significant in GG ($p < 0.05$).
}
\label{tab:q_si_corr}
\end{table}

\end{document}